\documentclass[sigconf,10pt]{acmart}

\usepackage[english]{babel}
\usepackage{blindtext}
\usepackage{enumitem}
\usepackage{makecell}
\usepackage[noend,ruled]{algorithm2e}

\usepackage{mathtools}

\renewcommand\footnotetextcopyrightpermission[1]{} 
\setcopyright{none}

\acmDOI{}

\acmISBN{}

\acmConference[MobiCom]{}
\acmYear{2022}
\copyrightyear{}

\acmPrice{}

\begin{document}

\title{Underwater Acoustic Ranging Between Smartphones}

\author{Tuochao Chen,$^\diamond$ Justin Chan,$^\diamond$ Shyamnath Gollakota}
\affiliation{%
  \institution{$^\diamond$Co-primary student authors}
    \institution{Paul G. Allen School of Computer Science \& Engineering}
  \institution{University of Washington, Seattle, WA, USA}
}

\newcommand{\xref}[1]{\S\ref{#1}}

\newcommand{\squishlist}{\begin{itemize}[itemsep=1pt,parsep=2pt,topsep=3pt,partopsep=0pt,leftmargin=0em, itemindent=1em,labelwidth=1em,labelsep=0.5em]}
\newcommand{\squishend}{\end{itemize}}

\newcommand{\squishenum}{\begin{enumerate}[itemsep=1pt,parsep=2pt,topsep=3pt,partopsep=0pt,leftmargin=0em,listparindent=1.5em,labelwidth=1em,labelsep=0.5em]}
\newcommand{\squishsubenum}{\begin{enumerate}[itemsep=1pt,parsep=2pt,topsep=0pt,partopsep=0pt,leftmargin=0em,listparindent=1.5em,labelwidth=1em,labelsep=0.5em]}
\newcommand{\squishenumend}{\end{enumerate}}
\SetKw{Continue}{Continue}

\maketitle
{\bf Abstract---} We present a novel  underwater system that can perform acoustic ranging between commodity smartphones. To achieve this, we  design a real-time  underwater ranging protocol that computes the  time-of-flight between smartphones. To address the severe underwater multipath, we  present a  dual-microphone optimization algorithm that can more reliably identify the direct path. Our  underwater evaluations show that our system has median errors of 0.48--0.86~m at  distances upto 35~m. Further, our system can operate across smartphone model pairs and works in the presence of  clock drifts. While existing  underwater localization research is targeted for custom  hydrophone hardware, we believe that our work breaks new ground by demonstrating a path  to bringing     underwater ranging  capabilities to  billions of existing smartphones, without additional hardware.

\section{Introduction}

Scuba diving is a popular underwater group activity that  millions participate in each year~\cite{stats}. Maintaining close proximity with a dive buddy or  between a dive instructor and the other divers is critical to ensure that the  divers are able to help each other in the event of an emergency such as failure of diving equipment, injury or being trapped by ropes or nets~\cite{failure,net}. This can be challenging in low visibility situations such as turbid waters or during a silt out, which can cause divers to become disoriented and separated from their dive buddy or  instructor~\cite{buddy,muddy}. Estimates show that around 86\% of scuba diving  fatalities occurred when the divers were diving solo or got separated from their buddy~\cite{death1}.

Ideally, we need a solution where the diver's hardware  can compute the distance from their buddy, even in turbid waters, and alert  the diver   if they go beyond a pre-set range.  While dive lights and other signaling hardware are used to get the attention of other dive members~\cite{light} and dive compasses  are used for navigation~\cite{compass}, they do not provide the distance information between divers. 
 Indeed, underwater localization is an  active area of research   for dive computers, sensor networks and robotics~\cite{tracking1, tracking2,tracking3}. These systems use anchor nodes such as  floating buoys with known locations  to estimate  distance  with accuracies of 0.6--3m~\cite{vickery1998acoustic, cario2021accurate, ullah2019efficient}. This prior work, however, requires powerful custom hardware that is not ubiquitous and does not have  economies of scale.

In this paper, we take on an under-explored    research question: can we achieve underwater  ranging between commodity smartphones? Smartphones are universally pervasive across the world and have the economy of scale that is lacking in custom  hydrophone hardware. Furthermore, smartphones are increasingly  being used with diving-proof  cases   for  underwater video logging  and as a replacement for  a dive computer~\cite{diveroid}. In principle, anyone with a smartphone should be able to download a software app before their diving activities and use it to track the distance from their buddy or dive instructor  during underwater activities. Smartphones also have displays and vibration motors that can be used to alert the diver when they go beyond a pre-set distance from their buddy. Achieving this, without the need for custom  hardware, can bring underwater ranging to billions of existing smartphones using only software and has potential use in  tens of millions of dives that occur each year.

\begin{figure}[t!]
\vskip -0.1in
    \includegraphics[width=.35\textwidth]{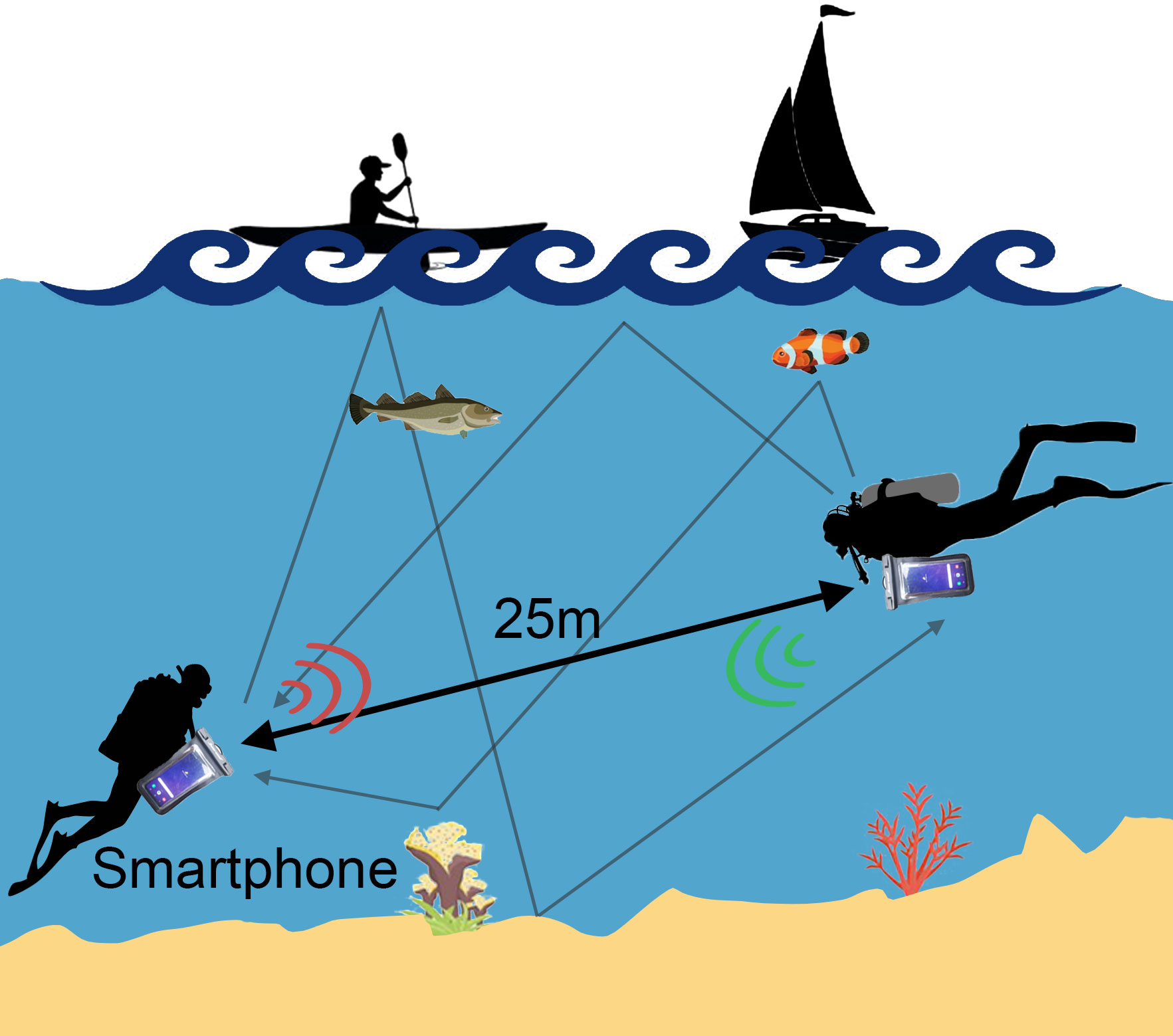}
    \vskip -0.15in
\caption{Underwater acoustic ranging  using smartphones. Severe reflections and echoes from the phone's waterproof case and the environment like the waterbed, surface, and aquatic life create a challenging multipath environment for time-of-flight ranging.}
 \vskip -0.15in
\label{fig:fig1}
\end{figure}

We introduce {\it AquaRanger}, a novel underwater  ranging system that can compute the distance between smartphones using time-of-arrival techniques. At a high level, we use the microphones and speakers that are ubiquitous on smartphones to enable acoustic ranging between these mobile devices. While in-air acoustic ranging has been well-explored in the research community~\cite{peng2007beepbeep,tracko}, acoustic signals propagate at around 4.5x higher speed in water (1500~m/s) than air (330~m/s), which  correspondingly reduces the time-of-flight resolution.  Commodity smartphones and underwater conditions impose three additional  technical challenges.
\squishlist
\item {\bf Water-proof cases create atypical multipath profiles.} A fundamental assumption that in-air acoustic tracking algorithms make is that when two devices are  in line-of-sight  and have no nearby reflecting surfaces, the multipath profile is clean with the  direct path being the strongest and  earliest path~\cite{mao2016cat,millisonic,peng2007beepbeep,tracko}. Water-proof smartphone cases   however  result in atypical multipath  due to interactions with the case material and echoes  within the case. As a result, even when the  two case-enclosed devices are operating in air, are in line-of-sight  and have no nearby reflecting surfaces, the direct path may no longer be the strongest path (Fig.~\ref{fig:case1}a,b).  Further, peaks in the underwater multipath profile can appear before the direct path due to noise and other factors (Fig.~\ref{fig:case1}c). So,  we can not rely on the  assumption that the first non-negligible peak in the multipath profile is the direct path. 

\item {\bf Underwater reflections create challenging multipath.} Underwater channels are known for their challenging multipath environments where the acoustic signals bounce back and forth between the waterbed and
the surface as well as from aquatic animals (e.g., fish) and plants. Further, particles suspended in the water can scatter the signal. The speed of sound also spreads these reflections across time causing a large delay spread~\cite{backscatterlocalization}. Smartphones have a limited  bandwidth  of 3-4~kHz  underwater (see~\xref{sec:dual}) and  a low sampling rate of 44.1~kHz compared to commodity hydrophones. This makes it challenging to disambiguate the direct  path with high resolution on a commodity smartphone.


\item {\bf Buffering  delays and synchronization issues.}  Unlike  hydrophones that are custom designed for underwater operations, smartphone microphones and speakers are primarily designed for speech. In addition to  the underwater frequency response of these sensors  varying  across smartphone hardware, the microphone and speaker buffers on each smartphone are not  synced with each other. These buffers are filled in independently by the OS and so we may not know the timestamps corresponding to the samples in the two buffers. Furthermore, some of these phones can use a different clock for the two buffers resulting in  slightly different  sampling rates. One-way in-air  acoustic tracking systems~\cite{cai2018accurate, youssef2006pinpoint, wang2019millisonic} synchronize the  clocks on the two smartphones in advance but can suffer from clock drifts over time. Two-way systems  like BeepBeep~\cite{peng2007beepbeep} and Swordfight~\cite{zhang2012swordfight} address this  problem but use an RF channel to synchronize between the phones by sharing the buffer timestamp information. However, radio signals  attenuate significantly underwater and cannot be used as an out-of-band synchronization channel.

\squishend

\begin{figure*}[t!]
    \includegraphics[width=\textwidth]{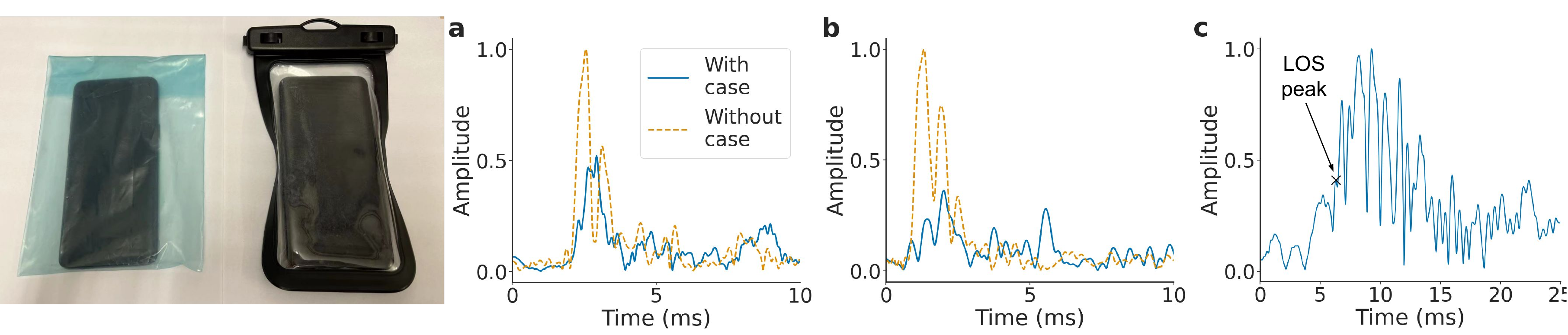}
    \vskip -0.15in
\caption{Acoustic channel estimates  between two smartphones in open air with  (a) the plastic case and (b) the  waterproof pouch. (c) shows an example underwater channel between the smartphones.}
 \vskip -0.15in
\label{fig:case1}
\end{figure*}

We address the above challenges and present a novel  underwater ranging system for commodity smartphones. At a high level, to address the atypical multipath  from  the water-proof case and the severe underwater reflections, we use  the two microphones on either ends of a smartphone. Smartphones use the bottom microphone  close to the user's mouth to capture their voice while the top microphone   closer to their ear is used to perform noise cancellation.  Our key intuition is that the time difference of arrival for the direct path at the bottom and top  microphones is upper bounded by the physical   distance between them. Thus, the sample offset between the direct path in the top and the bottom microphone channels  should be lower than the acoustic propagation time between them. Furthermore, given their spatial separation, the multipath created by the water-proof case is different at the two microphones. Finally, the hardware at the two microphones can  have a different noise profile. Thus, our approach  identifies the   direct path as the earliest non-negligible peaks in {\it both} channels whose sample offset satisfies the physical distance constraint between the two  microphones.  In~\xref{sec:dual}, we  present a light-weight algorithm to perform this dual-microphone  optimization  and identify the direct path.

We also design a  two-way underwater ranging protocol for smartphones that does not require  explicit  information exchange about timestamps.  We use a self-synchronization mechanism  that synchronizes the microphone and speaker buffers at each phone and corrects for the buffering delays introduced by the OS (see~\xref{sec:sync}). The phones auto-adjust their response time to account for the buffering delays to eliminate the need for explicitly exchanging this  information. We show that the errors introduced by our mechanism can be  minimized and  that our approach does not suffer from clock drifts between the two smartphones. 

We implemented our software system in real-time on the Android platform. We also designed a query-response based  protocol to support underwater ranging of multiple devices (see~\xref{sec:multi}).  We evaluated our system in four different underwater environments including a busy swimming pool in the presence of other swimmers, an outdoor fishing dock with a depth of 9~m, a waterfront of a park and a boathouse on a lake  with people fishing and kayaking close to the dock. 

Our findings are as follows:
\squishlist
\item  The median errors achieved by 
our system were 0.48, 0.80 and 0.86~m at 10, 20 and 35~m respectively. The 95\% errors were 0.83, 1.02 and 1.51~m. When the distance increased to 45~m, the median and 95\% errors increased to 1.67 and 2.57~m.
\item {In the busy swimming pool with other swimmers, the median and 95\% errors were 0.35 and 0.84~m. When the phone itself was moving across a 5--18~m distance, the median and 95\% errors were 0.45 and 1.12~m.}
\item  Testing with Samsung Galaxy S9, Google Pixel
3a, and OnePlus 8 Pro shows that the median and 95\% errors between phone model pairs were  0.28-0.54~m and 0.41-0.75~m at 20~m.
\squishend

\vskip 0.05in\noindent{\bf Contributions.} Existing underwater localization research has been focused on custom hardware for sensors, robotics and dive computers~\cite{tracking1, tracking2,tracking3}. In contrast, we present a  real-time underwater  system for ranging between commodity smartphones,   without any custom hardware. Our software-only system  makes multiple  contributions. First, we identify the atypical multipath profiles caused by water-proof smartphone cases that make it challenging to identify the direct path. Second, we introduce a dual-microphone optimization technique that can more reliably identify the direct path in the presence of underwater noise and multipath. Third, we design a two-way underwater ranging protocol that does not require exchanging timestamps and addresses the phone OS buffering delays. Finally, we evaluate our system in various underwater conditions and demonstrate its feasibility.

\section{AquaRanger}

We first present the characterization of the channel resulting from the water-proof case and  then describe our system.

\subsection{Multipath with water-proof case}\label{sec:case}


We  explore the effects of water-proof cases on the acoustic channel. We perform in-air experiments by placing two phones 10~cm away in an open space with no nearby reflectors.  We experiment with two different case material as shown in Fig.~\ref{fig:case1}. We run three groups of experiments: the phones without case,  phones with the first case, and phones with the second case. The first phone sends a 100~ms OFDM symbol between 1--5~kHz to the second phone where we perform channel estimation.  Figs.~\ref{fig:case1}a-b show the estimated channel with and without the two different case materials. The plots show that while the direct path is the strongest without the case, the multipath profile looks a lot more atypical with the cases. This is despite the  devices being close to each other and not having a close-by reflector. In addition, different case  materials   lead to different  channels, which makes it hard to predict and compensate for the water-proof case  since the phone can move around a bit  within it.

In addition to the in-air experiments, we also performed channel estimation underwater. We put the two phones with the first water-proof case into the water. The phones are separated by 20~m and are at a depth of 2~m in a 9~m deep water body. Fig.~\ref{fig:case1}c shows that the multi-path profile is  dense compared to the air scenario; the former is  affected by both the water-proof case and the severe underwater reflections. Further, the multipath profile has peaks that appear before the direct path due to noise and low SNR signals.  

\subsection{System Design}

Our   system uses time of arrival (ToA) techniques to achieve underwater ranging using smartphones. At a high level, the distance $d$ between the sender and receiver can be written as, $d=c\Delta t$, where $c$ is the propagation speed of the signal and $\Delta t$ is the time-of-flight from the sender to the receiver. 

While time-of-flight computation is critical for this approach, accurate ranging also requires knowing the propagation speed in the desired medium. The  speed of sound in underwater scenarios depends on multiple parameters and can be approximated using  Wilson's speed equation~\cite{wilson1960equation}:
$$c = 1449 + 4.6T - 0.055T^2 + 0.0003T^3 + 1.39(S - 35) + 0.017D$$
Here, $T$ is the temperature in degrees Celsius, $S$ is the salinity in PSU, and $D$ is the depth in meters. In practice, the depth limit for {recreational} divers is {40}~m~\cite{scubadepth,scubadepth2}. Prior work~\cite{kuperman2007underwater} shows that at these depths, the maximum change in the speed of sound across different seasons is around $30 m/s$. This results in a  $2\%$ relative error at $1500 m/s$, which is acceptable in our applications. 
Hence,  we can fix  the underwater acoustic speed to calculate distance. While depth $D$ has negligible impact at recreational scuba diving depths, one  can improve  accuracy by configuring the temperature and salinity values that are  known for different water bodies.

In the rest of this section, we first describe our ranging protocol that addresses synchronization issues and then describe our dual-microphone optimization. 

\subsubsection{AquaRanger's ranging  protocol.} \label{sec:sync}

The key challenge in performing time-of-arrival is that different devices have clocks that are not synchronized. Existing in-air acoustic ranging systems use two  pipelines for measuring time-of-flight: one-way ranging and two-way ranging. In  one-way ranging systems~\cite{cai2018accurate, youssef2006pinpoint, wang2019millisonic}, the two clocks at the sender and receiver are  synchronized in advance and the sender transmits its signal at the preset timestamps. However, due to the instability of local oscillator in smartphones, the sampling clock can drift  within a few seconds~\cite{sur2014autodirective}. To mitigate the drifting, prior work~\cite{cai2018accurate, youssef2006pinpoint} use mathematical approaches to compensate for the clock drifting between different devices~\cite{wang2019millisonic}. In our  underwater scenario,   accumulation of error due to clock drifting in  underwater environment  with varying temperature, pressure, humidity, can make it challenging to achieve robust clock drift compensation  for long-term  use~\cite{walls1992environmental}. Researchers have also  proposed  two-way ranging approaches that address clock synchronization between devices~\cite{curtis2014android, peng2007beepbeep,zhang2012swordfight}.  In these approaches~\cite{peng2007beepbeep,zhang2012swordfight}, the sender sends a signal and the receiver replies. By exchanging the timestamps of the receiving  and sending signals at the microphones of the two devices, the distance can be calculated. However, these  systems~\cite{peng2007beepbeep,zhang2012swordfight} exchange the timestamp information via Wi-Fi/Bluetooth that does not work underwater. While one could design and use an underwater communication system, the  available bandwidth is low resulting in significant communication overhead.

\vskip 0.05in\textbf{Our approach.} We present a  two-way ranging protocol for smartphones. Our approach has three main advantages: (1) it does not require information exchange about timestamps, (2) it  addresses the delays introduced by the mobile OS, and   {(3) the ranging error does not accumulate with time.} Our protocol  is illustrated in Fig.~\ref{fig:pipeline1}. At first, Phone A transmits a signal to phone B. {When Phone B detects the beginning of the signal, it  waits  for a preset known time interval $t_{reply}$ and then responds back to Phone A.  After detecting the reply signal, Phone A calculates the interval $t_{send}$ between the beginning of its own preamble and the reply preamble from Phone B. The time of flight $\Delta \tau$ can be calculated as \cite{peng2007beepbeep},} 
\begin{equation}
\label{eq:new}
\Delta \tau = (t_{send} - t_{reply} + \delta_1 + \delta_2)/2
\end{equation}
{
$t_{reply}$ is the time interval between Phone A's preamble  and Phone B's own reply preamble physically arriving at microphone hardware of  Phone B. $t_{sent}$ is the time interval between Phone A's own preamble and Phone B's preamble at  Phone A's microphone. 
Here $\delta_1$ and $\delta_2$ are the propagation delay from the speaker to the microphone on the same device.} The distance between the speaker and the microphone on the same phone is quite small and fixed, and so we can calibrate it for each smartphone model similar to \cite{peng2007beepbeep}. 

The key challenge is in addressing the buffer delays at each of the phones. Specifically, there are two key issues.
\squishlist
\item At each  smartphone, the microphone and speaker buffers  are not synced with  each other~\cite{android1}. These   buffers  are filled in   independently by the OS. As a result, we do not know the  timestamps corresponding to the  samples in the two buffers.
\item While many Android smartphones use the same clock for the microphone and speaker buffers,  data sheets state that they can use a different clock in some phones~\cite{android1}. So the sampling rates may also be slightly different between them.
\squishend

At a high level, we explore  the  low-level audio timing in the OS to achieve self-synchronization between the microphone and speaker buffers on each smartphone.  {The responding phone B, adjusts for its buffer delays between the microphone and speaker to ensure that the response is transmitted at $t_{reply}-\delta_2$ (in other words, ensure the response  arrives at its own microphones at $t_{reply}$). 
Phone A can also self-synchronize and estimate $t_{send}$, which it can use to compute the time of flight using  Eq.~\ref{eq:new}.}


\begin{figure}[t!]
    \includegraphics[width=.47\textwidth]{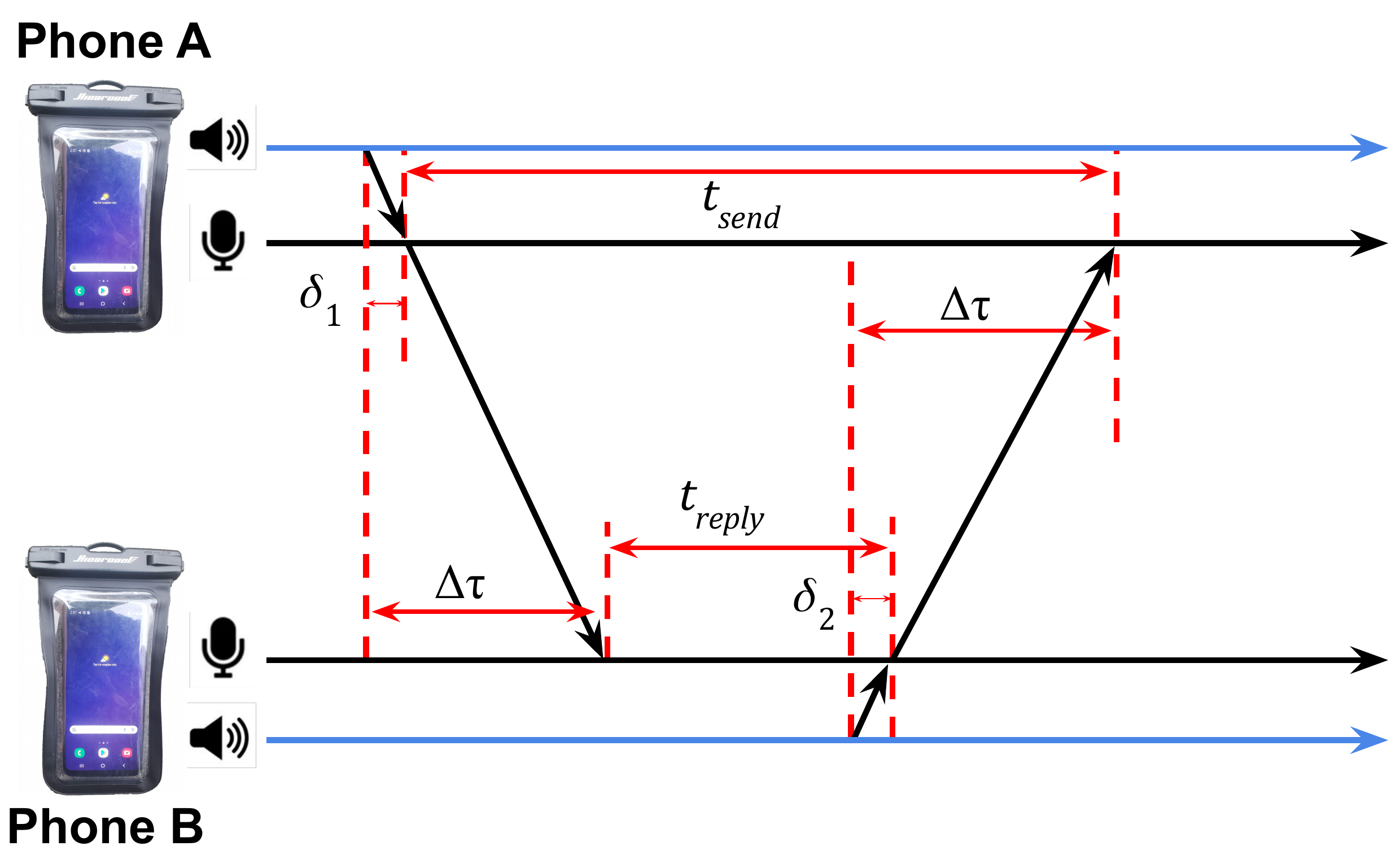}
    \vskip -0.15in
\caption{{\bf Various delays in a two-way ranging system.}}
\vskip -0.15in
\label{fig:pipeline1}
\end{figure}

\vskip 0.05in\textbf{Low-level audio timing.} 
Our goal is to ensure that the replying phone (Phone B) can send a preamble at a precise sample index in the future that  corresponds to  $t_{reply}$, after the signal from the sender arrives at the phone. To map the sample indices to time, we need to look into how Android transmits and records sound at a low level. The low-level OpenSL ES audio library in Android  exposes access to the speaker and microphone audio sample buffers. Specifically, the library provides the ability to directly write audio samples to a future speaker buffer even during speaker playback. During runtime, the library executes a \textit{CallBack} function 
when the microphone buffer is full or the speaker buffer is empty. In this way, we can acquire a continuous data stream for microphone data and another continuous data stream for speaker data.  Thus, the sample indices in the microphone  and speaker streams have a linear relationship with   timestamps:
\begin{equation}
\begin{aligned}
& t_s(n) = n/f_s^s + t_s^0 ,   
& t_m(m) = m/f_s^m + t_m^0 
\end{aligned}
\label{eq:linear}
\end{equation}
Here $m$ and $n$ are the sample indices in the microphone and speaker data streams. $t_s(n)$ is the timestamp when  sample $n$ in the  buffer is send out by the speaker and $t_m(m)$ is the timestamp when  sample $m$ arrives in the microphone  buffer. $t_s^0$ is the initial timestamp when the first sample in the stream is sent by the speaker, and $t_m^0$ is the initial timestamp when the first sample in the stream arrives at the microphone. $f_s^s$ and $f_s^m$ are the  sampling rates for the speaker and microphone, which may not be exactly  our desired sampling rate ($f_s$ = 44.1~kHz). We assume that $f_s^s = f_s/(1-\alpha)$ and $f_s^m = f_s/(1-\beta)$, where $|\alpha| \ll 1$ and $|\beta| \ll 1$.

\vskip 0.05in\textbf{Self-synchronizing speaker and microphone streams.}
As shown in Fig.~\ref{fig:pipeline2}, we do not know the exact timestamp $t_3$ when the preamble arrived at the microphone of Phone B. Instead, we only know the sample index $m_2$ of the recorded preamble in the microphone stream. At the speaker side, we also cannot directly know the exact timestamp $t_4$ when the speaker sends the signal, but we can control the sample index $n_2$  in the speaker stream.  {According to the definition of $t_{reply}$ in Eq.~\ref{eq:new}, we have $t_{reply} = t_5 - t_3 = t_4 + \delta_2 - t_3$ from  Fig.~\ref{fig:pipeline2}. Combining this with  Eq.~\ref{eq:linear}, we have}
\begin{equation}
\begin{aligned}
t_{reply} & = t_4 + \delta_2 - t_3= n_2/f_s^s + t_s^0 + \delta_2 - m_2/f_s^m - t_m^0 
\end{aligned}
\label{eq:reply}
\end{equation}
 The microphone  and speaker buffers work  separately, and there is no guarantee of the relative order between the two buffers. In other words, the initial offsets  $t_s^0$ and $t_m^0$ can be different each time we open the streams. To address this,  once we open the microphone and speaker data streams, we do not close them so as to  keep the offset, $t_s^0-t_m^0$, constant. We write zeros to the speaker stream when we are transmitting nothing to keep the buffer full. Further, after initializing the two streams, the speaker sends a calibration signal (green in Fig.~\ref{fig:pipeline2}) to estimate this offset. We write the calibration signal to the sample index $n_1$ in the speaker stream. Then the microphone stream would receive this calibration signal at sample index $m_1$. 
{The propagation time of the  calibration signal from the speaker to the microphone on Phone B is $(t_2-t_1)$ (i.e., $\delta_2$). By applying Eq.~\ref{eq:linear}, we get:}
\begin{equation}
\begin{aligned}
& t_2-t_1 = m_1/f_s^m + t_m^0 - n_1/f_s^s - t_s^0 = \delta_2 
\end{aligned}
\label{eq:delta}
\end{equation}
\begin{figure}[t!]
    \includegraphics[width=.47\textwidth]{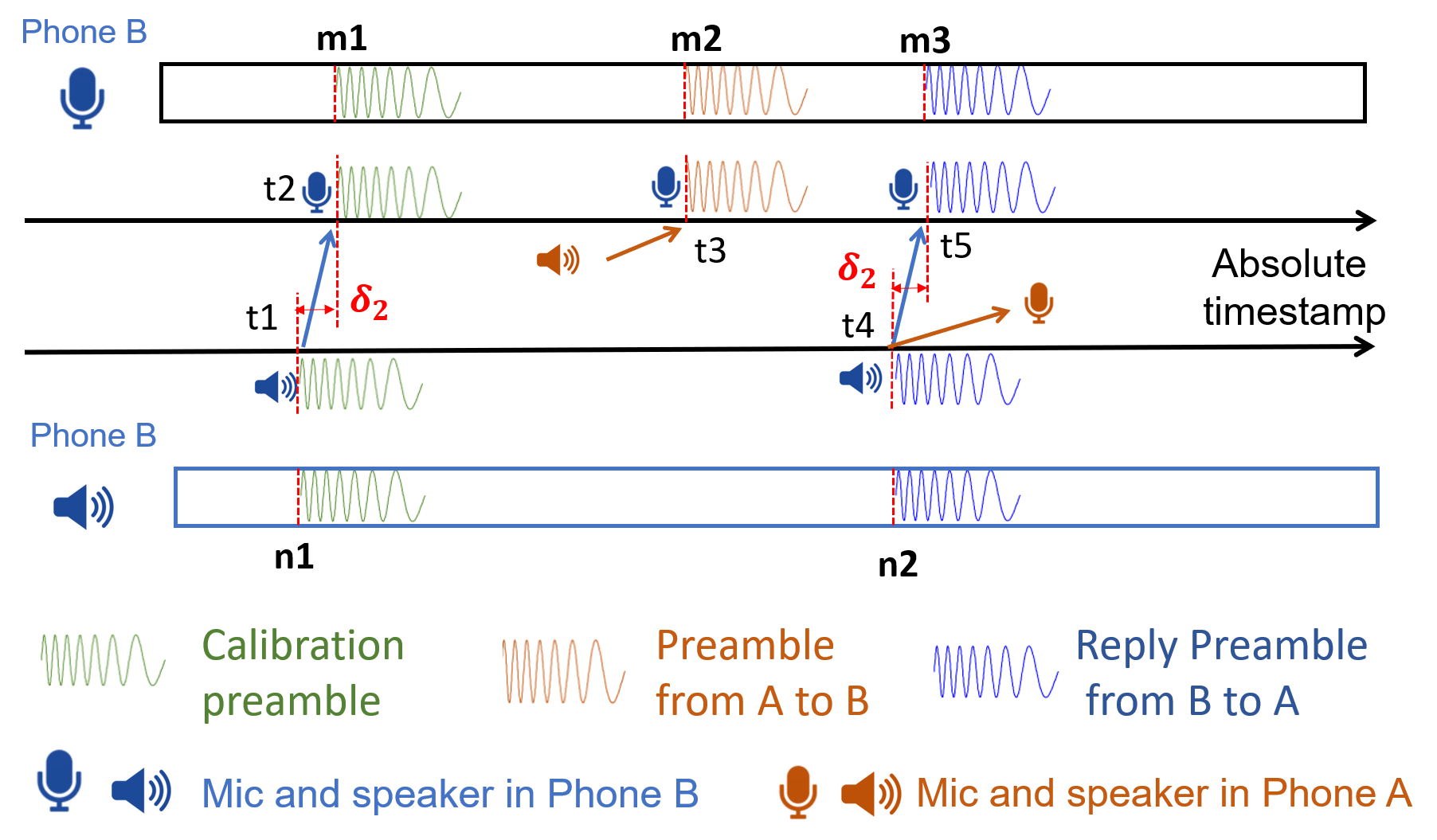}
    \vskip -0.15in
\caption{{\bf Mapping  buffer samples to absolute time.}}
\vskip -0.15in
\label{fig:pipeline2}
\end{figure}
{Now, we compute  the  offset $(n1 - m1)$ between the  microphone and speaker after initial calibration, which can be used for $(t_s^0-t_m^0)$  compensation. }
After initial calibration, our goal is when phone B detects the start of the signal from phone A at the sample index $m_2$ in Fig.~\ref{fig:pipeline2},  phone B will write the reply signal at the sample index $n_2$ in the speaker stream, such that they are separated in time by the desired gap, $t_{reply}^0$, after adjusting for buffer delays. So, by compensating the indices offset acquired from calibration, we set $n_2$ to, 
\begin{equation}
\begin{aligned}
n_2 = m_2 + (n_1 - m_1) + fs\cdot t_{reply}^0, 
\end{aligned}
\label{eq:n2}
\end{equation}
Here $f_s$ is the desired sampling rate. However the real reply interval $t_{reply}$ is shown in Eq.~\ref{eq:reply}. The difference between the real and desired reply times is,
\begin{equation}
\begin{aligned}
& t_{reply} - t_{reply}^0 = n_2/f_s^s + t_s^0 - m_2/f_s^m - t_m^0 - t_{reply}^0 + \delta_2\\
\label{eq:diff}
\end{aligned}
\end{equation}
{By combining Eqs.~\ref{eq:delta} and~\ref{eq:diff} and using the  relationship between $f_s$, $f_s^m$, $f_s^s$, we can rewrite the above equation as,}
\begin{equation}
\begin{aligned}
& t_{reply} - t_{reply}^0 
& = - \alpha t_{reply}^0 + \frac{(m_2-m_1)(\beta - \alpha)}{f_s}
\end{aligned}
\end{equation}

\begin{figure}[t!]
    \includegraphics[width=.23\textwidth]{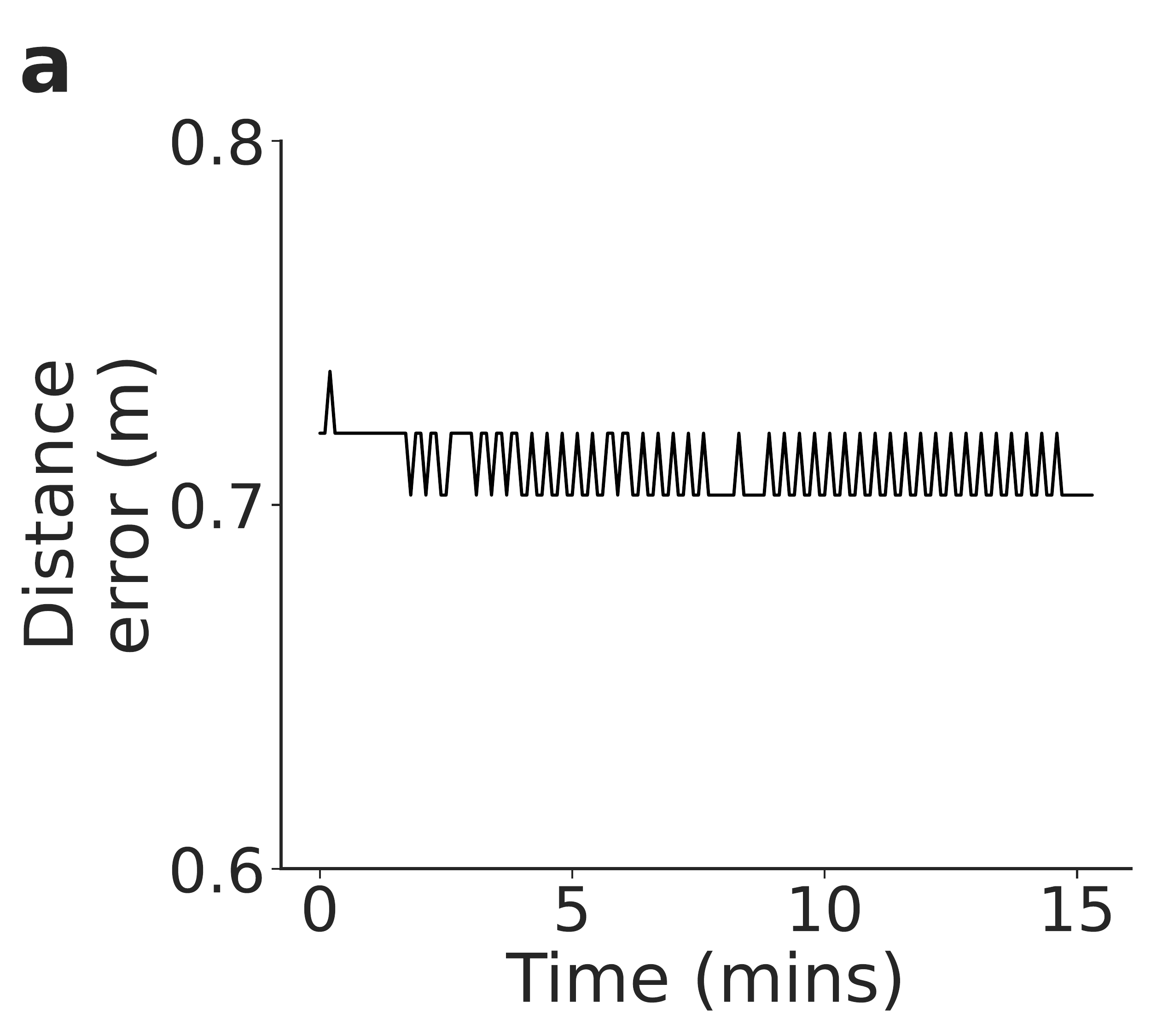}
    \includegraphics[width=.23\textwidth]{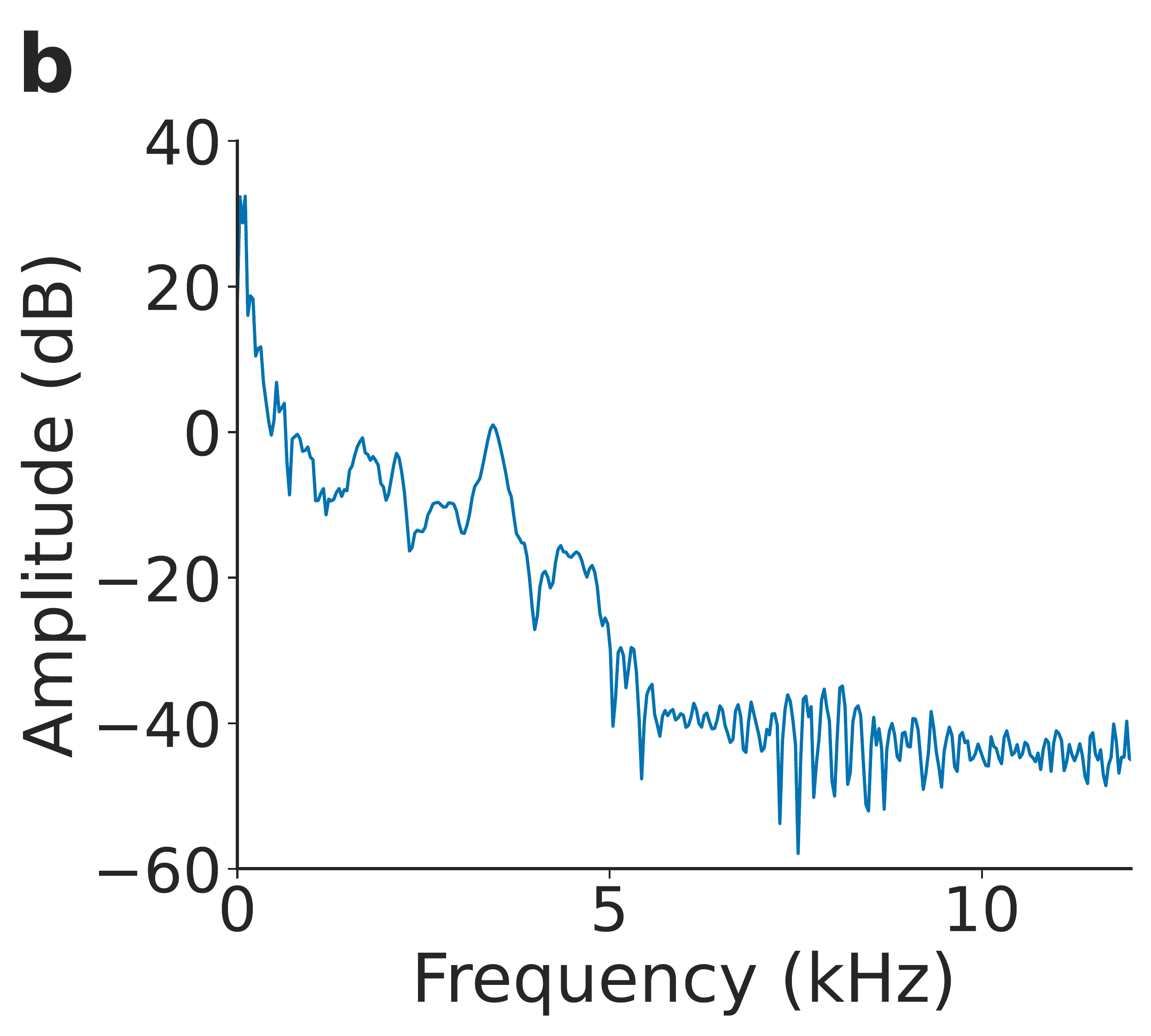}
    \vskip -0.15in
\caption{{\bf (a) Ranging error does not drift over 15~mins. (b) Underwater frequency response between  phones.}}
\vskip -0.15in
\label{fig:drift}
\end{figure}


The key observation here is that the main error source is from the difference between the actual sampling rate and the nominal sampling rate. 
{$\alpha$ for  smartphones is around 1-80 ppm~\cite{guggenberger2015analysis}. Since we perform ranging every second, the timing error in the first item is 1-80 $\mu$s (0.15-12~cm), which is lower than our target accuracies. As for the second term, while  $\beta-\alpha$ is quite small even when the two sensors do not share a common clock, if the calibration is done only once after initialization (green signal in Fig.~\ref{fig:pipeline2}) the time interval between initial calibration and subsequent received signal for Phone B, $m_2 - m_1$ will increase with time, which may result in error accumulation. Our solution to this problem is to use the reply signal from the device for continuous calibration. In this approach,  $m_2-m_1$ is limited to the interval between sending the current reply signal and receiving the next signal (in our design,  ranging is done at a around 1~Hz frequency, so $(m_2-m_1)/fs$ is also around 1s). In this way, the second error item is negligible and would not accumulate for a long period. Fig.~\ref{fig:drift}a shows the  evaluation of our underwater ranging  over a duration of around  15~minutes while the ranging is done once every 6~s. {This evaluation was done between two different smartphone models: the Google Pixel 3A, and Samsung Galaxy S10.} The plot shows that our system does not suffer from error accumulation.  }

\subsubsection{Dual-microphone joint synchronization}\label{sec:dual}
The above description assumes that the  phones can accurately estimate the beginning of the received signals. This is challenging because of the multipath introduce by the water-proof case in underwater environments. \xref{sec:case} showed that the direct path can be severely attenuated and that we can not rely on the assumption that the highest peak or the first non-negligible peak in the multipath profile is the direct path. As shown in Fig.~\ref{fig:dual_chan}, there can be some peaks before the direct path with amplitude greater than the average noise level (red point with "Wrong peak" in Fig.~\ref{fig:dual_chan}). 

\begin{figure}[t!]
    \includegraphics[width=.47\textwidth]{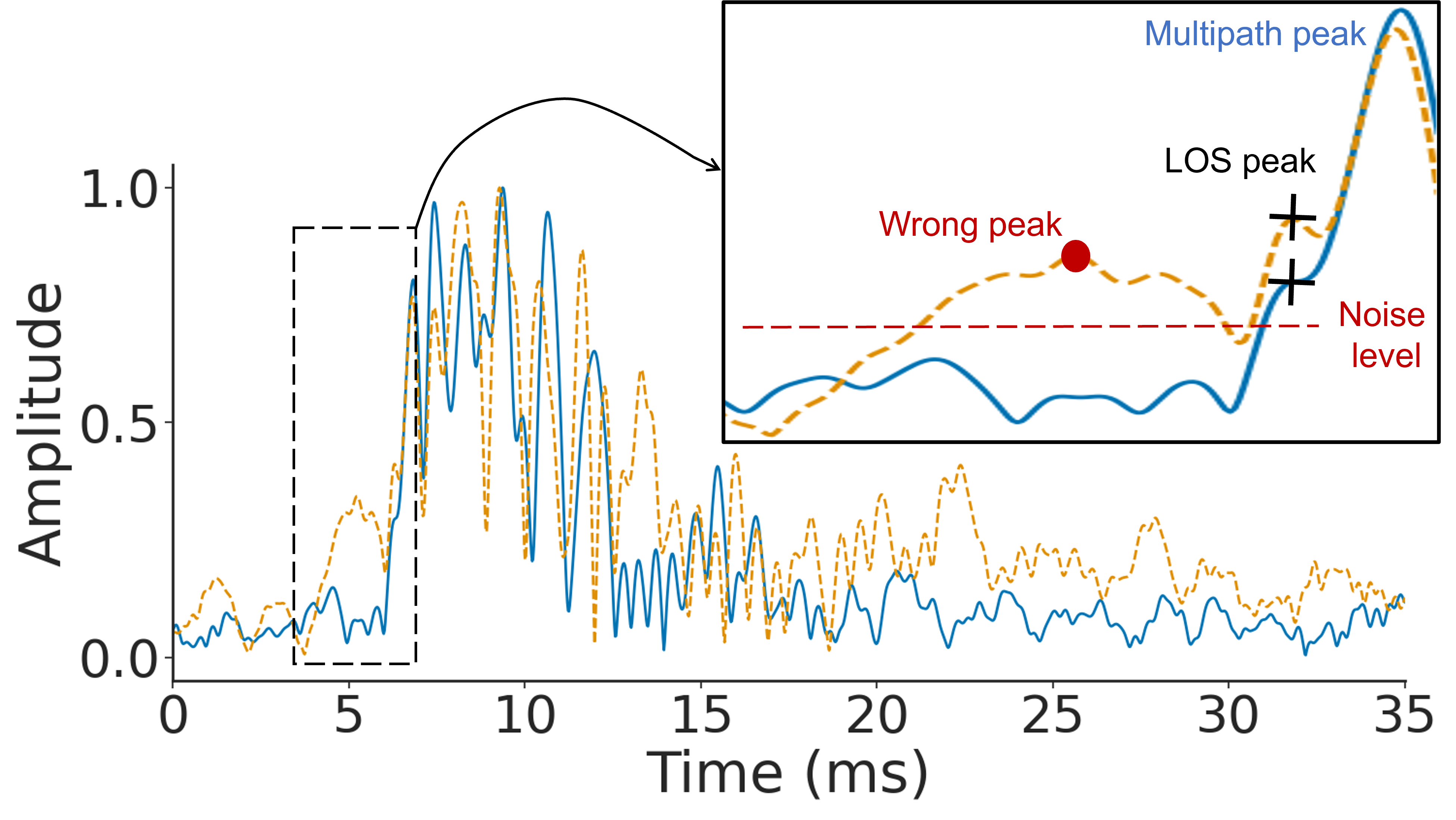}
    \vskip -0.15in
\caption{The yellow and blue curves correspond to the channel estimates at the two microphones.}
 \vskip -0.15in
\label{fig:dual_chan}
\end{figure}

To reduce the probability of picking  these wrong peaks, we use the two microphones on the smartphone. The basic idea of our joint synchronization algorithm  is that the time difference of arrival at the bottom and top  microphones is physically constrained by the distance between them. Thus, the sample offset between the direct path at the top microphone channel and the direct path at the bottom microphone should be lower than the acoustic propagation time between the two microphones (two black cross symbols in Fig.~\ref{fig:dual_chan}). Furthermore, the multipath created by the water-proof case is different at the two microphones. Finally, the two microphones may have a  different noise profile. 

Thus, our algorithm  identifies the direct path as the earliest non-negligible peaks in both channels whose sample offset satisfies the physical distance constrain  between the two  microphones. Specifically, we denote the estimated channel for the  bottom microphone as $h_1(n)$ and  the top microphones as $h_2(m)$, where $n$ and $m$ are the channel tap numbers. Then, we normalized $h_1$ and $h_2$ to be between  0 and 1. We then check whether the sample $n$ in channel $h_1$ is a peak. Next, we estimate the channel noise level for the two microphone channels by calculating the average power in the last 100 channel taps, respectively denoted as $w_1$ and $w_2$. 
\begin{align*}
&\min_{\tau_{LOS}}\quad \tau_{LOS} = (n+m)/2, \quad\forall m,n \in [0, L)\\
& \begin{array}{r@{\quad}r@{}l@{\quad}l}
s.t.&  & h_1(n) > w_1 + \lambda,\\
& & h_2(m) > w_2 + \lambda, \\
& & IsPeak(n, h_1) \cap IsPeak(m, h_2),  \\
& & |n - m| \leq d/c,
\end{array}
\end{align*}
The above equations shows our optimization problem. Here $\tau_{LOS}$ is the delay of the direct path, $n$ and $m$ are  the tap numbers in the channels $h_1$ and $h_2$. $w_1$ and $w_2$ are the estimated noise levels in these two channels. $\lambda$ is a conservative parameter (we set it empirically to 0.2). $d$ is the physical  distance between the two microphone,  $c$ is the  speed of sound  and $L$ is the entire length of the channel (1260 samples). We solve this optimization problem using  Algorithm.~\ref{algo:alg3}. 

\begin{algorithm}[t!]
\caption{Dual-microphone  sync  algorithm} 
\label{algo:alg3}
\DontPrintSemicolon
\For{$n \leftarrow 0$ to $L-1$} {
\If{$h_1(n) < w_1+\lambda$ $\cup$ $\neg$ IsPeak(n, $h_1$)} {
\Continue
}
\For{$m \leftarrow (n - d/c)$ to $(n + d/c)$} {
\If{$h_2(m) > w_2+\lambda$ $\cap$  IsPeak(m, $h_2$)} {
$\tau_{LOS} \gets (m+n)/2$\\
\Return $\tau_{LOS}$
}
}
}
\end{algorithm}

In the rest of this section, we  describe our preamble design and signal processing pipeline for  channel estimation.

{\bf Signal processing pipeline.} 
We use OFDM symbols between 1-5~kHz as the preamble. We use this frequency band given the underwater response of mobile devices (see Fig.~\ref{fig:drift}b). We fill the OFDM bins  with a ZC sequence~\cite{wen2006cazac} which is  phase-modulated and  orthogonal to its delayed version~\cite{zhang2020endophasia}.  ZC-modulated OFDM symbols can achieve much better performance than their well-known counterpart, chirps~\cite{sesia2011lte, cai2018accurate}. We then concatenate eight such identical OFDM symbols and multiply each with a PN sequence with different signs ([-1, 1, 1, 1, 1, 1, -1, 1]), to  increase robustness to  noise~\cite{nasir2010performance}.   
Between  each OFDM symbol, we insert a cyclic prefix to avoid inter-symbol interference.

Our preamble synchronization algorithm at the receiver is composed of there steps: cross-correlation, auto-correlation, and channel estimation. Since the pattern of preamble is known in advance, we can perform cross-correlation between the microphone stream and the preamble pattern. In the presence of a preamble, this results in a correlation peak. However, the height of the correlation peak could vary a lot as the SNR decreases at long distances. Meanwhile, some underwater spiky noise like  bubbles would also cause high peaks  in the cross-correlation. Such noise may lead to a  plenty of false positives.  To address this we use auto-correlation. Since our preamble has  8 OFDM symbols that are encoded with a 8-bit PN sequence, we split the received signal into 8 segments corresponds to the 8 OFDM symbols, multiply each segment by the PN sequence, and apply  correlation among them~\cite{nasir2010performance}. Auto-correlation is helpful because  the spiky noise rarely has such complex encoded pattern (PN sequence) and since the 8 received OFDM symbols suffer from nearly the same multi-path, the correlation value between two received OFDM symbols would be much higher than the correlation value between the received and transmitted symbols. We set a threshold of 0.35 in our design.
 
Due to the severe underwater multi-path profiles, the side-lobe height in the correlation curve is usually higher than the direct path. Hence,  coarse synchronization error based on only correlation is usually hundreds of samples, corresponding to over 6~m  error. To achieve more fine-grained synchronization, we apply channel estimation, where we leverage the channel profiles to identify the direct path. 


There are several channel estimation methods: LS~\cite{kewen2010research}, MMSE~\cite{alihemmati2005channel}, and other MUSIC-like estimator~\cite{cai2018accurate, xue2020push}. MMSE estimators can minimize the mean-square channel estimation error, but  requires some prior knowledge about noise variance and channel covariance~\cite{kewen2010research}. While MUSIC-like estimators could achieve super-resolution channel profiles, the signal space decomposition is difficult due to the extremely dense underwater channel and it has a high computational complexity. Therefore, we use an LS channel estimator.

Specifically, based on the coarse synchronization of cross-correlation and auto-correlation, we segment out 8 received OFDM symbols $y_1, y_2, \dots, y_8$ from the microphone stream. Then we apply  FFTs on these 8 symbols to get $Y_1, Y_2, \dots, Y_8$. We denote the FFT of the original OFDM symbol before multiplication with PN sequence by $X$ and denote the PN sequence by $PN_1, PN_2, \dots, PN_8$.  The channel model can be written as $Y_i(k) = H(k)(PN_i\cdot X(k)) + N_i(k)$, where $k$ represents the $k^{th}$ frequency bin. The estimated channel is $\hat{H}(k) = \frac{1}{8}\sum_{i=1}^8 \frac{1}{PN_i}\cdot X(k)^{-1} Y_i(k)$. Finally, we apply an IFFT to convert the estimated channel to the time-domain. 

\begin{figure}[t!]
    \includegraphics[width=.47\textwidth]{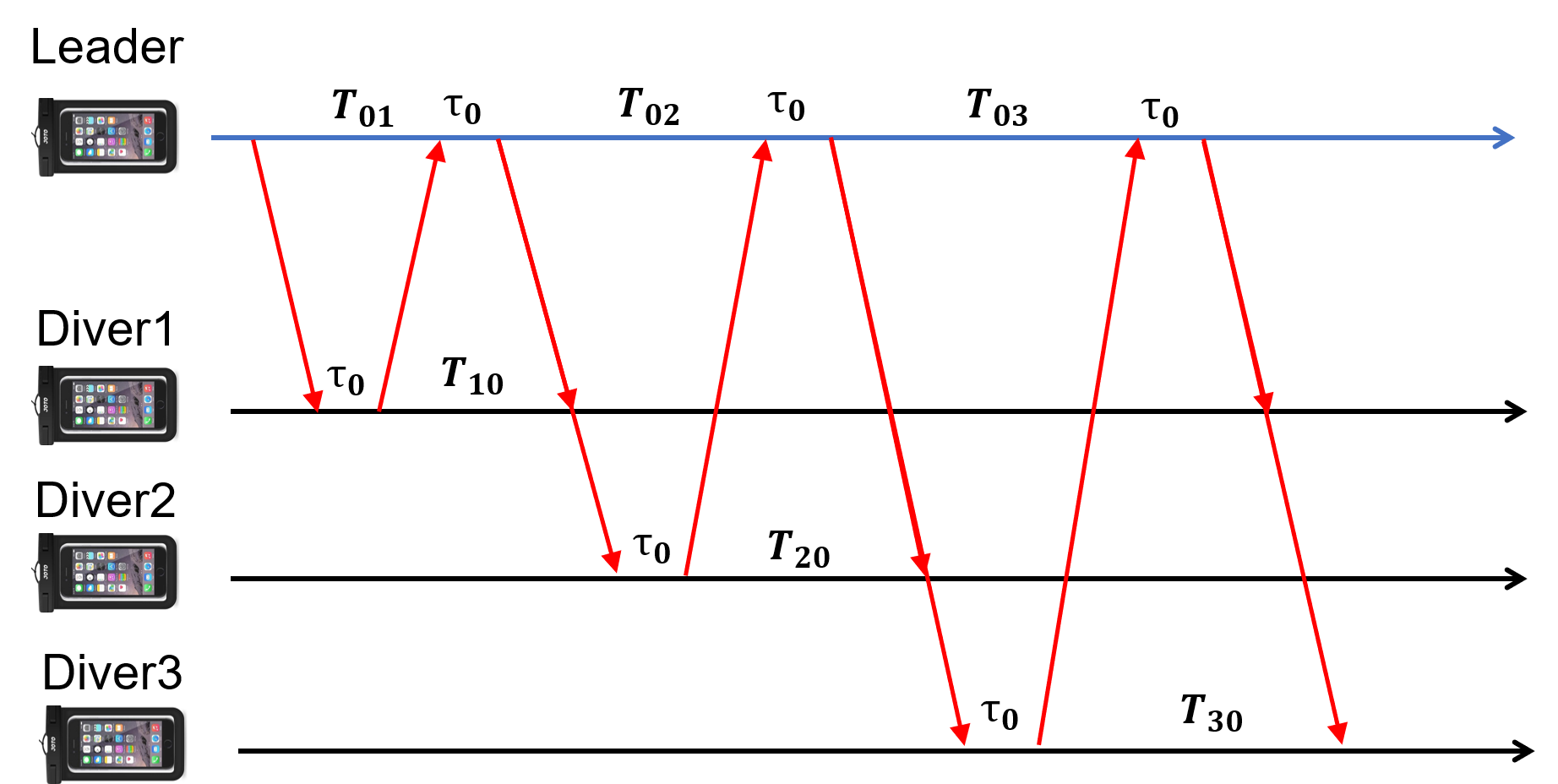}
    \vskip -0.15in
\caption{{\bf Supporting multiple diver devices. Device $i-1$ uses the the leader  query for device $i$ and its own prior response to estimate its distance from the leader.}}
\vskip -0.15in
\label{fig:users}
\end{figure}

\subsection{Supporting multiple divers}\label{sec:multi}
 To support ranging of multiple devices in a dive group at the same time, we design a query-response based system where each of the devices is queried by the leader device in a round-robin manner.  The dive instructor in the group is assigned as the leader device which is the  sender and the other divers  are the repliers. Before diving into the water, each device is  designated a unique ID. 

While our ranging system is designed to compute the distance value at the sender device (i.e., the leader), adjacent query messages from the leader device can also be  used to compute the distances at the replier devices (i.e., other divers). Specifically, say the 
 reply time for each replier and sender is $\tau_0$.  The leader first sends a preamble concatenated with the first diver's ID. When the phones  receive the preamble, they  check the ID information. Diver 1  sends back its preamble after a reply time $\tau_0$, while the devices at the other divers stay silent. When the leader receives the reply preamble, the leader can estimate the distance from the first diver. Then the leader sends the second preamble with the second diver's ID after the reply time $\tau_0$. When diver 1 overhears this  preamble with the second diver's ID, the first diver can also compute the distance to the leader as,  $d_{10} = \frac{T_{10} - \tau_0}{2c}$. Meanwhile, the second diver would send back its reply preamble after $\tau_0$ and the other phones remain silent. When the leader receives the reply from the second diver, it would calculate the distances to it and send the third preamble after $\tau_0$. In this way, after sending the preamble to each diver, the leader can acquire the distance of each diver. Meanwhile each diver also knows its  distance to the leader by processing the 
 query for the next diver. We repeat this process in a round-robin manner to compute the distance at the divers and the leader device. Note that given the time-division nature of supporting multiple divers, this proportionally reduces the rate at which the distance information can be computed for each diver. One may consider designing code division multiplexing mechanisms to query all the divers concurrently but this reduces the energy in each of diver queries and  the range. Thus, our design is focused on the time-division based query/response protocol to support multiple diver devices. 
 

\begin{table}[!t]
\centering
\begin{tabular}{|c|c|}
\hline
{\bf \makecell{Operation}} & {\bf \makecell{Runtime (ms)}}\\ \hline
Cross-correlation & 59.8 ± 10.6 \\ \hline
Auto-correlation & 28.4 ± 4.7 \\ \hline
Channel estimation & 1.9 ± 0.6 \\ \hline \hline
Total (for 500~ms buffer) & 90.1 ± 11.2 \\ \hline
\end{tabular}
\caption{{\bf  Statistics  across 100 measurements. Runtime is measured for computation over a 500~ms buffer.}}
\vskip -0.2in
\end{table}

\begin{figure*}[t!]
    \includegraphics[width=\textwidth]{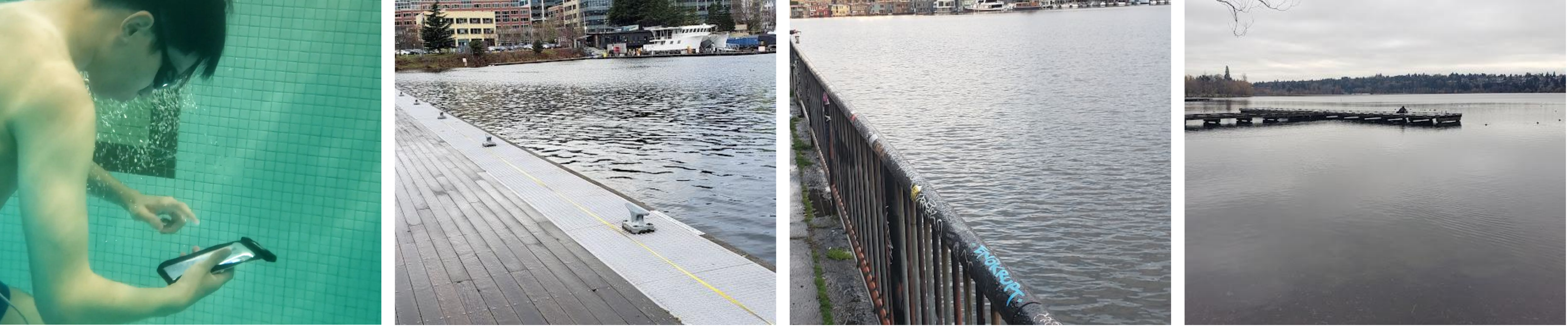}
    \vskip -0.15in
\caption{Different underwater  scenarios. (a) Swimming pool: a busy environment with human mobility and occlusion. (b) Dock: an outdoor location with boats, seaplanes, and animals. (c) Viewpoint: At a park with a shallow water depth of 1.5~m. (d) Boathouse: A busy dock with people fishing and kayaking.}
\vskip -0.15in
\label{fig:locs}
\end{figure*}

{We encode the ID of a user using a simple frequency-based  mechanism. Specifically, we map the user's ID to a unique frequency from  the 16 frequencies: $1500,1535,1570,...,2060~Hz$. The user's ID is appended to the preamble as a 100~ms  frequency tone at their assigned frequency. We empirically select this frequency range as we find that the SNR of the tones are strongest in this range. We select a spacing of 35~Hz to account for frequency shifts that could occur from the Doppler effect as a result of mobility. To decode the user ID, we compute the SNR at all 16 frequency bins and select the bin with the maximum SNR. }

\section{Implementation and Evaluation}

Our ranging system has been implemented to run in the user space on Android devices. Our implementation performs the cross-correlation, auto-correlation and channel estimation algorithm within an average runtime of 90.1 ± 11.2~ms over a 500~ms audio sample buffer. We use the fftw3 library to perform FFTs in realtime. We use the OpenSL ES audio interface on Android which provides low-level access to the speaker and microphone audio buffers. In our system, the replying phone is configured to send a reply, one second after it has detected a preamble from the sender.  


\subsection{Results}
We evaluated our  system in the four  environments in Fig~\ref{fig:locs}. 

\squishlist
\item \textit{Swimming pool.} The length of the water here is around 23~m. The depth of the swimming pool varied from around 1 to 2.5~m. This is a busy location with people swimming laps or engaging in other recreational water activities.
\item \textit{Dock.} This outdoor location has a length of around 50~m with a depth of 9~m. Boats and seaplanes would frequently sail or dock at this location with aquatic plants and animals.
\item \textit{Viewpoint.} By the waterfront of a park with a length of 40~m. The water had a depth of around 1 to 1.5~m. This is a busy location with boats and strong currents. 
\item \textit{Boathouse.} Fishing dock by the lake with a horizontal distance of 30~m. The lake had a depth of 5~m. This is a busy location with people fishing and kayaking close to the dock.
\squishend

\subsubsection{Accuracy versus device separation.} 
We first present our evaluation of the range of our system along the dock of a lake with an average water depth of 9~m. We performed  experiments using two Samsung Galaxy S9 phones set to transmit at the maximum speaker volume. The phones were set to transmit using the speaker at the bottom of the device, and receive using the microphones at the bottom and top of the device. We designate the two phones as being either a sender or replier. In this evaluation, the experiment was repeated in each location every six seconds, where the sender was configured to send a message to the replier. The replier is then set to continuously run the preamble detection algorithm and respond with its reply after it has received a preamble from the sender. At each distance, the sender and replier are set to exchange messages up to a maximum of 60 times. The measurements were divided into roughly three sessions, where after 20  measurements, the phones were removed from the water and submerged again.

\begin{figure}[t!]
    \includegraphics[width=.48\textwidth]{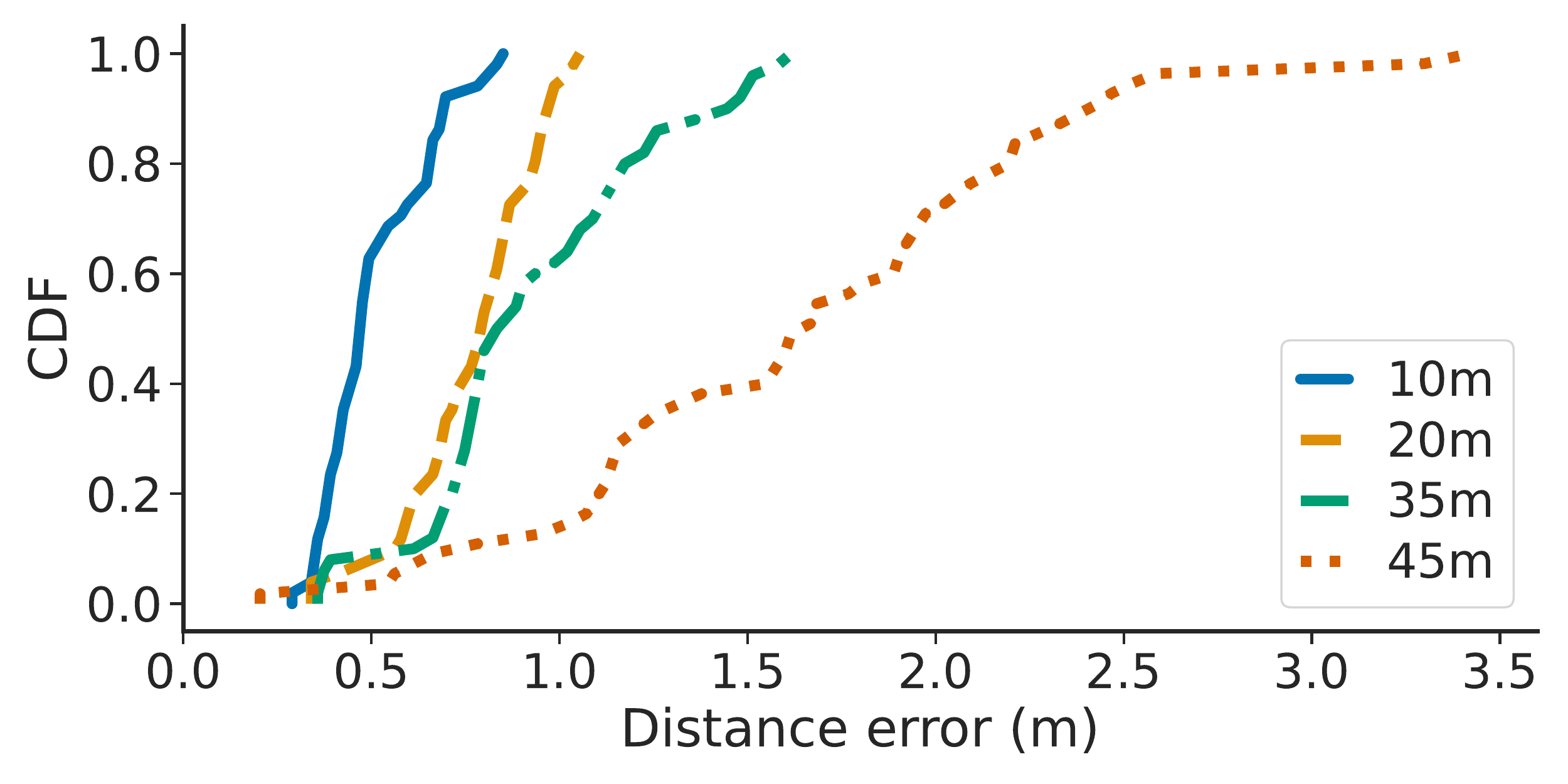}
    \vskip -0.15in
\caption{Ranging accuracy versus device separation. CDF of absolute error as a function of separation. }
\vskip -0.15in
\label{fig:dist}
\end{figure}

The phones were enclosed in a  pouch (Hiearcool waterproof phone pouch~\cite{pouch}) and attached to a selfie stick and telescopic extension pole, which was used to submerge the phones to a depth of 2.5~m. The selfie stick and extension pole were attached using waterproof tape and zip ties. This setup was chosen so that the phone's position and angle could be controlled. We used a tape measure to mark different measurement distances up to a maximum of 45~m along the dock. We set  1500~m/s as the speed of sound underwater. 

Fig.~\ref{fig:dist} shows the CDF of the absolute error obtained by our system for four distances of 10, 20, 35 and 45~m. The plot shows that the median error of our system is 0.48~m at 10~m, 0.80~m at 20~m, 0.86~m at 35~m, and increases to 1.67~m at 45~m. The corresponding 95th percentile errors were 0.83~m, 1.02~m, 1.51~m and 2.57~m. The error in distance increases with separation between the devices because the signal strength is lower at larger separation. This results in an increased probability of confusing the direct path with the multipath and noise. Thus, the conservative range of our system is around 35~m. We note however that this scaling of error with separation is desirable in practice since higher resolution is required when the devices are close by than when they are farther. A 2.5~m error when the separation is 45~m is still meaningful and provides useful information.

We also note that in this experiments, we use a preamble with a length of 316~ms. When we use a longer preamble with a length of 479~ms, the median error of the system decreases to 1.63~m at 45~m respectively. This shows a tradeoff where increasing the preamble length can reduce the error at longer ranges but will reduce the rate at which ranging can be achieved by our system. We note that in both settings, we can provide a location value at a rate of 1~Hz, which is sufficient for our target underwater ranging applications.


\begin{figure}[t!]
    \includegraphics[width=.48\textwidth]{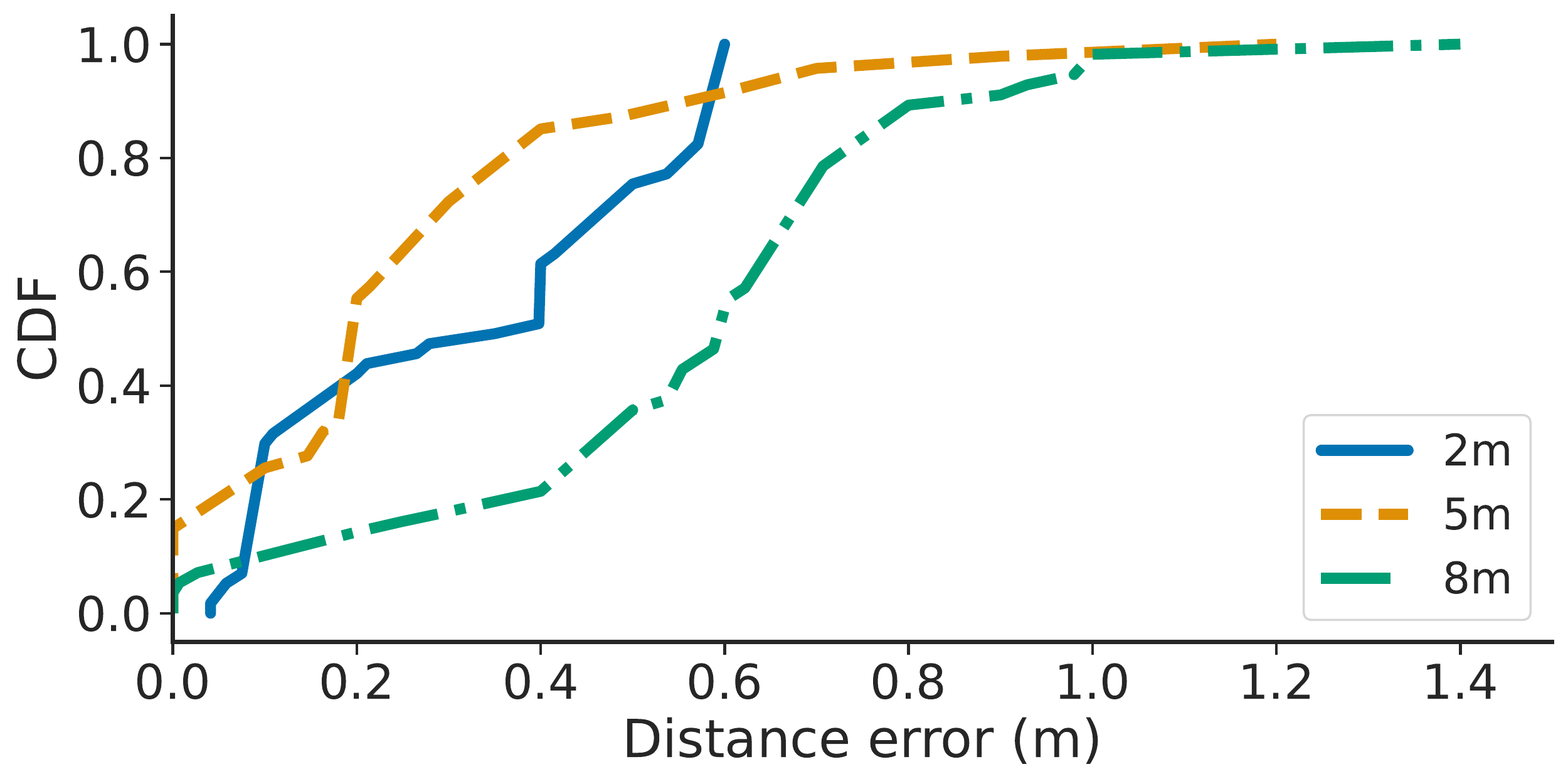}
    \vskip -0.15in
\caption{Effect of device depth. Errors for different depths with  devices  separated horizontally by 18~m.}
\vskip -0.15in
\label{fig:depth}
\end{figure}
\subsubsection{Accuracy versus depth.} 
We next present our evaluation of the ranging accuracy when the smartphones are placed at different water depths at the dock location which had a total depth of 9~m.  We lowered the smartphones into the water using ropes marked at approximately 2, 5, and 8~m. The phones were weighed down with a bag of pebbles to ensure the ropes were vertically straight. The phones were positioned at a  horizontal distance of 18~m. Unlike previous experiments, the rope would cause the phone to rotate and sway slowly, making it difficult to control the angle of the phone during the measurement.  We repeat the measurements three times at each depth. Fig.~\ref{fig:depth} shows the CDF of the absolute error at different depths. The plot  shows that the median and 95th percentile error is lowest at 0.28 and 0.73~m for the 5~m depth, which is around the midpoint depth of the dock. In contrast, the median distance error at 2 and 8~m are 0.33 and 0.58~m respectively. This likely is because multipath reflections  can be stronger when the devices are close to the surface or floor of the water body. In these experiments the phones could sway and rotate and hence were not static.

\begin{figure}[t!]
    \includegraphics[width=.23\textwidth]{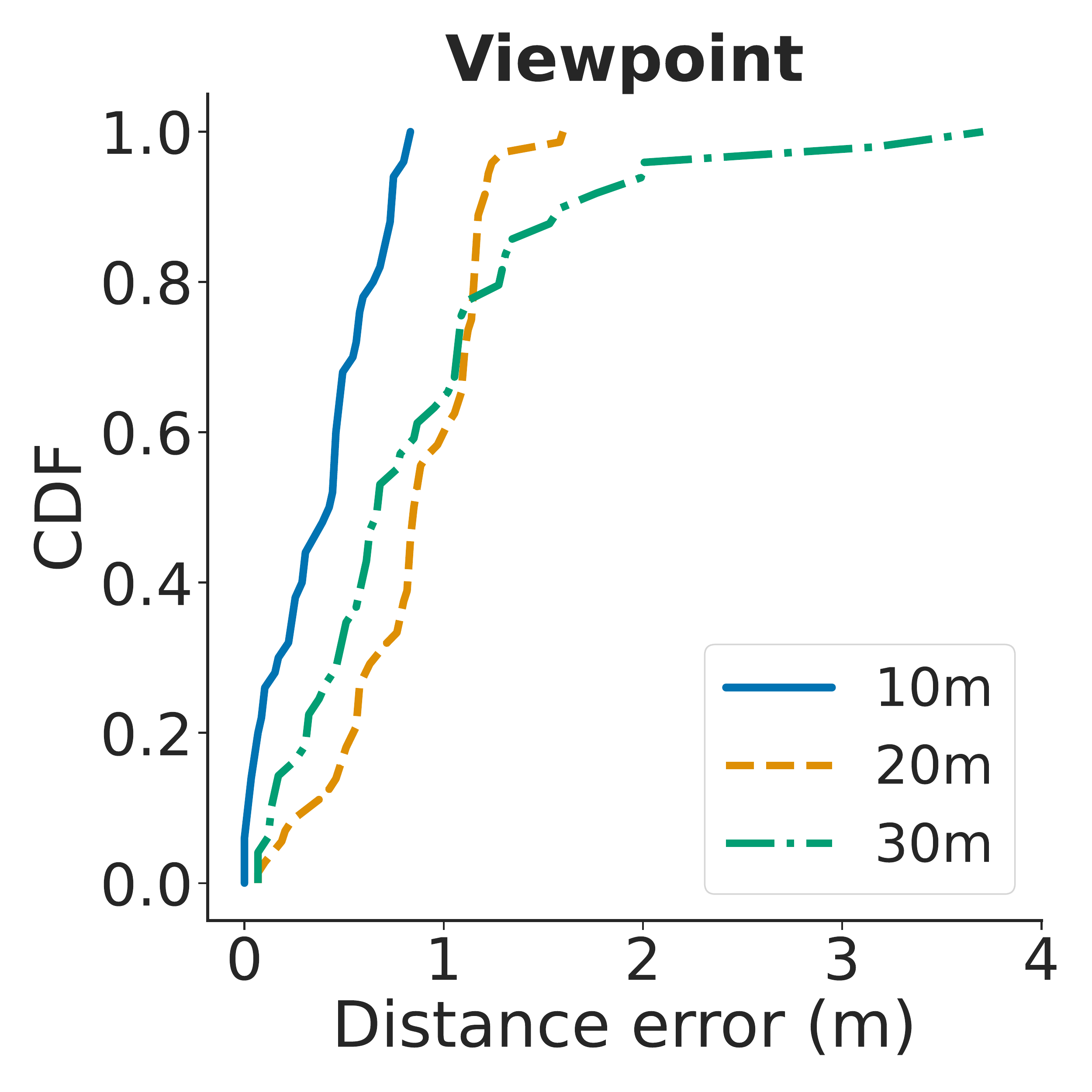}
    \includegraphics[width=.23\textwidth]{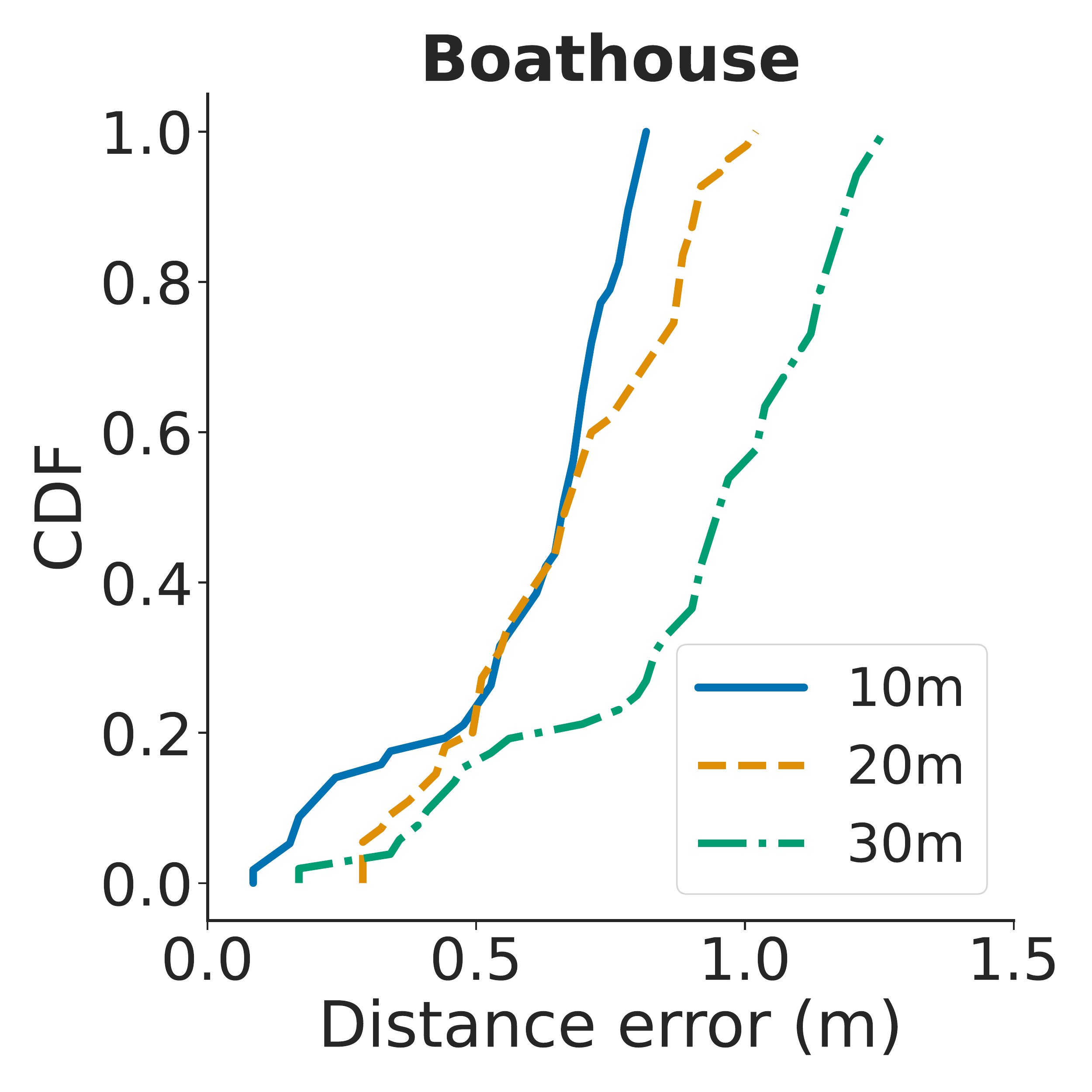}
    \vskip -0.15in
\caption{Effect of shallow waters. Errors  at two different locations with a  depth of 1.5 and 5~m respectively.}
\vskip -0.15in
\label{fig:locs2}
\end{figure}
\subsubsection{Evaluation in shallower waters.} The above evaluations were in a water body that had a depth of around 9~m. Shallower waters may contribute to more severe multi-path environments because of multiple reflections from the surface and floor. To evaluate the performance of our system in such environments, we test our system in the viewpoint and boathouse scenarios. At the viewpoint, the total water depth ranged from 1 to 1.5~m and the phones were submerged to a depth of 0.5~m. At the boathouse the total water depth was  5~m and the phones were submerged to a depth of 2.5~m. We measured the  errors  at distances of 10, 20, and 30~m.  Fig.~\ref{fig:locs2}  shows the CDF of the distance error at these locations. {The plot shows that the ranging error in the shallow water setting of the viewpoint had the highest 95th percentile error across all natural water bodies tested, with an error of 2.06~m at 30~m. In comparison the error at the boathouse and dock location with deeper waters had a 95th percentile error of 1.22 and 1.48~m respectively.} We note that each of these underwater environments have a different multipath and noise profile where  shallower waters likely have more severe multipath.  


\begin{figure*}[t!]
    \includegraphics[width=.23\textwidth]{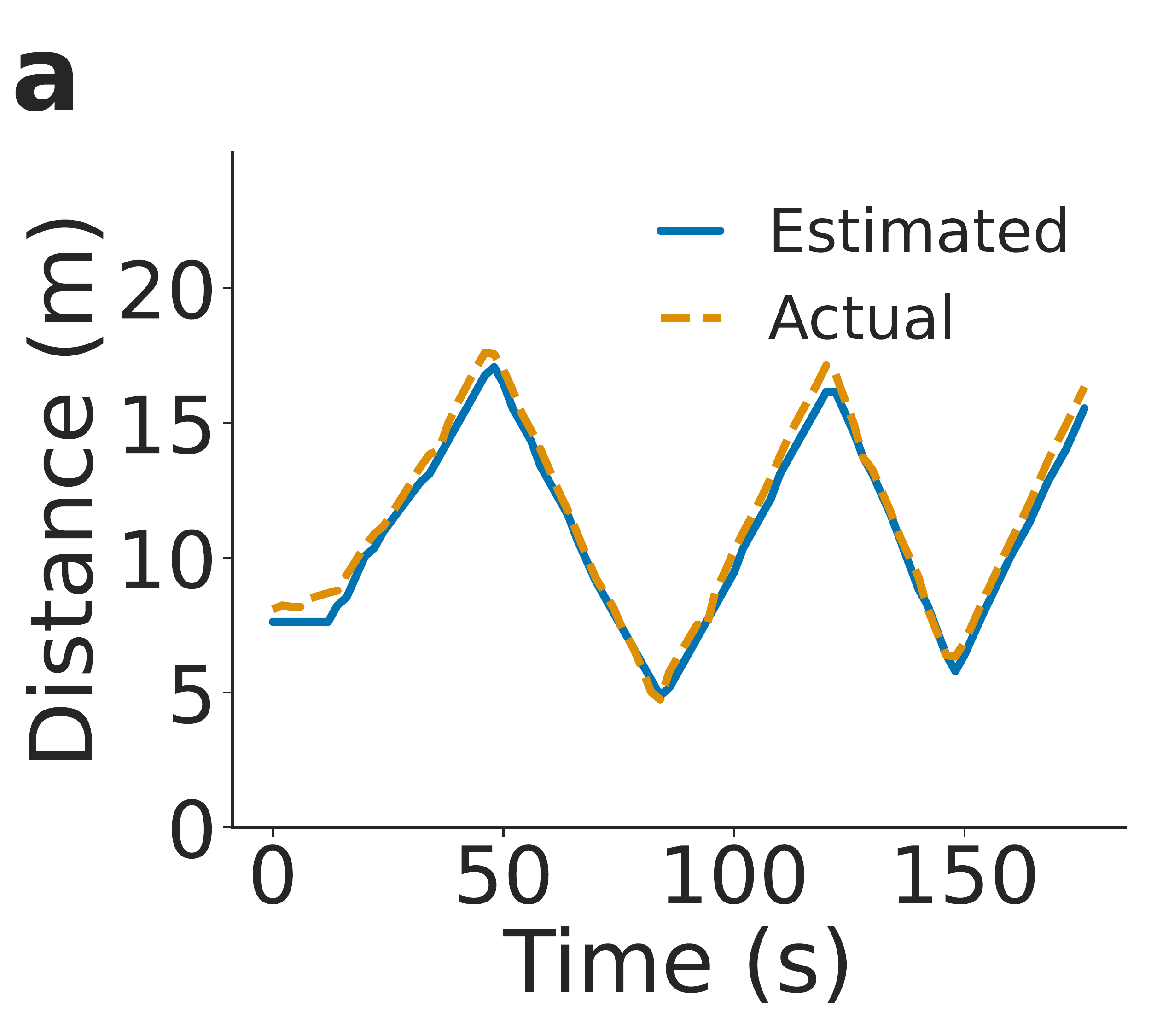}
    \includegraphics[width=.23\textwidth]{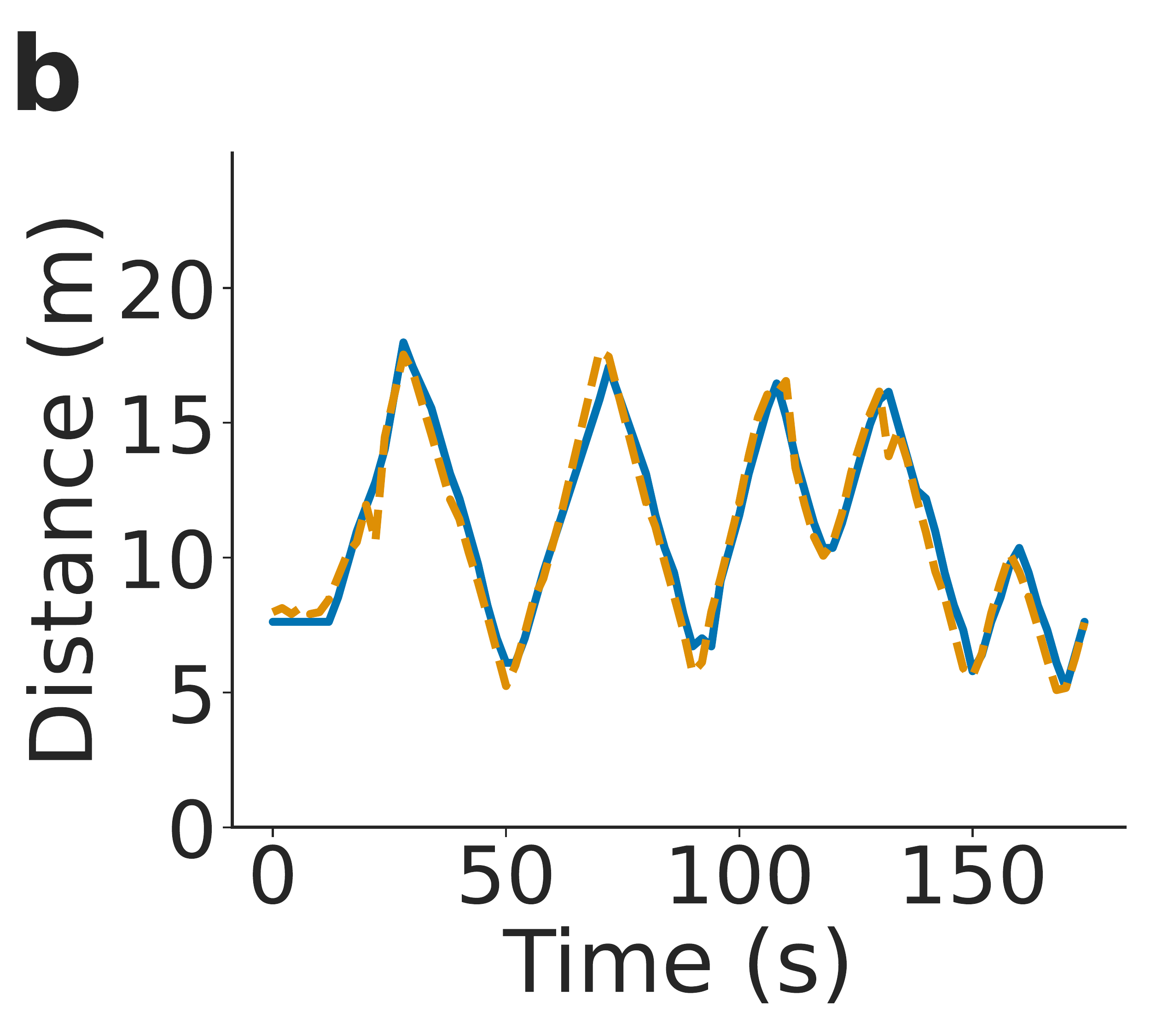}
    \includegraphics[width=.23\textwidth]{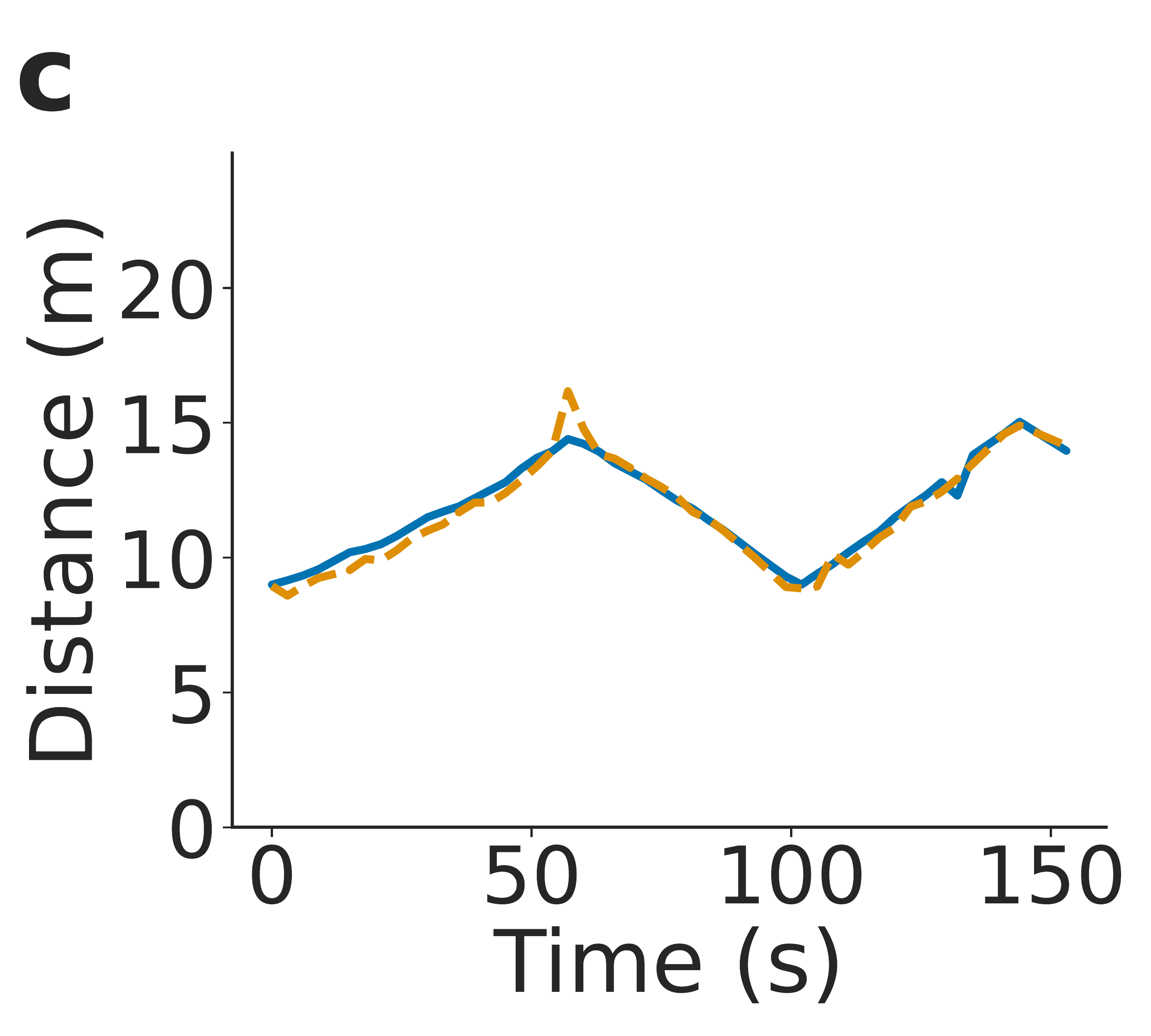}
    \includegraphics[width=.23\textwidth]{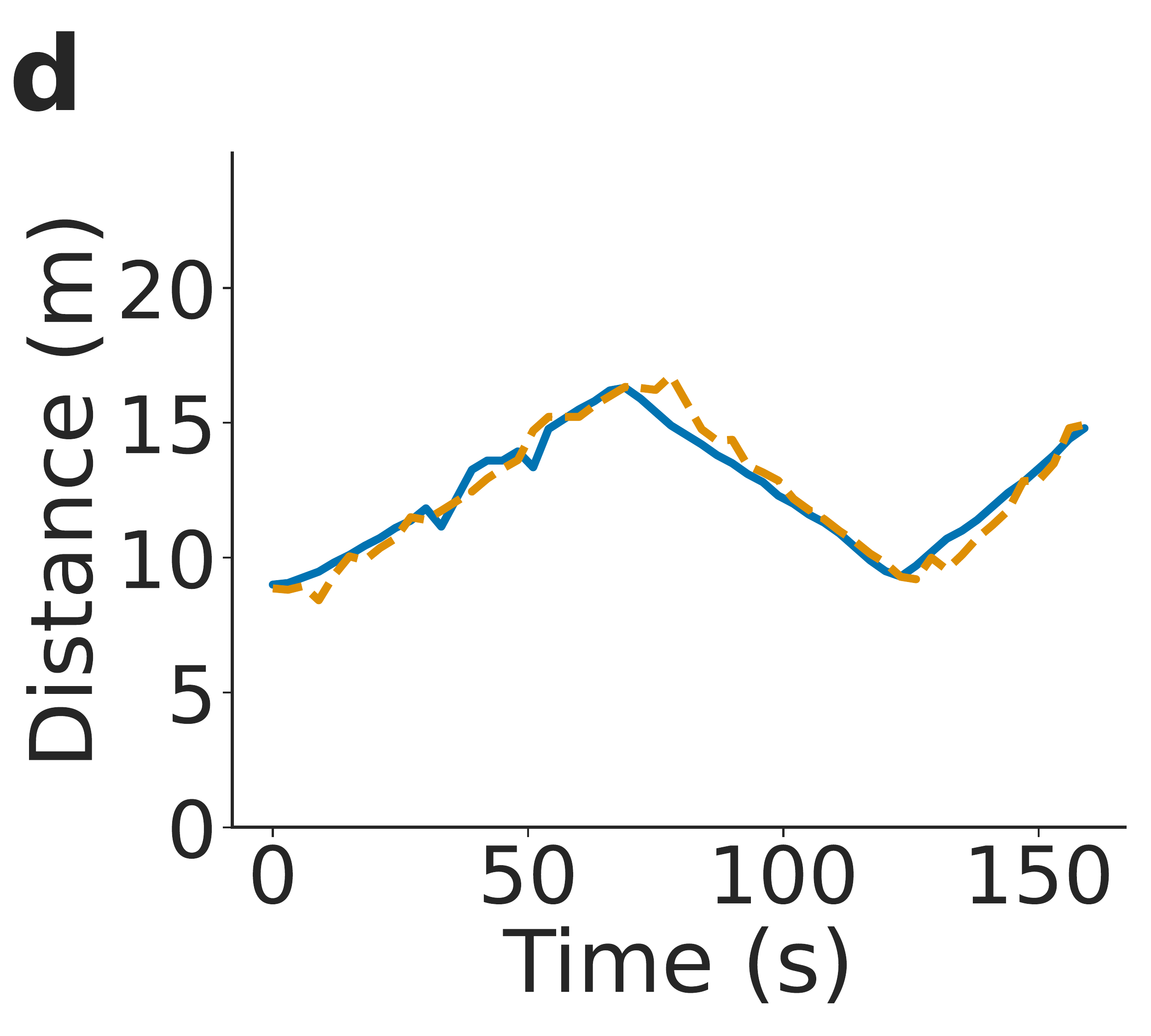}
    \vskip -0.15in
\caption{Trajectories computed by our smartphone ranging system when the phone moves at different speeds. Note that underwater motion speeds are lower than that in in-air due to much higher water resistance.}
\vskip -0.15in
\label{fig:track}
\end{figure*}

\subsubsection{Tracking device motion.}
We evaluate whether our system can be used to track the distance  of the smartphone over time. In this experiment, we moved the smartphone along different 1D trajectories parallel to the coast at the dock location for measurement durations of {153 to 176~s}. {The smartphone was moved at two times at a slow speed of 15 cm/s, and two times at a faster speed of 32 and 56 cm/s.} We note that it was challenging to move the extension pole with the phone at higher speeds due to the resistance of the water. At higher speeds, the range measurements of the phone underwater would lag behind the ground truth recordings of the the pole above water.  We obtained the ground truth distance measurement by recording a video of the smartphone's position with respect to a tape measure placed along the dock. The video recording was then synchronized with the measurements obtained from the smartphones underwater, based on the phone's timestamp that was displayed on the screen. The video frames corresponding to each ranging estimate was extracted, and the reading on the tape measure was recorded.  We compute a range value using our system once every second. Fig.~\ref{fig:track} compares the trajectories estimated by our system with the ground truth measurements. {The median and 95th percentile error across  the two slower motion patterns was 0.35 and 0.93~m. In comparison, the median and 95th percentile errors across the faster motion patterns was slightly higher at 0.51 and 1.17~m. These results suggest that our ranging frequency of 1~Hz is sufficient for providing tracking capabilities at a range of  speeds.}

\subsubsection{Human mobility in the environment.} While we performed the above evaluations in natural outdoor lakes that have aquatic life, there were no humans swimming in these water bodies. So to understand the effect of other human mobility in the environment we  evaluate our system in a busy swimming pool. We performed measurements at a fixed horizontal distance of 6.1~m over a period of eight minutes. During this time, there were a number of swimmers swimming around in the pool and in adjacent lanes. The depth of the pool was 2.5~m and the phones were held by hand at a depth of 1~m. Note that since the experiments were performed over eight minutes, there were slight movements in the hand over the test duration.  Fig.~\ref{fig:mobility}a shows the CDF of the distance error at the approximate ground truth value of 6.1~m. The median and 95th percentile error were  0.35 and 0.84~m respectively. These low errors are likely because while a mobile human in the environment changes the multipath in the environment, since the objective of our algorithm is to zone in on the direct  path by combining the information across the two microphones, it can ignore these multipath variations and focus on the direct path that does not change.

\subsubsection{Effect of human occlusion.}  Next, we  measure the effect of human occlusion from the smartphone user. We perform these experiments again in the swimming pool, where  we periodically block the speaker and microphone of one of the phones with our  body. The experiment was performed at a horizontal distance of 6.1~m over 8 minutes.  Fig.~\ref{fig:mobility}b shows the CDF of the distance error with and without occlusion at the same distance. The presence of an occlusion increased the median error from 0.31 to 0.87~m. The 95th percentile error further increased from 0.89 to 2.41~m. This is expected because  occlusions can sometimes reduce the amplitude of  the direct  path to be close to noise. 
 A key assumption that     ranging systems~\cite{peng2007beepbeep,millisonic,mao2016cat,tracko} make is that  there exists a non-zero  direct path. This is also true for existing underwater localization systems that require custom buoys and hardware~\cite{tracking1, tracking2,tracking3,tracking4,tracking5,tracking6,tracking7,tracking8}. In our system, the direct path does not need to be the strongest path but it should be non-zero. When such a path does not exist, ranging based systems are limited in their accuracy.

\subsubsection{With different smartphone models.} 
We also evaluated our system with different smartphone model pairs. As different smartphones use different speakers and microphones, the frequency selectivity of an acoustic signal can vary between devices. In this experiment, we evaluated three Android smartphones models: Samsung Galaxy S9, Google Pixel 3a, and OnePlus 8 Pro. Fig.~\ref{fig:phones_rot}a shows the CDFs of  errors between different device pairs. The plot shows that the median error for a given device pair ranges from 0.28 to 0.54~m. Across all device pairs and measurements, the median and 95th percentile error is 0.41 and 0.75~m respectively.

\subsubsection{Effect of phone orientation.} 
We evaluate the effect of phone orientation on the  accuracy at the dock at a horizontal distance of 20~m and a depth of 2.5~m. We first positioned the speaker and microphone of both phones to directly face each other so their azimuth $\phi$ and polar angle $\theta$ is set to 0° and 180° respectively. We measure the  error when the sender phone is rotated to different azimuth and polar angles. We first  rotate the sender phone in the azimuth angle to 90° and 180° while keeping the polar angle constant. We then reposition the phone so its speaker and microphone faces upwards with  $\phi=0^{\circ}$, $\theta=0^{\circ}$. Fig.~\ref{fig:phones_rot}b shows the CDF of the distance error for these different rotation configurations. The median  error ranges from 0.54 to 1.25~m across different orientations. The 95th percentile  error ranges from 0.78 to 1.62~m. Note that when the phone faces upwards it had the largest error likely because the phones are closer to the water surface  resulting is higher multipath when pointing towards the surface.

\begin{figure}[t!]
    \includegraphics[width=.23\textwidth]{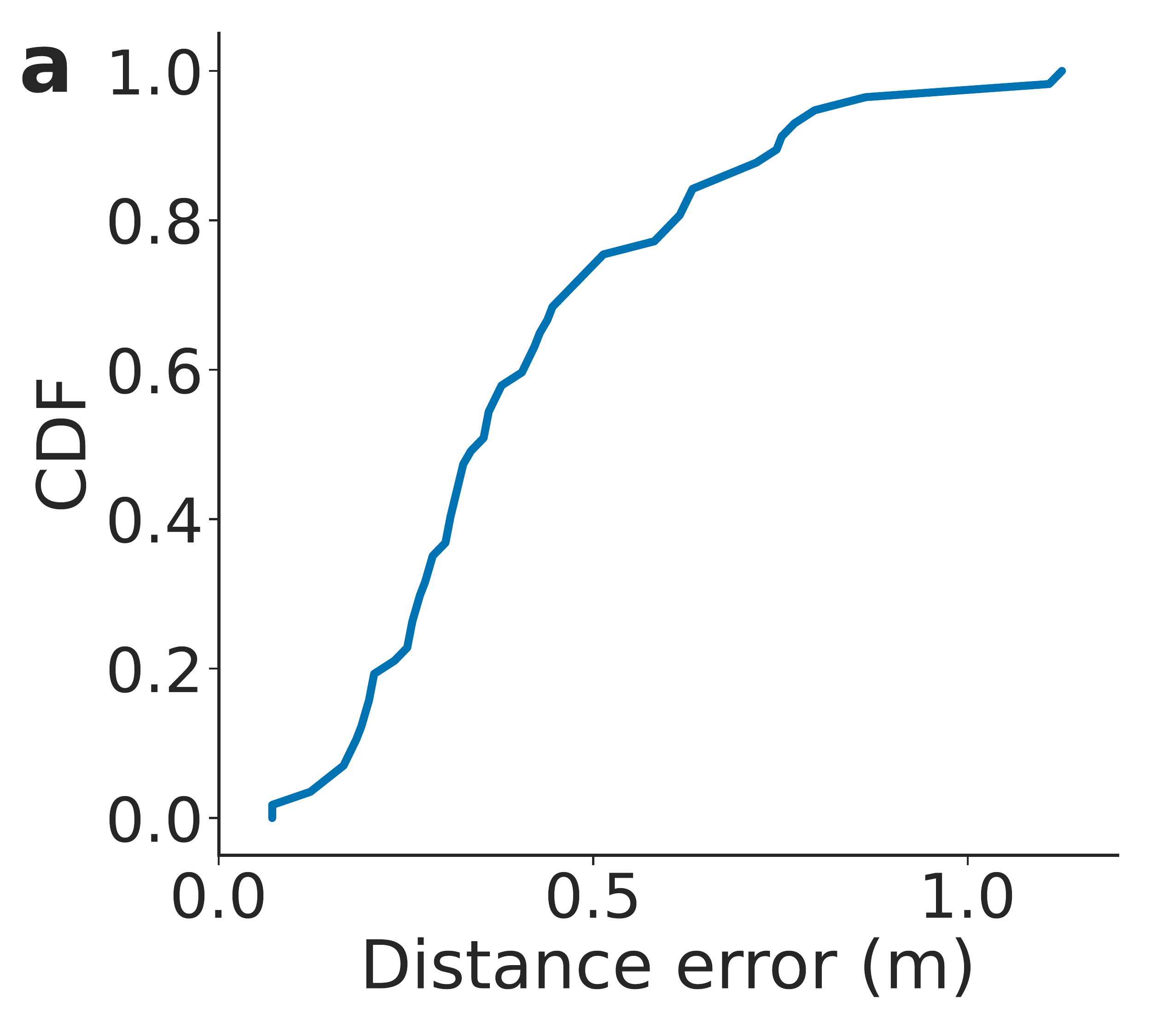}
    \includegraphics[width=.23\textwidth]{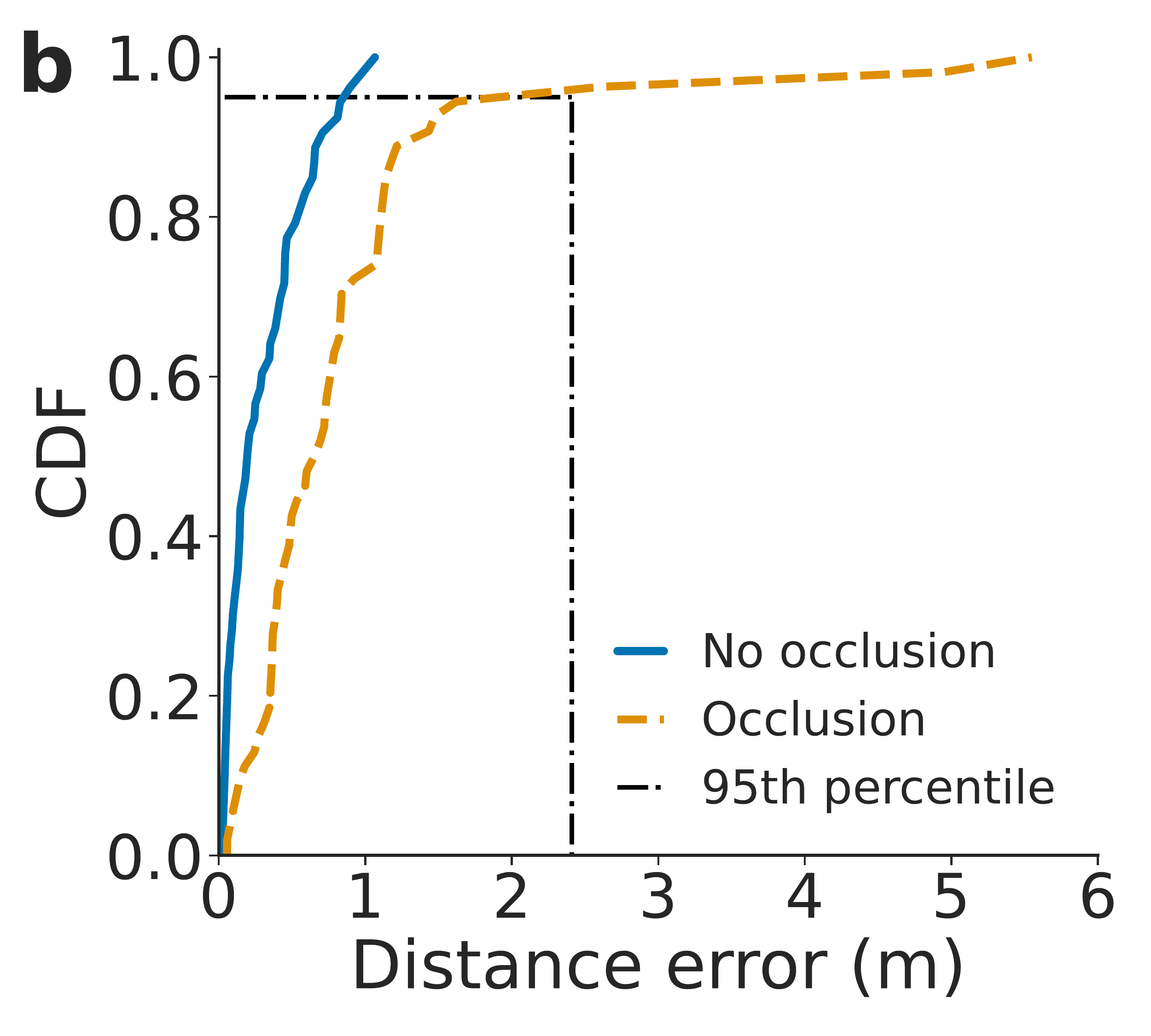}
    \vskip -0.15in
\caption{Effect of (a) other human  mobility in environment, and (b) occlusion between the devices.}
\vskip -0.15in
\label{fig:mobility}
\end{figure}

\begin{figure}[t!]
    \includegraphics[width=.23\textwidth]{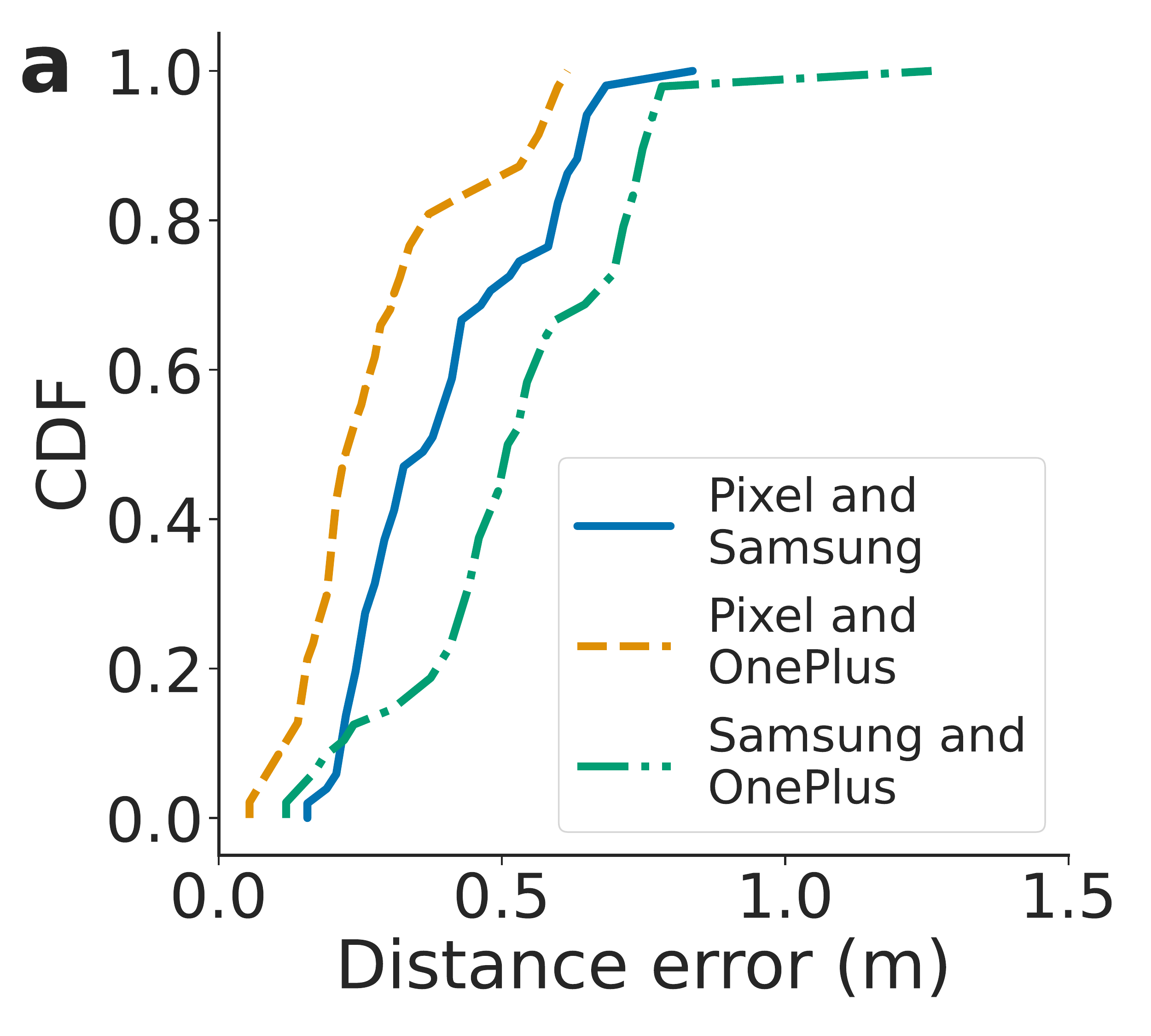}
    \includegraphics[width=.23\textwidth]{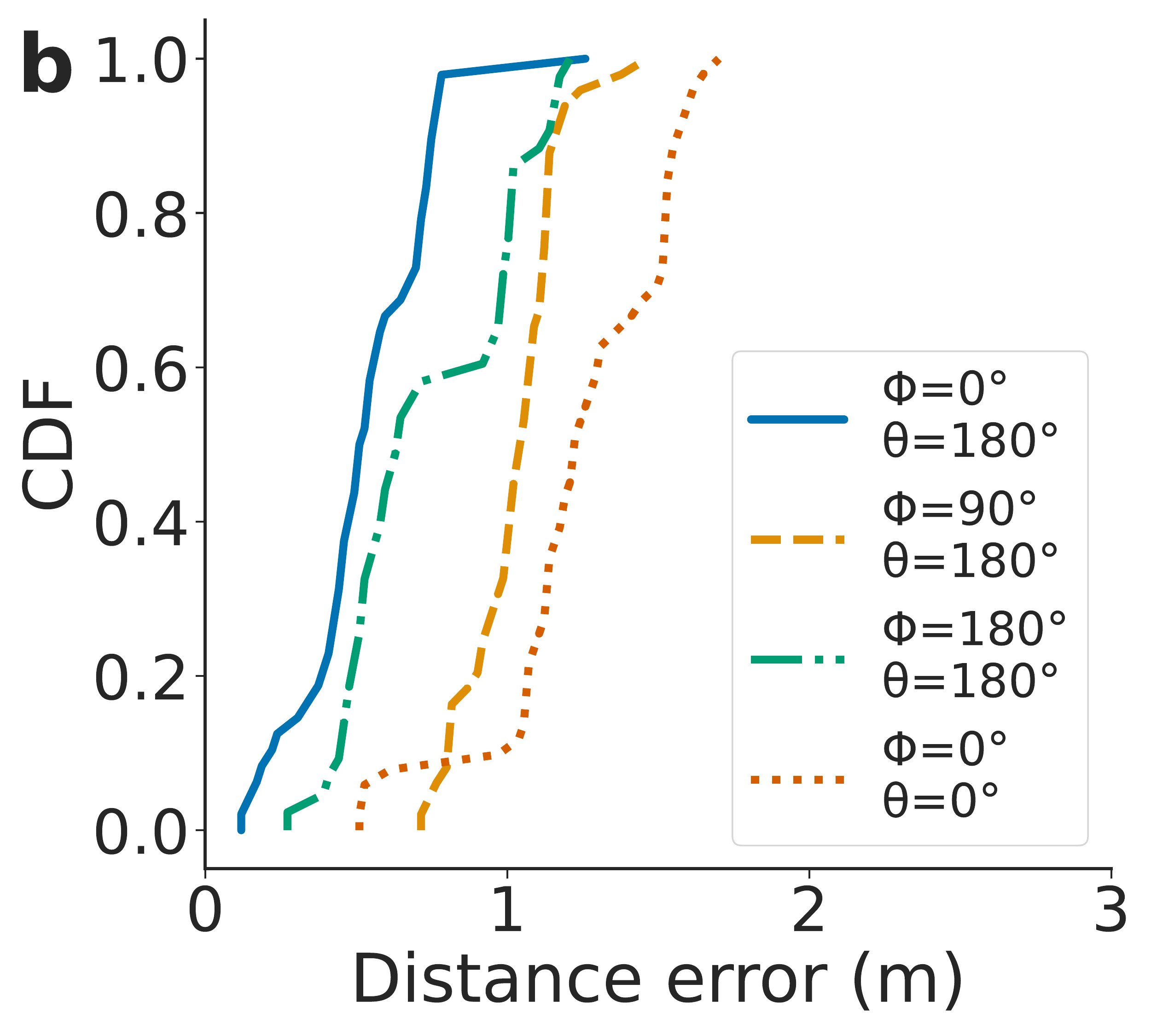}
    \vskip -0.15in
\caption{Effect of (a) various   smartphone model pairs, and (b)  orientations with phones separated by  20~m.}
\vskip -0.15in
\label{fig:phones_rot}
\end{figure}

\subsubsection{Ranging of multiple devices.} 
 {We evaluate if our system can reliably support ranging updates to multiple devices. To evaluate this, we consider a network deployment with one leader device and three diver devices placed at the dock location. The leader and one diver device was submerged using the extension pole, and the two remaining diver devices were submerged using a rope. All diver devices were submerged to a depth of approximately 3~m.  Devices 1  and 2 were horizontally spaced from the leader device by 7~m, while device 3 had a spacing of 9~m. Fig.~\ref{fig:multiid}a shows the CDF of distance error for the three devices. The median error for each device ranged from 0.21 to 0.47~m, while the 95th percentile error ranged from 0.68 to 0.79~m.  Additionally, we evaluate how well our system is able to accurately decode the device ID appended to the end of each preamble. For this experiment, we sent between 300--480 random preamble and ID pairs at distances up to 45~m and measured the decoding accuracy of our system across all transmitted IDs. Fig.~\ref{fig:multiid}b shows that our system is able to correctly decode the ID at a $> 95\%$ accuracy at distances up to 45~m. These results   show the feasibility of  our protocol working with  multiple users.}

\subsubsection{Effect of multiple microphones.} {Here we analyze the effect of using both the top and bottom microphones for ranging versus using only a single microphone in isolation. Fig.~\ref{fig:dual_compass}a shows the 95th percentile distance error for these scenarios at distances of up to 45~m. The figure reveals the following: firstly, utilizing both microphones yields lower ranging errors at all distances. This can reduce error by as much as 4.52~m at a distance of 45~m. Secondly, when a single microphone is used in isolation, there is no clear relationship between microphone position and distance error. This is likely due to the different multipath and noise profiles at different distances. These results show that identifying the direct path using two microphones is a more accurate strategy than using a single microphone alone.}

\begin{figure}[t!]
    \includegraphics[width=.23\textwidth]{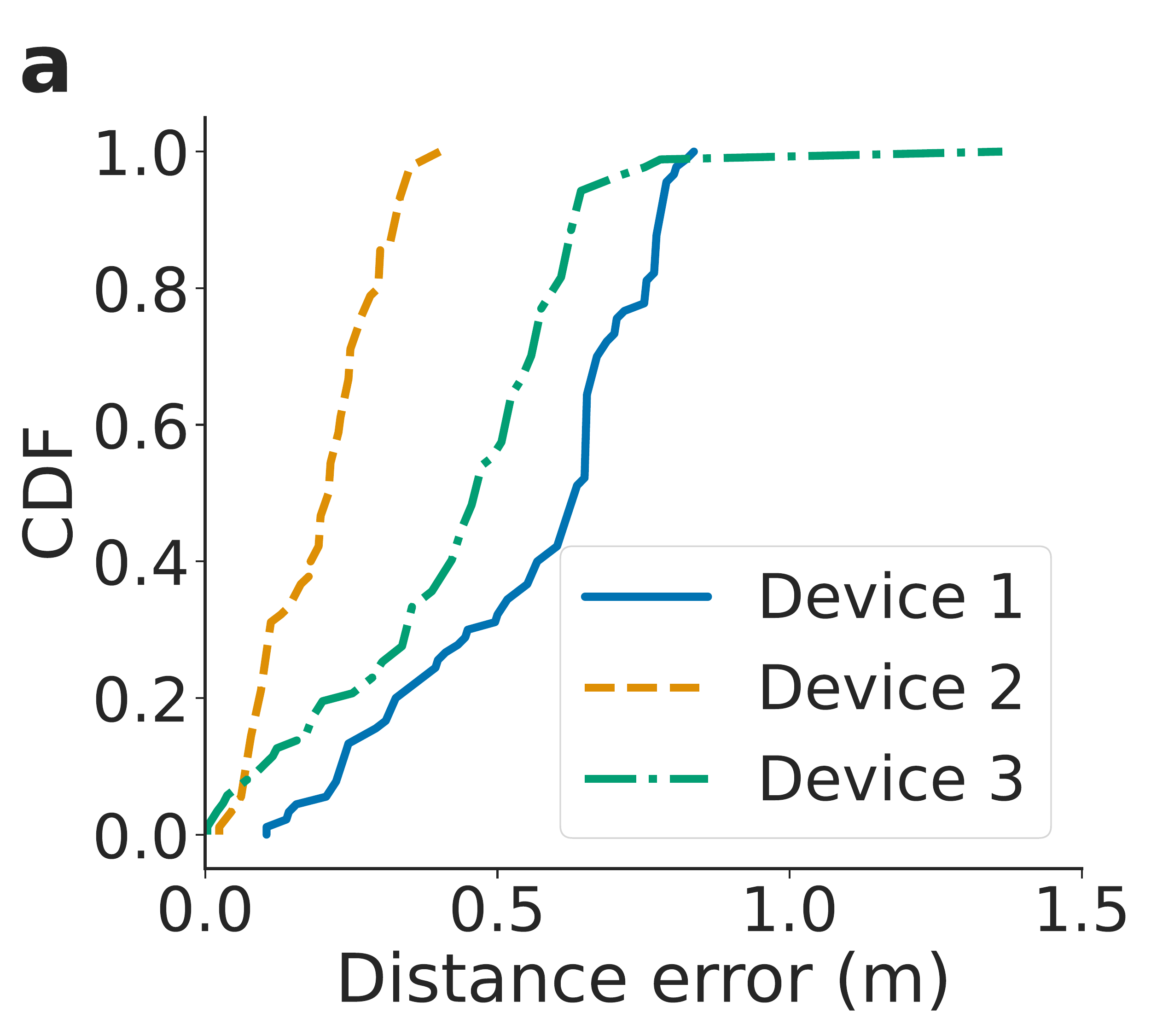}
    \includegraphics[width=.23\textwidth]{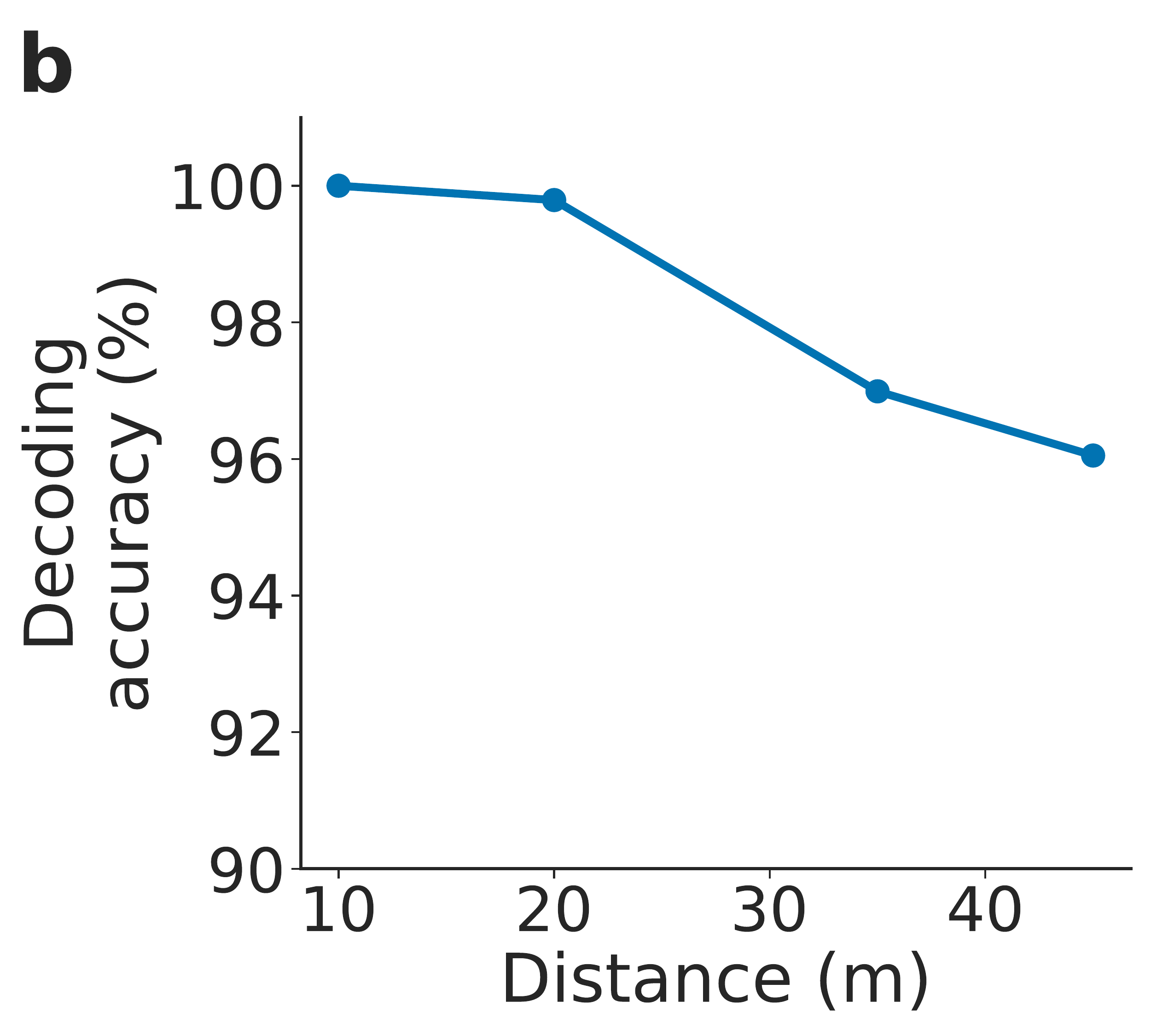}
    \vskip -0.15in
\caption{Ranging of multiple devices. (a) Errors for each of the three   devices. (b) ID decoding accuracies. }
 \vskip -0.15in
\label{fig:multiid}
\end{figure}

\section{Related work}
 While there have been significant  efforts to achieve  underwater communication~\cite{isonar,sigcomm22,cog-usrp,usrp2,underwater-cognitive1,underwater-cognitive2,underwater-cognitive3,waterbackscatter} using OFDM  modulation  and rate adaptation~\cite{submission} and  sensing~\cite{sensing1,sensing2,sensing3}, here we focus on the prior work in localization.

\vskip 0.05in\noindent{\bf Underwater tracking.} There has been significant interest in achieving underwater tracking for diving computers, sensors  and robotics~\cite{tracking1, tracking2,tracking3,tracking4,tracking5,tracking6,tracking7,tracking8,sensor1}. While our work builds on this prior art, we are the first to demonstrate an end-to-end  underwater acoustic ranging system  between smartphones. Below we describe these prior works in detail.

\cite{gpsdiver1} proposes the design of a custom diving computer that can communicate with a surface buoy  for underwater mapping and tracking. 
\cite{diver1} uses  hydrophone devices on the surface as beacons with known locations. The diver then uses a hand-held display connected to an acoustic communication module.  \cite{ekf1} proposes the use of an underwater pinger that emits acoustic signals periodically. The pinger is synchronized with  a very high precision clock that is synchronized to GPS, prior to deployment. The acoustic signals are  tracked from the surface by using a system of four buoys  that measures the times of arrival of the acoustic signals emitted by the pinger. The surface buoys may be equipped with GPS that can be translated to underwater GPS~\cite{gps1}.  \cite{reverse1} proposes the reverse process where    the time difference of arrivals is   measured at a sensor to detect range differences from the sensor to four anchor nodes. These range differences are averaged over multiple beacon intervals  to estimate the 3-D sensor location. \cite{single1}  uses  a single beacon and compute the position based on time of arrival measurements. \cite{directional1} proposes the use of directional beacons for localization. 
 \cite{loc_simulation1} presents the results of experiments designed in a virtual environment used to simulate real acoustic underwater localization systems. \cite{loc_ml1,loc_ml2} propose to use machine learning to improve localization for custom time-difference of arrival  based hydrophones.  In contrast to this prior work that is designed for custom hardware, our work achieves a novel underwater ranging  system between  smartphones.   

 A recent short paper presents two quick experiments underwater using smartphone hydrophones~\cite{phonerange}. This prior paper however does not provide details about the smartphone used, algorithms or the underwater scenarios it was tested in. {In addition, it does not consider any practice issues like clock drifting, underwater multipath, OS buffering or  mobility. } In contrast, we design  real-time algorithms, explore the challenges of using smartphones underwater and provide extensive evaluation  in various underwater scenarios.

Recent works also propose the use of an autonomous marine surface~\cite{surface1}
 or underwater vehicles~\cite{underwatervehicle1} to increase diver safety by enabling navigation and reliable monitoring from the surface. ~\cite{underwater-hri1} proposes the use of autonomous underwater robots to track divers in real time and to re-identify them. \cite{hri2} proposes the detection and tracking of a diver with a high-frequency  sonar. \cite{currents1,currents2} design algorithms to achieve underwater tracking in the presence of ocean current uncertainty on the surface buoy. \cite{soundspeed1} designs algorithms to account for variations in sound  velocity that is dependent on water temperature, pressure and salinity.    \cite{backscatterlocalization} explores  the challenges of achieving localization for underwater backscatter sensors.  Our work instead is focused on  enabling underwater  ranging  between commodity smartphones.

\begin{figure}[t!]
    \includegraphics[width=.23\textwidth]{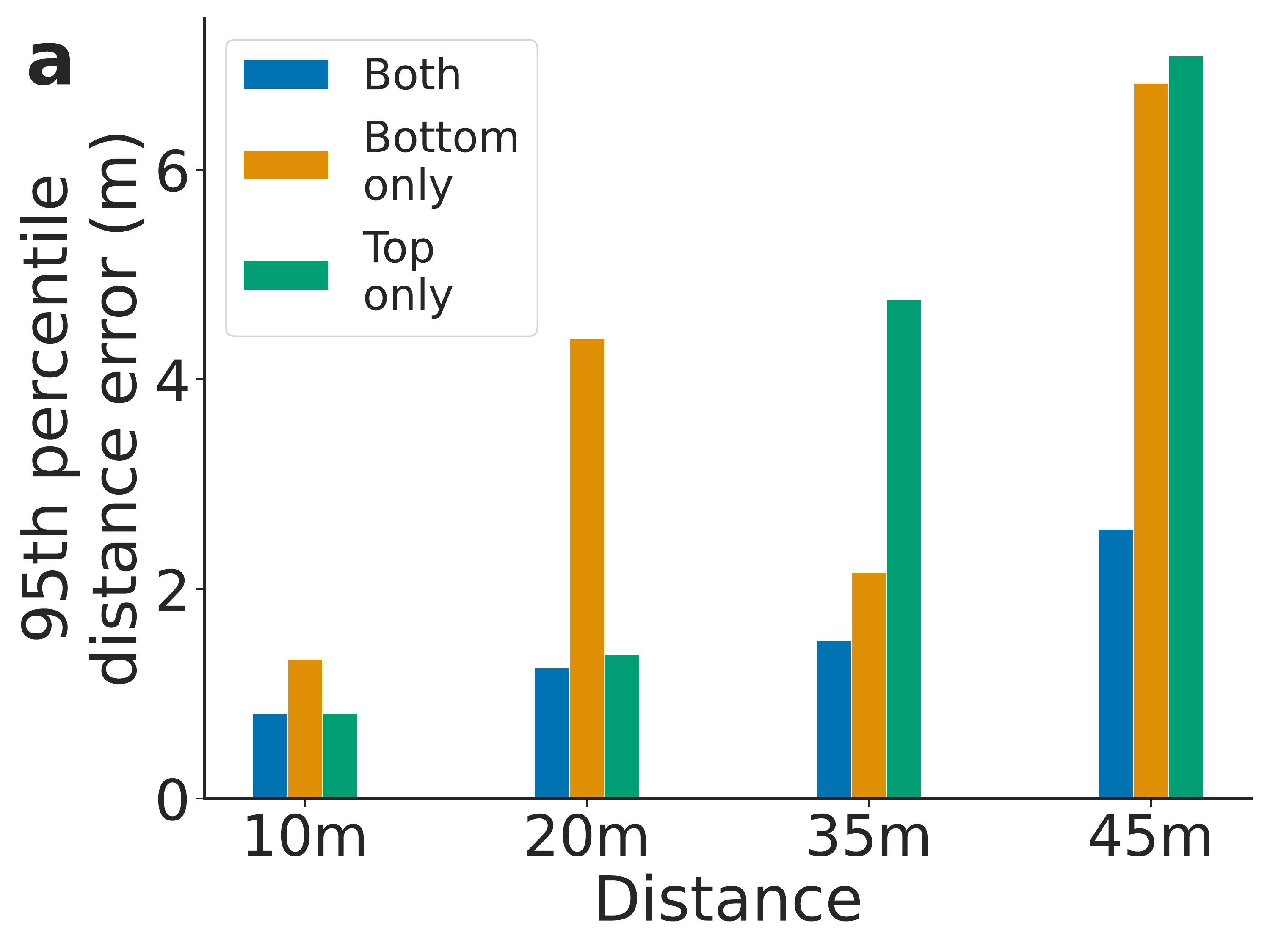}
    \includegraphics[width=.23\textwidth]{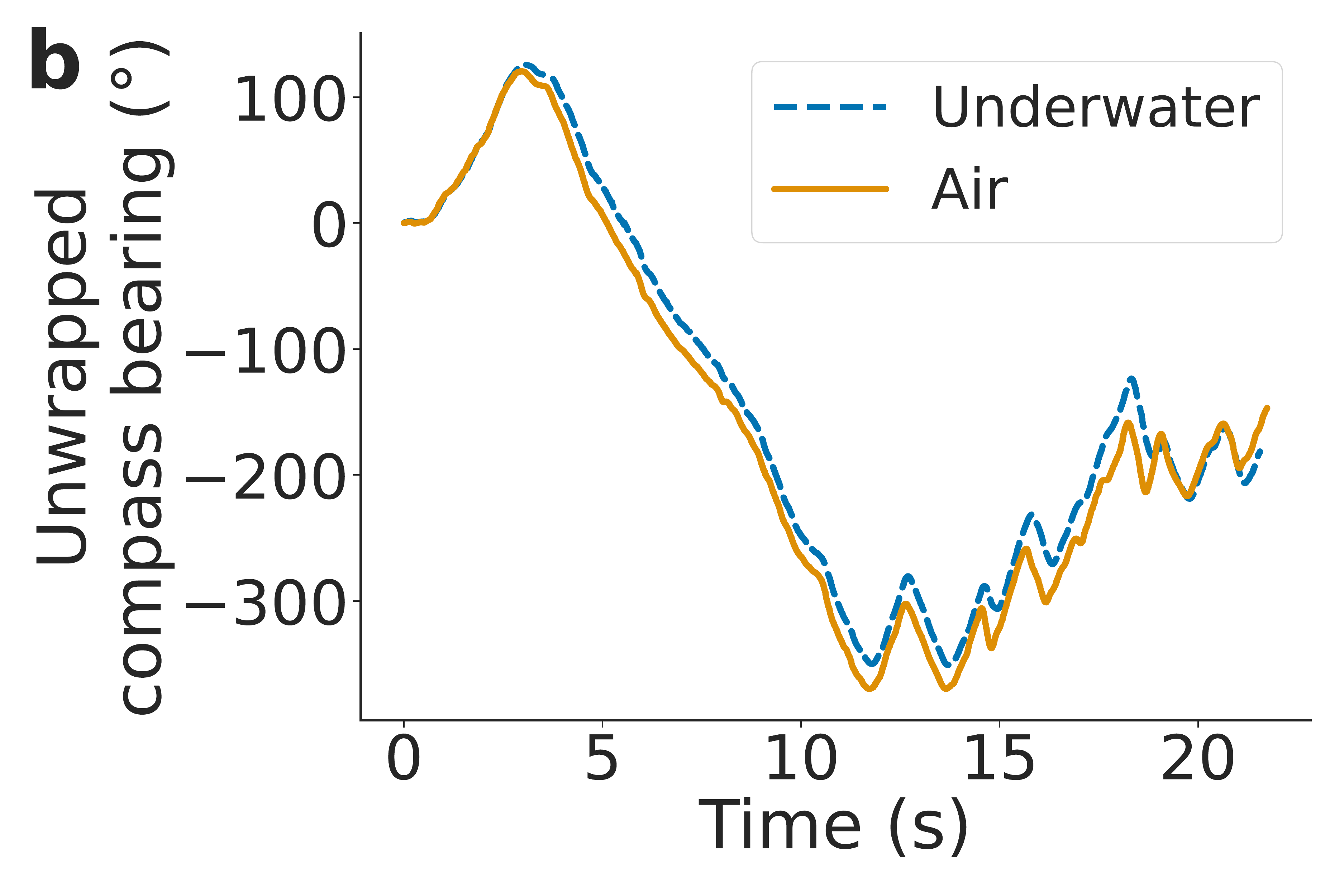}
    \vskip -0.15in
\caption{(a) 95\% errors using both microphones, the bottom and top microphone only. (b) Smartphone compass bearings measured underwater and in air.}
\vskip -0.15in
\label{fig:dual_compass}
\end{figure}
 
 \vskip 0.05in\noindent{\bf In-air acoustic tracking.}  Our work builds on prior work that achieve in-air acoustic device ranging, tracking,  localization and health applications~\cite{apneaapp,opioid,infection,fingerio,infant,cardiac,attack}.  BeepBeep~\cite{peng2007beepbeep} and Swordfight~\cite{zhang2012swordfight} achieves  ranging  between phones. These require a secondary RF channel between the two phones to exchange synchronization information between the two devices. Radio signals at  2.4~GHz   can attenuate as much as 169~dB per meter in seawater~\cite{wifi,shrimp}.       Sonoloc~\cite{erdelyi2018sonoloc} can achieve distributed acoustic localization in-air between multiple devices but  requires more than ten  devices to achieve reasonable accuracies.  \cite{infocom2018} achieves  acoustic in-air tracking by assuming that there is no significant multipath; underwater scenarios in contrast are known for their severe multipath challenges.  {ALPS~\cite{alps} and Tracko~\cite{tracko} use  a combination of Bluetooth and ultrasonic; Bluetooth however cannot be used underwater.}  CAT~\cite{mao2016cat}, Millisonic~\cite{millisonic} and SoundTrak~\cite{zhang2017soundtrak} have been proposed to improve the acoustic accuracies. CAT combines  IMU sensor data and frequency modulated continuous wave (FMCW)  localization to address in-air multipath.   SoundTrak~\cite{zhang2017soundtrak} is designed for a smart watch and a customized finger ring using phase tracking where the area of movement is limited to a $20cm\times16cm\times11cm$ area in-air. Millisonic~\cite{millisonic}  uses  a microphone array and  time difference of arrival  algorithms. While we build on this prior work on in-air acoustic  tracking, our focus is on achieving underwater ranging between  smartphones.

\section{Discussion and Conclusion}

We present a novel real-time  underwater system that  can achieve  acoustic ranging between commodity smartphones. We evaluate our design in various underwater settings and demonstrate its efficacy. We believe that this work explores a new underwater research direction by bringing ranging capabilities to commodity smartphones. Here we discuss design considerations and  avenues for future work.

\vskip 0.05in\noindent{\it Using phone compass underwater.} If the compass bearings of a smartphone placed underwater correspond to those of a smartphone in open air, it can be used for underwater navigation. To check this, we attached two phones to the bottom and top of the extension pole. The smartphone at the bottom end of the pole was submerged directly underwater to a depth of 3~m while the smartphone at the top end of the pole remained in air. We twisted the pole on its axis in random clockwise and anticlockwise motions and recorded the compass bearing measurements on both smartphones. Fig.~\ref{fig:dual_compass}b shows the unwrapped compass bearing angles measured by the smartphone underwater and in the air, showing a match between the two measurements.

\vskip 0.05in\noindent{\it Frequency range.}  Our smartphone ranging system operates  at 1-5~kHz   which is audible to human hearing. We note that multiple  acoustic underwater modems both in the industry and the academic community~\cite{waterbackscatter,lowfreq-modem,lowfreq-modem1} also  operate in the sub-20~kHz human-audible frequency range.
\vskip 0.05in\noindent{\it Providing user feedback underwater.}  An important question is how do we provide feedback to the diver when they are go beyond a pre-set distance from their buddy?  To prevent the need for the diver to keep checking on the phone, we can use the vibration motor to provide feedback to the diver, which would be an interesting underwater U/I design to explore.

\vskip 0.05in\noindent{\it Relative distance to absolute location.} Ranging provides the relative distance from one device to the other but not the absolute location. Computing the absolute location requires three devices with known locations to perform triangulation. These devices can either be floating buoys  or other  underwater smartphones in known locations. Designing this is an interesting research avenue for future work.



\begin{thebibliography}{92}


\ifx \showCODEN    \undefined \def \showCODEN     #1{\unskip}     \fi
\ifx \showDOI      \undefined \def \showDOI       #1{#1}\fi
\ifx \showISBNx    \undefined \def \showISBNx     #1{\unskip}     \fi
\ifx \showISBNxiii \undefined \def \showISBNxiii  #1{\unskip}     \fi
\ifx \showISSN     \undefined \def \showISSN      #1{\unskip}     \fi
\ifx \showLCCN     \undefined \def \showLCCN      #1{\unskip}     \fi
\ifx \shownote     \undefined \def \shownote      #1{#1}          \fi
\ifx \showarticletitle \undefined \def \showarticletitle #1{#1}   \fi
\ifx \showURL      \undefined \def \showURL       {\relax}        \fi
\providecommand\bibfield[2]{#2}
\providecommand\bibinfo[2]{#2}
\providecommand\natexlab[1]{#1}
\providecommand\showeprint[2][]{arXiv:#2}

\bibitem[\protect\citeauthoryear{??}{mud}{2004}]%
        {muddy}
 \bibinfo{year}{2004}\natexlab{}.
\newblock \bibinfo{title}{Muddy Waters: Techniques for Low-Vis Diving}.
\newblock
  \bibinfo{howpublished}{\url{https://dtmag.com/thelibrary/muddy-waters-techniques-low-vis-diving/}}.
    (\bibinfo{year}{2004}).
\newblock


\bibitem[\protect\citeauthoryear{??}{dea}{2010}]%
        {death1}
 \bibinfo{year}{2010}\natexlab{}.
\newblock \bibinfo{title}{Why divers die? Cyprus Federation of underwater
  activities}.
\newblock \bibinfo{howpublished}{\url{http://www.cfua.org/Divers-Death.htm}}.
  (\bibinfo{year}{2010}).
\newblock


\bibitem[\protect\citeauthoryear{??}{net}{2011}]%
        {net}
 \bibinfo{year}{2011}\natexlab{}.
\newblock \bibinfo{title}{Diver drowns in fishing net}.
\newblock
  \bibinfo{howpublished}{\url{https://scubaboard.com/community/threads/diver-drowns-in-fishing-net-bc-canada.391393/}}.
    (\bibinfo{year}{2011}).
\newblock


\bibitem[\protect\citeauthoryear{??}{bud}{2018}]%
        {buddy}
 \bibinfo{year}{2018}\natexlab{}.
\newblock \bibinfo{title}{Buddy Diving For Safety and Support}.
\newblock
  \bibinfo{howpublished}{\url{https://www.divein.com/diving/buddy-diving-for-safety/}}.
    (\bibinfo{year}{2018}).
\newblock


\bibitem[\protect\citeauthoryear{??}{low}{2018}]%
        {lowfreq-modem}
 \bibinfo{year}{2018}\natexlab{}.
\newblock \bibinfo{title}{Evo Logics 7/17 communication and positioning
  devices. https://evologics.de/acoustic-modem/7-17}.
\newblock   (\bibinfo{year}{2018}).
\newblock


\bibitem[\protect\citeauthoryear{??}{com}{2020}]%
        {compass}
 \bibinfo{year}{2020}\natexlab{}.
\newblock \bibinfo{title}{Best Dive Compass for Scuba Diving}.
\newblock
  \bibinfo{howpublished}{\url{https://oceanscubadive.com/best-dive-compass/}}.
   (\bibinfo{year}{2020}).
\newblock


\bibitem[\protect\citeauthoryear{??}{lig}{2020}]%
        {light}
 \bibinfo{year}{2020}\natexlab{}.
\newblock \bibinfo{title}{Best Scuba Diving Lights}.
\newblock
  \bibinfo{howpublished}{\url{https://oceanscubadive.com/best-dive-compass/}}.
   (\bibinfo{year}{2020}).
\newblock


\bibitem[\protect\citeauthoryear{??}{fai}{2020}]%
        {failure}
 \bibinfo{year}{2020}\natexlab{}.
\newblock \bibinfo{title}{Scuba Equipment Issues Get This Diver in Deep
  Trouble}.
\newblock
  \bibinfo{howpublished}{\url{https://www.scubadiving.com/scuba-equipment-issues-get-this-diver-in-deep-trouble}}.
    (\bibinfo{year}{2020}).
\newblock


\bibitem[\protect\citeauthoryear{??}{and}{2021}]%
        {android1}
 \bibinfo{year}{2021}\natexlab{}.
\newblock \bibinfo{title}{Audio Latency.
  https://developer.android.com/ndk/guides/audio/audio-latency}.
\newblock   (\bibinfo{year}{2021}).
\newblock


\bibitem[\protect\citeauthoryear{??}{scu}{2021a}]%
        {scubadepth}
 \bibinfo{year}{2021}\natexlab{a}.
\newblock \showarticletitle{Become a Certified Scuba Diver FAQs}.
\newblock  (\bibinfo{year}{2021}).
\newblock
\showURL{%
\url{https://www.padi.com/help/scuba-certification-faq}}


\bibitem[\protect\citeauthoryear{??}{div}{2021}]%
        {diveroid}
 \bibinfo{year}{2021}\natexlab{}.
\newblock \bibinfo{title}{Diveroid: turn your smartphone into an all-in-one
  dive gear}.
\newblock \bibinfo{howpublished}{\url{https://diveroid.com/}}.
  (\bibinfo{year}{2021}).
\newblock


\bibitem[\protect\citeauthoryear{??}{sta}{2021}]%
        {stats}
 \bibinfo{year}{2021}\natexlab{}.
\newblock \bibinfo{title}{Global Scuba Diving Equipment Market - Growth,
  Trends, COVID-19 Impact, and Forecasts (2022 - 2027), Mordor Intelligence}.
\newblock   (\bibinfo{year}{2021}).
\newblock


\bibitem[\protect\citeauthoryear{??}{scu}{2021b}]%
        {scubadepth2}
 \bibinfo{year}{2021}\natexlab{b}.
\newblock \showarticletitle{How Deep Can You Scuba Dive?}
\newblock  (\bibinfo{year}{2021}).
\newblock
\showURL{%
\url{https://www.scubadiving.com/why-is-130-feet-depth-limit-for-recreational-scuba-diving}}


\bibitem[\protect\citeauthoryear{??}{pou}{2021}]%
        {pouch}
 \bibinfo{year}{2021}\natexlab{}.
\newblock \showarticletitle{Universal Waterproof Case, Waterproof Phone Pouch
  Compatible for iPhone 13 12 11 Pro Max XS Max XR X 8 7 Samsung Galaxy s10/s9
  Google Pixel 2 HTC Up to 7.0", IPX8 Cellphone Dry Bag -2 Pack}.
\newblock  (\bibinfo{year}{2021}).
\newblock
\showURL{%
\url{https://www.amazon.com/gp/product/B08S3SG5KF/ref=ppx_yo_dt_b_asin_title_o02_s00?ie=UTF8&psc=1}}


\bibitem[\protect\citeauthoryear{Alcocer, Oliveira, and Pascoal}{Alcocer
  et~al\mbox{.}}{2004}]%
        {ekf1}
\bibfield{author}{\bibinfo{person}{Alex Alcocer}, \bibinfo{person}{Paulo
  Oliveira}, {and} \bibinfo{person}{Antonio Pascoal}.}
  \bibinfo{year}{2004}\natexlab{}.
\newblock \showarticletitle{Study and implementation of an EKF GIB-based
  underwater positioning system}.
\newblock \bibinfo{journal}{{\em IFAC Proceedings Volumes\/}}
  \bibinfo{volume}{37} (\bibinfo{date}{07} \bibinfo{year}{2004}),
  \bibinfo{pages}{383--390}.
\newblock
\showDOI{%
\url{https://doi.org/10.1016/S1474-6670(17)31762-7}}


\bibitem[\protect\citeauthoryear{Alcocer, Oliveira, and Pascoal}{Alcocer
  et~al\mbox{.}}{2006}]%
        {gps1}
\bibfield{author}{\bibinfo{person}{Alex Alcocer}, \bibinfo{person}{Paulo
  Oliveira}, {and} \bibinfo{person}{Antonio Pascoal}.}
  \bibinfo{year}{2006}\natexlab{}.
\newblock \showarticletitle{Underwater acoustic positioning systems based on
  buoys with GPS}.
\newblock  (\bibinfo{date}{01} \bibinfo{year}{2006}).
\newblock


\bibitem[\protect\citeauthoryear{Alihemmati and Kalantari}{Alihemmati and
  Kalantari}{2005}]%
        {alihemmati2005channel}
\bibfield{author}{\bibinfo{person}{R Alihemmati} {and} \bibinfo{person}{ME
  Kalantari}.} \bibinfo{year}{2005}\natexlab{}.
\newblock \showarticletitle{On channel estimation and equalization in OFDM
  based broadband fixed wireless MAN networks}. In \bibinfo{booktitle}{{\em The
  7th International Conference on Advanced Communication Technology, 2005,
  ICACT 2005.}}, Vol.~\bibinfo{volume}{1}. IEEE, \bibinfo{pages}{224--229}.
\newblock


\bibitem[\protect\citeauthoryear{Anjangi, Gibson, Ignatius, Pendharkar, Kurian,
  Low, and Chitre}{Anjangi et~al\mbox{.}}{2020}]%
        {diver1}
\bibfield{author}{\bibinfo{person}{Prasad Anjangi}, \bibinfo{person}{Amy
  Gibson}, \bibinfo{person}{Manu Ignatius}, \bibinfo{person}{Chinmay
  Pendharkar}, \bibinfo{person}{Anne Kurian}, \bibinfo{person}{Alan Low}, {and}
  \bibinfo{person}{Mandar Chitre}.} \bibinfo{year}{2020}\natexlab{}.
\newblock \showarticletitle{Diver Communication and Localization System using
  Underwater Acoustics}. In \bibinfo{booktitle}{{\em Global Oceans 2020:
  Singapore – U.S. Gulf Coast}}. \bibinfo{pages}{1--8}.
\newblock
\showDOI{%
\url{https://doi.org/10.1109/IEEECONF38699.2020.9389462}}


\bibitem[\protect\citeauthoryear{Barberá, Bernal-Polo, and
  Herrero-Perez}{Barberá et~al\mbox{.}}{2021}]%
        {sensor1}
\bibfield{author}{\bibinfo{person}{Humberto Barberá}, \bibinfo{person}{Pablo
  Bernal-Polo}, {and} \bibinfo{person}{David Herrero-Perez}.}
  \bibinfo{year}{2021}\natexlab{}.
\newblock \showarticletitle{Sensor Modeling for Underwater Localization Using a
  Particle Filter}.
\newblock \bibinfo{journal}{{\em Sensors\/}}  \bibinfo{volume}{21}
  (\bibinfo{date}{02} \bibinfo{year}{2021}), \bibinfo{pages}{1549}.
\newblock
\showDOI{%
\url{https://doi.org/10.3390/s21041549}}


\bibitem[\protect\citeauthoryear{Bayat, Crasta, Aguiar, and Pascoal}{Bayat
  et~al\mbox{.}}{2015}]%
        {currents2}
\bibfield{author}{\bibinfo{person}{Behzad Bayat}, \bibinfo{person}{Naveen
  Crasta}, \bibinfo{person}{A.~Pedro Aguiar}, {and} \bibinfo{person}{Antonio
  Pascoal}.} \bibinfo{year}{2015}\natexlab{}.
\newblock \showarticletitle{Range-Based Underwater Vehicle Localization in the
  Presence of Unknown Ocean Currents: Theory and Experiment}.
\newblock \bibinfo{journal}{{\em Control Systems Technology, IEEE Transactions
  on\/}} (\bibinfo{date}{04} \bibinfo{year}{2015}).
\newblock
\showDOI{%
\url{https://doi.org/10.1109/TCST.2015.2420636}}


\bibitem[\protect\citeauthoryear{Bicen, Sahin, and Akan}{Bicen
  et~al\mbox{.}}{2012}]%
        {underwater-cognitive2}
\bibfield{author}{\bibinfo{person}{A.~Ozan Bicen}, \bibinfo{person}{A.~Behzat
  Sahin}, {and} \bibinfo{person}{Ozgur~B. Akan}.}
  \bibinfo{year}{2012}\natexlab{}.
\newblock \showarticletitle{Spectrum-Aware Underwater Networks: Cognitive
  Acoustic Communications}.
\newblock \bibinfo{journal}{{\em IEEE Vehicular Technology Magazine\/}}
  \bibinfo{volume}{7}, \bibinfo{number}{2} (\bibinfo{year}{2012}),
  \bibinfo{pages}{34--40}.
\newblock
\showDOI{%
\url{https://doi.org/10.1109/MVT.2012.2190176}}


\bibitem[\protect\citeauthoryear{Borden and DeArruda}{Borden and
  DeArruda}{2012}]%
        {lowfreq-modem1}
\bibfield{author}{\bibinfo{person}{Joe Borden} {and} \bibinfo{person}{Jeffery
  DeArruda}.} \bibinfo{year}{2012}\natexlab{}.
\newblock \showarticletitle{Long range acoustic underwater communication with a
  compact AUV}. In \bibinfo{booktitle}{{\em 2012 Oceans}}.
  \bibinfo{pages}{1--5}.
\newblock
\showDOI{%
\url{https://doi.org/10.1109/OCEANS.2012.6405091}}


\bibitem[\protect\citeauthoryear{Cai, Hu, Ma, Peng, and Liu}{Cai
  et~al\mbox{.}}{2018}]%
        {cai2018accurate}
\bibfield{author}{\bibinfo{person}{Chao Cai}, \bibinfo{person}{Menglan Hu},
  \bibinfo{person}{Xiaoqiang Ma}, \bibinfo{person}{Kai Peng}, {and}
  \bibinfo{person}{Jiangchuan Liu}.} \bibinfo{year}{2018}\natexlab{}.
\newblock \showarticletitle{Accurate ranging on acoustic-enabled IoT devices}.
\newblock \bibinfo{journal}{{\em IEEE Internet of Things Journal\/}}
  \bibinfo{volume}{6}, \bibinfo{number}{2} (\bibinfo{year}{2018}),
  \bibinfo{pages}{3164--3174}.
\newblock


\bibitem[\protect\citeauthoryear{Cario, Casavola, Gagliardi, Lupia, and
  Severino}{Cario et~al\mbox{.}}{2021a}]%
        {cario2021accurate}
\bibfield{author}{\bibinfo{person}{Gianni Cario}, \bibinfo{person}{Alessandro
  Casavola}, \bibinfo{person}{Gianfranco Gagliardi}, \bibinfo{person}{Marco
  Lupia}, {and} \bibinfo{person}{Umberto Severino}.}
  \bibinfo{year}{2021}\natexlab{a}.
\newblock \showarticletitle{Accurate Localization in Acoustic Underwater
  Localization Systems}.
\newblock \bibinfo{journal}{{\em Sensors\/}} \bibinfo{volume}{21},
  \bibinfo{number}{3} (\bibinfo{year}{2021}), \bibinfo{pages}{762}.
\newblock


\bibitem[\protect\citeauthoryear{Cario, Casavola, Gagliardi, Lupia, and
  Severino}{Cario et~al\mbox{.}}{2021b}]%
        {loc_simulation1}
\bibfield{author}{\bibinfo{person}{Gianni Cario}, \bibinfo{person}{Alessandro
  Casavola}, \bibinfo{person}{Gianfranco Gagliardi}, \bibinfo{person}{Marco
  Lupia}, {and} \bibinfo{person}{Umberto Severino}.}
  \bibinfo{year}{2021}\natexlab{b}.
\newblock \showarticletitle{Accurate Localization in Acoustic Underwater
  Localization Systems}.
\newblock \bibinfo{journal}{{\em Sensors\/}}  \bibinfo{volume}{21}
  (\bibinfo{date}{01} \bibinfo{year}{2021}).
\newblock
\showDOI{%
\url{https://doi.org/10.3390/s21030762}}


\bibitem[\protect\citeauthoryear{Casey, Guimond, and Hu}{Casey
  et~al\mbox{.}}{2007}]%
        {single1}
\bibfield{author}{\bibinfo{person}{Thomas Casey}, \bibinfo{person}{Brian
  Guimond}, {and} \bibinfo{person}{James Hu}.} \bibinfo{year}{2007}\natexlab{}.
\newblock \showarticletitle{Underwater Vehicle Positioning Based on Time of
  Arrival Measurements from a Single Beacon}. In \bibinfo{booktitle}{{\em
  OCEANS 2007}}. \bibinfo{pages}{1--8}.
\newblock
\showDOI{%
\url{https://doi.org/10.1109/OCEANS.2007.4449186}}


\bibitem[\protect\citeauthoryear{Chan, Raju, Nandakumar, Bly, and
  Gollakota}{Chan et~al\mbox{.}}{2019a}]%
        {infection}
\bibfield{author}{\bibinfo{person}{Justin Chan}, \bibinfo{person}{Sharat Raju},
  \bibinfo{person}{Rajalakshmi Nandakumar}, \bibinfo{person}{Randall Bly},
  {and} \bibinfo{person}{Shyamnath Gollakota}.}
  \bibinfo{year}{2019}\natexlab{a}.
\newblock \showarticletitle{Detecting middle ear fluid using smartphones}.
\newblock \bibinfo{journal}{{\em Science Translational Medicine\/}}
  \bibinfo{volume}{11} (\bibinfo{date}{05} \bibinfo{year}{2019}),
  \bibinfo{pages}{eaav1102}.
\newblock
\showDOI{%
\url{https://doi.org/10.1126/scitranslmed.aav1102}}


\bibitem[\protect\citeauthoryear{Chan, Rea, Gollakota, and Sunshine}{Chan
  et~al\mbox{.}}{2019b}]%
        {attack}
\bibfield{author}{\bibinfo{person}{Justin Chan}, \bibinfo{person}{Thomas Rea},
  \bibinfo{person}{Shyamnath Gollakota}, {and} \bibinfo{person}{Jacob
  Sunshine}.} \bibinfo{year}{2019}\natexlab{b}.
\newblock \showarticletitle{Contactless cardiac arrest detection using smart
  devices}.
\newblock \bibinfo{journal}{{\em npj Digital Medicine\/}}  \bibinfo{volume}{2}
  (\bibinfo{date}{06} \bibinfo{year}{2019}), \bibinfo{pages}{52}.
\newblock
\showDOI{%
\url{https://doi.org/10.1038/s41746-019-0128-7}}


\bibitem[\protect\citeauthoryear{Chen, Chan, and Gollakota}{Chen
  et~al\mbox{.}}{2022}]%
        {sigcomm22}
\bibfield{author}{\bibinfo{person}{Tuochao Chen}, \bibinfo{person}{Justin
  Chan}, {and} \bibinfo{person}{Shyamnath Gollakota}.}
  \bibinfo{year}{2022}\natexlab{}.
\newblock \showarticletitle{Underwater Messaging Using Mobile Devices}. In
  \bibinfo{booktitle}{{\em Proceedings of the ACM SIGCOMM 2022 Conference}}
  {\em (\bibinfo{series}{SIGCOMM '22})}. \bibinfo{publisher}{Association for
  Computing Machinery}, \bibinfo{address}{New York, NY, USA},
  \bibinfo{pages}{545–559}.
\newblock
\showISBNx{9781450394208}
\showDOI{%
\url{https://doi.org/10.1145/3544216.3544258}}


\bibitem[\protect\citeauthoryear{Cheng, Shu, Liang, and Du}{Cheng
  et~al\mbox{.}}{2008}]%
        {reverse1}
\bibfield{author}{\bibinfo{person}{Xiuzhen Cheng}, \bibinfo{person}{Haining
  Shu}, \bibinfo{person}{Qilian Liang}, {and} \bibinfo{person}{David Hung-Chang
  Du}.} \bibinfo{year}{2008}\natexlab{}.
\newblock \showarticletitle{Silent Positioning in Underwater Acoustic Sensor
  Networks}.
\newblock \bibinfo{journal}{{\em IEEE Transactions on Vehicular Technology\/}}
  \bibinfo{volume}{57}, \bibinfo{number}{3} (\bibinfo{year}{2008}),
  \bibinfo{pages}{1756--1766}.
\newblock
\showDOI{%
\url{https://doi.org/10.1109/TVT.2007.912142}}


\bibitem[\protect\citeauthoryear{Curtis, Banavar, Zhang, Spanias, and
  Weber}{Curtis et~al\mbox{.}}{2014}]%
        {curtis2014android}
\bibfield{author}{\bibinfo{person}{Paul Curtis}, \bibinfo{person}{Mahesh~K
  Banavar}, \bibinfo{person}{Sai Zhang}, \bibinfo{person}{Andreas Spanias},
  {and} \bibinfo{person}{Vitor Weber}.} \bibinfo{year}{2014}\natexlab{}.
\newblock \showarticletitle{Android acoustic ranging}. In
  \bibinfo{booktitle}{{\em IISA 2014, The 5th International Conference on
  Information, Intelligence, Systems and Applications}}. IEEE,
  \bibinfo{pages}{118--123}.
\newblock


\bibitem[\protect\citeauthoryear{de~Langis and Sattar}{de~Langis and
  Sattar}{2020}]%
        {underwater-hri1}
\bibfield{author}{\bibinfo{person}{Karin de Langis} {and}
  \bibinfo{person}{Junaed Sattar}.} \bibinfo{year}{2020}\natexlab{}.
\newblock \showarticletitle{Realtime Multi-Diver Tracking and Re-identification
  for Underwater Human-Robot Collaboration}. In \bibinfo{booktitle}{{\em 2020
  IEEE International Conference on Robotics and Automation (ICRA)}}.
  \bibinfo{pages}{11140--11146}.
\newblock
\showDOI{%
\url{https://doi.org/10.1109/ICRA40945.2020.9197308}}


\bibitem[\protect\citeauthoryear{DeMarco, West, and Howard}{DeMarco
  et~al\mbox{.}}{2013}]%
        {hri2}
\bibfield{author}{\bibinfo{person}{Kevin~J. DeMarco},
  \bibinfo{person}{Michael~E. West}, {and} \bibinfo{person}{Ayanna~M. Howard}.}
  \bibinfo{year}{2013}\natexlab{}.
\newblock \showarticletitle{Sonar-Based Detection and Tracking of a Diver for
  Underwater Human-Robot Interaction Scenarios}. In \bibinfo{booktitle}{{\em
  2013 IEEE International Conference on Systems, Man, and Cybernetics}}.
  \bibinfo{pages}{2378--2383}.
\newblock
\showDOI{%
\url{https://doi.org/10.1109/SMC.2013.406}}


\bibitem[\protect\citeauthoryear{Dong, Mi, and Zhou}{Dong
  et~al\mbox{.}}{2016}]%
        {wifi}
\bibfield{author}{\bibinfo{person}{Yuhan Dong}, \bibinfo{person}{Xuelong Mi},
  {and} \bibinfo{person}{Yiqing Zhou}.} \bibinfo{year}{2016}\natexlab{}.
\newblock \showarticletitle{Spatial Channel Model for Underwater Wireless
  Optical Communication Links: [Extended Abstract]}. In
  \bibinfo{booktitle}{{\em Proceedings of the 11th ACM International Conference
  on Underwater Networks \& Systems}} {\em (\bibinfo{series}{WUWNet '16})}.
  \bibinfo{publisher}{Association for Computing Machinery},
  \bibinfo{address}{New York, NY, USA}, Article \bibinfo{articleno}{18},
  \bibinfo{numpages}{2}~pages.
\newblock
\showISBNx{9781450346375}
\showDOI{%
\url{https://doi.org/10.1145/2999504.3001113}}


\bibitem[\protect\citeauthoryear{Erd{\'e}lyi, Le, Bhattacharjee, Druschel, and
  Ono}{Erd{\'e}lyi et~al\mbox{.}}{2018}]%
        {erdelyi2018sonoloc}
\bibfield{author}{\bibinfo{person}{Viktor Erd{\'e}lyi},
  \bibinfo{person}{Trung-Kien Le}, \bibinfo{person}{Bobby Bhattacharjee},
  \bibinfo{person}{Peter Druschel}, {and} \bibinfo{person}{Nobutaka Ono}.}
  \bibinfo{year}{2018}\natexlab{}.
\newblock \showarticletitle{Sonoloc: Scalable positioning of commodity mobile
  devices}.
\newblock  (\bibinfo{year}{2018}).
\newblock


\bibitem[\protect\citeauthoryear{Flores, Hossein~Motlagh, Zuniga, Liyanage,
  Passananti, Tarkoma, Youssef, and Nurmi}{Flores et~al\mbox{.}}{2020a}]%
        {sensing2}
\bibfield{author}{\bibinfo{person}{Huber Flores}, \bibinfo{person}{Naser
  Hossein~Motlagh}, \bibinfo{person}{Agustin Zuniga}, \bibinfo{person}{Mohan
  Liyanage}, \bibinfo{person}{Monica Passananti}, \bibinfo{person}{Sasu
  Tarkoma}, \bibinfo{person}{Moustafa Youssef}, {and} \bibinfo{person}{Petteri
  Nurmi}.} \bibinfo{year}{2020}\natexlab{a}.
\newblock \bibinfo{title}{Toward Large-Scale Autonomous Monitoring and Sensing
  of Underwater Pollutants}.
\newblock   (\bibinfo{date}{05} \bibinfo{year}{2020}).
\newblock


\bibitem[\protect\citeauthoryear{Flores, Zuniga, Motlagh, Liyanage, Passananti,
  Tarkoma, Youssef, and Nurmi}{Flores et~al\mbox{.}}{2020b}]%
        {sensing1}
\bibfield{author}{\bibinfo{person}{Huber Flores}, \bibinfo{person}{Agustin
  Zuniga}, \bibinfo{person}{Naser~Hossein Motlagh}, \bibinfo{person}{Mohan
  Liyanage}, \bibinfo{person}{Monica Passananti}, \bibinfo{person}{Sasu
  Tarkoma}, \bibinfo{person}{Moustafa Youssef}, {and} \bibinfo{person}{Petteri
  Nurmi}.} \bibinfo{year}{2020}\natexlab{b}.
\newblock \showarticletitle{PENGUIN: Aquatic Plastic Pollution Sensing Using
  AUVs}. In \bibinfo{booktitle}{{\em Proceedings of the 6th ACM Workshop on
  Micro Aerial Vehicle Networks, Systems, and Applications}} {\em
  (\bibinfo{series}{DroNet '20})}. \bibinfo{publisher}{Association for
  Computing Machinery}, \bibinfo{address}{New York, NY, USA}, Article
  \bibinfo{articleno}{5}, \bibinfo{numpages}{6}~pages.
\newblock
\showISBNx{9781450380102}
\showDOI{%
\url{https://doi.org/10.1145/3396864.3399704}}


\bibitem[\protect\citeauthoryear{Foresti and Gentili}{Foresti and
  Gentili}{2000}]%
        {sensing3}
\bibfield{author}{\bibinfo{person}{G.L. Foresti} {and}
  \bibinfo{person}{Stefania Gentili}.} \bibinfo{year}{2000}\natexlab{}.
\newblock \showarticletitle{A Vision Based System for Object Detection in
  Underwater Images.}
\newblock \bibinfo{journal}{{\em IJPRAI\/}}  \bibinfo{volume}{14}
  (\bibinfo{date}{03} \bibinfo{year}{2000}), \bibinfo{pages}{167--188}.
\newblock
\showDOI{%
\url{https://doi.org/10.1142/S021800140000012X}}


\bibitem[\protect\citeauthoryear{Ghaffarivardavagh, Afzal, Rodriguez, and
  Adib}{Ghaffarivardavagh et~al\mbox{.}}{2020}]%
        {backscatterlocalization}
\bibfield{author}{\bibinfo{person}{Reza Ghaffarivardavagh},
  \bibinfo{person}{Sayed~Saad Afzal}, \bibinfo{person}{Osvy Rodriguez}, {and}
  \bibinfo{person}{Fadel Adib}.} \bibinfo{year}{2020}\natexlab{}.
\newblock \showarticletitle{Underwater Backscatter Localization: Toward a
  Battery-Free Underwater GPS}. In \bibinfo{booktitle}{{\em Proceedings of the
  19th ACM Workshop on Hot Topics in Networks}} {\em (\bibinfo{series}{HotNets
  '20})}. \bibinfo{publisher}{Association for Computing Machinery},
  \bibinfo{address}{New York, NY, USA}, \bibinfo{pages}{125–131}.
\newblock
\showISBNx{9781450381451}
\showDOI{%
\url{https://doi.org/10.1145/3422604.3425950}}


\bibitem[\protect\citeauthoryear{Guggenberger, Lux, and
  B{\"o}sz{\"o}rmenyi}{Guggenberger et~al\mbox{.}}{2015}]%
        {guggenberger2015analysis}
\bibfield{author}{\bibinfo{person}{Mario Guggenberger},
  \bibinfo{person}{Mathias Lux}, {and} \bibinfo{person}{Laszlo
  B{\"o}sz{\"o}rmenyi}.} \bibinfo{year}{2015}\natexlab{}.
\newblock \showarticletitle{An analysis of time drift in hand-held recording
  devices}. In \bibinfo{booktitle}{{\em International Conference on Multimedia
  Modeling}}. Springer, \bibinfo{pages}{203--213}.
\newblock


\bibitem[\protect\citeauthoryear{Jang and Adib}{Jang and Adib}{2019}]%
        {waterbackscatter}
\bibfield{author}{\bibinfo{person}{Junsu Jang} {and} \bibinfo{person}{Fadel
  Adib}.} \bibinfo{year}{2019}\natexlab{}.
\newblock \showarticletitle{Underwater Backscatter Networking}. In
  \bibinfo{booktitle}{{\em Proceedings of the ACM Special Interest Group on
  Data Communication}} {\em (\bibinfo{series}{SIGCOMM '19})}.
  \bibinfo{publisher}{Association for Computing Machinery},
  \bibinfo{address}{New York, NY, USA}, \bibinfo{pages}{187–199}.
\newblock
\showISBNx{9781450359566}
\showDOI{%
\url{https://doi.org/10.1145/3341302.3342091}}


\bibitem[\protect\citeauthoryear{Jin, Holz, and Hornb{\ae}k}{Jin
  et~al\mbox{.}}{2015}]%
        {tracko}
\bibfield{author}{\bibinfo{person}{Haojian Jin}, \bibinfo{person}{Christian
  Holz}, {and} \bibinfo{person}{Kasper Hornb{\ae}k}.}
  \bibinfo{year}{2015}\natexlab{}.
\newblock \showarticletitle{Tracko: Ad-hoc mobile 3d tracking using bluetooth
  low energy and inaudible signals for cross-device interaction}. In
  \bibinfo{booktitle}{{\em Proceedings of the 28th Annual ACM Symposium on User
  Interface Software \& Technology}}. ACM, \bibinfo{pages}{147--156}.
\newblock


\bibitem[\protect\citeauthoryear{Kewen et~al\mbox{.}}{Kewen
  et~al\mbox{.}}{2010}]%
        {kewen2010research}
\bibfield{author}{\bibinfo{person}{Liu Kewen} {et~al\mbox{.}}}
  \bibinfo{year}{2010}\natexlab{}.
\newblock \showarticletitle{Research of MMSE and LS channel estimation in OFDM
  systems}. In \bibinfo{booktitle}{{\em The 2nd international conference on
  information science and engineering}}. IEEE, \bibinfo{pages}{2308--2311}.
\newblock


\bibitem[\protect\citeauthoryear{Kuch, Buttazzo, Azzopardi, Sayer, and
  Sieber}{Kuch et~al\mbox{.}}{2012}]%
        {gpsdiver1}
\bibfield{author}{\bibinfo{person}{Benjamin Kuch}, \bibinfo{person}{Giorgio
  Buttazzo}, \bibinfo{person}{Elaine Azzopardi}, \bibinfo{person}{Martin
  Sayer}, {and} \bibinfo{person}{Arne Sieber}.}
  \bibinfo{year}{2012}\natexlab{}.
\newblock \showarticletitle{GPS diving computer for underwater tracking and
  mapping}.
\newblock \bibinfo{journal}{{\em Underwater Technology The International
  Journal of the Society for Underwater\/}}  \bibinfo{volume}{189}
  (\bibinfo{date}{07} \bibinfo{year}{2012}), \bibinfo{pages}{189--194}.
\newblock
\showDOI{%
\url{https://doi.org/10.3723/ut.30.189}}


\bibitem[\protect\citeauthoryear{Kuperman and Roux}{Kuperman and Roux}{2007}]%
        {kuperman2007underwater}
\bibfield{author}{\bibinfo{person}{William~A Kuperman} {and}
  \bibinfo{person}{Philippe Roux}.} \bibinfo{year}{2007}\natexlab{}.
\newblock \bibinfo{title}{Underwater acoustics}.
\newblock   (\bibinfo{year}{2007}), \bibinfo{numpages}{149--204}~pages.
\newblock


\bibitem[\protect\citeauthoryear{Lazik, Rajagopal, Shih, Sinopoli, and
  Rowe}{Lazik et~al\mbox{.}}{2015}]%
        {alps}
\bibfield{author}{\bibinfo{person}{Patrick Lazik}, \bibinfo{person}{Niranjini
  Rajagopal}, \bibinfo{person}{Oliver Shih}, \bibinfo{person}{Bruno Sinopoli},
  {and} \bibinfo{person}{Anthony Rowe}.} \bibinfo{year}{2015}\natexlab{}.
\newblock \showarticletitle{ALPS: A bluetooth and ultrasound platform for
  mapping and localization}. In \bibinfo{booktitle}{{\em Proceedings of the
  13th ACM conference on embedded networked sensor systems}}. ACM,
  \bibinfo{pages}{73--84}.
\newblock


\bibitem[\protect\citeauthoryear{Lin, Yu, Xiong, Zhang, Wang, Wu, and Luo}{Lin
  et~al\mbox{.}}{2021}]%
        {shrimp}
\bibfield{author}{\bibinfo{person}{Chi Lin}, \bibinfo{person}{Yongda Yu},
  \bibinfo{person}{Jie Xiong}, \bibinfo{person}{Yichuan Zhang},
  \bibinfo{person}{Lei Wang}, \bibinfo{person}{Guowei Wu}, {and}
  \bibinfo{person}{Zhongxuan Luo}.} \bibinfo{year}{2021}\natexlab{}.
\newblock \showarticletitle{Shrimp: A Robust Underwater Visible Light
  Communication System}. In \bibinfo{booktitle}{{\em Proceedings of the 27th
  Annual International Conference on Mobile Computing and Networking}}.
  \bibinfo{publisher}{Association for Computing Machinery},
  \bibinfo{address}{New York, NY, USA}, \bibinfo{pages}{134–146}.
\newblock
\showISBNx{9781450383424}
\showURL{%
\url{https://doi-org.offcampus.lib.washington.edu/10.1145/3447993.3448616}}


\bibitem[\protect\citeauthoryear{Luo, Zhao, Guo, Liu, Chen, and Ni}{Luo
  et~al\mbox{.}}{2008}]%
        {directional1}
\bibfield{author}{\bibinfo{person}{Hanjiang Luo}, \bibinfo{person}{Yiyang
  Zhao}, \bibinfo{person}{Zhongwen Guo}, \bibinfo{person}{Siyuan Liu},
  \bibinfo{person}{Pengpeng Chen}, {and} \bibinfo{person}{Lionel~M. Ni}.}
  \bibinfo{year}{2008}\natexlab{}.
\newblock \showarticletitle{UDB: Using Directional Beacons for Localization in
  Underwater Sensor Networks}. In \bibinfo{booktitle}{{\em 2008 14th IEEE
  International Conference on Parallel and Distributed Systems}}.
  \bibinfo{pages}{551--558}.
\newblock
\showDOI{%
\url{https://doi.org/10.1109/ICPADS.2008.31}}


\bibitem[\protect\citeauthoryear{Maki, Matsuda, Sakamaki, Ura, and Kojima}{Maki
  et~al\mbox{.}}{2013}]%
        {tracking6}
\bibfield{author}{\bibinfo{person}{Tara Maki}, \bibinfo{person}{Takumi
  Matsuda}, \bibinfo{person}{Takashi Sakamaki}, \bibinfo{person}{Tamaki Ura},
  {and} \bibinfo{person}{Junichi Kojima}.} \bibinfo{year}{2013}\natexlab{}.
\newblock \showarticletitle{Navigation Method for Underwater Vehicles Based on
  Mutual Acoustical Positioning With a Single Seafloor Station}.
\newblock \bibinfo{journal}{{\em Oceanic Engineering, IEEE Journal of\/}}
  \bibinfo{volume}{38} (\bibinfo{date}{01} \bibinfo{year}{2013}),
  \bibinfo{pages}{167--177}.
\newblock
\showDOI{%
\url{https://doi.org/10.1109/JOE.2012.2210799}}


\bibitem[\protect\citeauthoryear{Mao, He, and Qiu}{Mao et~al\mbox{.}}{2016}]%
        {mao2016cat}
\bibfield{author}{\bibinfo{person}{Wenguang Mao}, \bibinfo{person}{Jian He},
  {and} \bibinfo{person}{Lili Qiu}.} \bibinfo{year}{2016}\natexlab{}.
\newblock \showarticletitle{CAT: high-precision acoustic motion tracking}. In
  \bibinfo{booktitle}{{\em Proceedings of the 22nd Annual International
  Conference on Mobile Computing and Networking}}. ACM,
  \bibinfo{pages}{69--81}.
\newblock


\bibitem[\protect\citeauthoryear{Mishachandar and Vairamuthu}{Mishachandar and
  Vairamuthu}{2021}]%
        {underwater-cognitive3}
\bibfield{author}{\bibinfo{person}{B Mishachandar} {and} \bibinfo{person}{S
  Vairamuthu}.} \bibinfo{year}{2021}\natexlab{}.
\newblock \showarticletitle{An underwater cognitive acoustic network strategy
  for efficient spectrum utilization}.
\newblock \bibinfo{journal}{{\em Applied Acoustics\/}}  \bibinfo{volume}{175}
  (\bibinfo{year}{2021}), \bibinfo{pages}{107861}.
\newblock
\showISSN{0003-682X}
\showDOI{%
\url{https://doi.org/10.1016/j.apacoust.2020.107861}}


\bibitem[\protect\citeauthoryear{Miskovic, Nad, and Rendulic}{Miskovic
  et~al\mbox{.}}{2015}]%
        {surface1}
\bibfield{author}{\bibinfo{person}{Nikola Miskovic}, \bibinfo{person}{Dula
  Nad}, {and} \bibinfo{person}{Ivor Rendulic}.}
  \bibinfo{year}{2015}\natexlab{}.
\newblock \showarticletitle{Tracking Divers: An Autonomous Marine Surface
  Vehicle to Increase Diver Safety}.
\newblock \bibinfo{journal}{{\em IEEE Robotics Automation Magazine\/}}
  \bibinfo{volume}{22}, \bibinfo{number}{3} (\bibinfo{year}{2015}),
  \bibinfo{pages}{72--84}.
\newblock
\showDOI{%
\url{https://doi.org/10.1109/MRA.2015.2448851}}


\bibitem[\protect\citeauthoryear{Monteiro and Marti}{Monteiro and
  Marti}{2020}]%
        {phonerange}
\bibfield{author}{\bibinfo{person}{Mart{\'i}n Monteiro} {and}
  \bibinfo{person}{Arturo~C. Marti}.} \bibinfo{year}{2020}\natexlab{}.
\newblock \showarticletitle{Using smartphones as hydrophones: two experiments
  in underwater acoustics}.
\newblock \bibinfo{journal}{{\em arXiv: Physics Education\/}}
  (\bibinfo{year}{2020}).
\newblock


\bibitem[\protect\citeauthoryear{Moon, Yu, and Choi}{Moon
  et~al\mbox{.}}{2009}]%
        {tracking8}
\bibfield{author}{\bibinfo{person}{Hyun Moon}, \bibinfo{person}{Chang~Ho Yu},
  {and} \bibinfo{person}{Jae~Weon Choi}.} \bibinfo{year}{2009}\natexlab{}.
\newblock \showarticletitle{Performance of Sensor Localization in Underwater
  Wireless Sensor Networks}.
\newblock
\showDOI{%
\url{https://doi.org/10.13140/2.1.5126.4649}}


\bibitem[\protect\citeauthoryear{Munafo, Sliwka, and Alves}{Munafo
  et~al\mbox{.}}{2015}]%
        {tracking4}
\bibfield{author}{\bibinfo{person}{Andrea Munafo}, \bibinfo{person}{Jan
  Sliwka}, {and} \bibinfo{person}{Joao Alves}.}
  \bibinfo{year}{2015}\natexlab{}.
\newblock \showarticletitle{Dynamic placement of a constellation of surface
  buoys for enhanced underwater positioning}. \bibinfo{pages}{1--6}.
\newblock
\showDOI{%
\url{https://doi.org/10.1109/OCEANS-Genova.2015.7271663}}


\bibitem[\protect\citeauthoryear{Nandakumar, Gollakota, and
  Sunshine}{Nandakumar et~al\mbox{.}}{2019}]%
        {opioid}
\bibfield{author}{\bibinfo{person}{Rajalakshmi Nandakumar},
  \bibinfo{person}{Shyamnath Gollakota}, {and} \bibinfo{person}{Jacob
  Sunshine}.} \bibinfo{year}{2019}\natexlab{}.
\newblock \showarticletitle{Opioid overdose detection using smartphones}.
\newblock \bibinfo{journal}{{\em Science Translational Medicine\/}}
  \bibinfo{volume}{11} (\bibinfo{date}{01} \bibinfo{year}{2019}),
  \bibinfo{pages}{eaau8914}.
\newblock
\showDOI{%
\url{https://doi.org/10.1126/scitranslmed.aau8914}}


\bibitem[\protect\citeauthoryear{Nandakumar, Gollakota, and Watson}{Nandakumar
  et~al\mbox{.}}{2015}]%
        {apneaapp}
\bibfield{author}{\bibinfo{person}{Rajalakshmi Nandakumar},
  \bibinfo{person}{Shyamnath Gollakota}, {and} \bibinfo{person}{Nathaniel
  Watson}.} \bibinfo{year}{2015}\natexlab{}.
\newblock \showarticletitle{Contactless Sleep Apnea Detection on Smartphones}.
  In \bibinfo{booktitle}{{\em Proceedings of the 13th Annual International
  Conference on Mobile Systems, Applications, and Services}} {\em
  (\bibinfo{series}{MobiSys '15})}. \bibinfo{publisher}{Association for
  Computing Machinery}, \bibinfo{address}{New York, NY, USA},
  \bibinfo{pages}{45–57}.
\newblock
\showISBNx{9781450334945}
\showDOI{%
\url{https://doi.org/10.1145/2742647.2742674}}


\bibitem[\protect\citeauthoryear{Nandakumar, Iyer, Tan, and
  Gollakota}{Nandakumar et~al\mbox{.}}{2016}]%
        {fingerio}
\bibfield{author}{\bibinfo{person}{Rajalakshmi Nandakumar},
  \bibinfo{person}{Vikram Iyer}, \bibinfo{person}{Desney Tan}, {and}
  \bibinfo{person}{Shyamnath Gollakota}.} \bibinfo{year}{2016}\natexlab{}.
\newblock \showarticletitle{Fingerio: Using active sonar for fine-grained
  finger tracking}. In \bibinfo{booktitle}{{\em Proceedings of the 2016 CHI
  Conference on Human Factors in Computing Systems}}. ACM,
  \bibinfo{pages}{1515--1525}.
\newblock


\bibitem[\protect\citeauthoryear{Nasir, Durrani, and Kennedy}{Nasir
  et~al\mbox{.}}{2010}]%
        {nasir2010performance}
\bibfield{author}{\bibinfo{person}{Ali~A Nasir}, \bibinfo{person}{Salman
  Durrani}, {and} \bibinfo{person}{Rodney~A Kennedy}.}
  \bibinfo{year}{2010}\natexlab{}.
\newblock \showarticletitle{Performance of coarse and fine timing
  synchronization in OFDM receivers}. In \bibinfo{booktitle}{{\em 2010 2nd
  International Conference on Future Computer and Communication}},
  Vol.~\bibinfo{volume}{2}. IEEE, \bibinfo{pages}{V2--412}.
\newblock


\bibitem[\protect\citeauthoryear{Nađ, Mandić, and Miskovic}{Nađ
  et~al\mbox{.}}{2020}]%
        {underwatervehicle1}
\bibfield{author}{\bibinfo{person}{Đula Nađ}, \bibinfo{person}{Filip
  Mandić}, {and} \bibinfo{person}{Nikola Miskovic}.}
  \bibinfo{year}{2020}\natexlab{}.
\newblock \showarticletitle{Using Autonomous Underwater Vehicles for Diver
  Tracking and Navigation Aiding}.
\newblock \bibinfo{journal}{{\em Journal of Marine Science and Engineering\/}}
  \bibinfo{volume}{8} (\bibinfo{date}{06} \bibinfo{year}{2020}),
  \bibinfo{pages}{413}.
\newblock
\showDOI{%
\url{https://doi.org/10.3390/jmse8060413}}


\bibitem[\protect\citeauthoryear{Noh, Torres, and Gerla}{Noh
  et~al\mbox{.}}{2015}]%
        {usrp2}
\bibfield{author}{\bibinfo{person}{Youngtae Noh}, \bibinfo{person}{Dustin
  Torres}, {and} \bibinfo{person}{Mario Gerla}.}
  \bibinfo{year}{2015}\natexlab{}.
\newblock \showarticletitle{Software-defined underwater acoustic networking
  platform and its applications}.
\newblock \bibinfo{journal}{{\em Ad Hoc Networks\/}}  \bibinfo{volume}{34}
  (\bibinfo{date}{01} \bibinfo{year}{2015}).
\newblock
\showDOI{%
\url{https://doi.org/10.1016/j.adhoc.2015.01.010}}


\bibitem[\protect\citeauthoryear{ongoing}{ongoing}{2022}]%
        {submission}
\bibfield{author}{\bibinfo{person}{ongoing}.} \bibinfo{year}{2022}\natexlab{}.
\newblock \showarticletitle{Underwater Messaging Using Mobile Devices}.
\newblock \bibinfo{journal}{{\em under submission\/}} (\bibinfo{year}{2022}).
\newblock


\bibitem[\protect\citeauthoryear{Peng, Shen, Zhang, Li, and Tan}{Peng
  et~al\mbox{.}}{2007}]%
        {peng2007beepbeep}
\bibfield{author}{\bibinfo{person}{Chunyi Peng}, \bibinfo{person}{Guobin Shen},
  \bibinfo{person}{Yongguang Zhang}, \bibinfo{person}{Yanlin Li}, {and}
  \bibinfo{person}{Kun Tan}.} \bibinfo{year}{2007}\natexlab{}.
\newblock \showarticletitle{Beepbeep: a high accuracy acoustic ranging system
  using cots mobile devices}. In \bibinfo{booktitle}{{\em Proceedings of the
  5th international conference on Embedded networked sensor systems}}. ACM,
  \bibinfo{pages}{1--14}.
\newblock


\bibitem[\protect\citeauthoryear{Pottinger}{Pottinger}{2012}]%
        {loc_ml2}
\bibfield{author}{\bibinfo{person}{Mark~Gerard Pottinger}.}
  \bibinfo{year}{2012}\natexlab{}.
\newblock \showarticletitle{LOCALISATION OF UNDERWATER SENSOR NODES IN CONFINED
  SPACES}.
\newblock


\bibitem[\protect\citeauthoryear{Rauchenstein, Vishnu, Li, and
  Deng}{Rauchenstein et~al\mbox{.}}{2018}]%
        {loc_ml1}
\bibfield{author}{\bibinfo{person}{Lynn~T. Rauchenstein},
  \bibinfo{person}{Abhinav Vishnu}, \bibinfo{person}{Xinya Li}, {and}
  \bibinfo{person}{Zhiqun Deng}.} \bibinfo{year}{2018}\natexlab{}.
\newblock \showarticletitle{Improving Underwater Localization Accuracy with
  Machine Learning}.
\newblock \bibinfo{journal}{{\em Review of Scientific Instruments\/}}
  \bibinfo{volume}{89}, \bibinfo{number}{7} (\bibinfo{date}{7}
  \bibinfo{year}{2018}).
\newblock
\showDOI{%
\url{https://doi.org/10.1063/1.5012687}}


\bibitem[\protect\citeauthoryear{Restuccia, Demirors, and Melodia}{Restuccia
  et~al\mbox{.}}{2017}]%
        {isonar}
\bibfield{author}{\bibinfo{person}{Francesco Restuccia},
  \bibinfo{person}{Emrecan Demirors}, {and} \bibinfo{person}{Tommaso Melodia}.}
  \bibinfo{year}{2017}\natexlab{}.
\newblock \showarticletitle{ISonar: Software-Defined Underwater Acoustic
  Networking for Amphibious Smartphones}. In \bibinfo{booktitle}{{\em
  Proceedings of the International Conference on Underwater Networks \&
  Systems}} {\em (\bibinfo{series}{WUWNET'17})}.
  \bibinfo{publisher}{Association for Computing Machinery},
  \bibinfo{address}{New York, NY, USA}, Article \bibinfo{articleno}{15},
  \bibinfo{numpages}{9}~pages.
\newblock
\showISBNx{9781450355612}
\showDOI{%
\url{https://doi.org/10.1145/3148675.3148710}}


\bibitem[\protect\citeauthoryear{Sesia, Toufik, and Baker}{Sesia
  et~al\mbox{.}}{2011}]%
        {sesia2011lte}
\bibfield{author}{\bibinfo{person}{Stefania Sesia}, \bibinfo{person}{Issam
  Toufik}, {and} \bibinfo{person}{Matthew Baker}.}
  \bibinfo{year}{2011}\natexlab{}.
\newblock \bibinfo{booktitle}{{\em LTE-the UMTS long term evolution: from
  theory to practice}}.
\newblock \bibinfo{publisher}{John Wiley \& Sons}.
\newblock


\bibitem[\protect\citeauthoryear{Sklivanitis, Demirors, Batalama, Melodia, and
  Pados}{Sklivanitis et~al\mbox{.}}{2014}]%
        {cog-usrp}
\bibfield{author}{\bibinfo{person}{George Sklivanitis},
  \bibinfo{person}{Emrecan Demirors}, \bibinfo{person}{Stella~N. Batalama},
  \bibinfo{person}{Tommaso Melodia}, {and} \bibinfo{person}{Dimitris~A.
  Pados}.} \bibinfo{year}{2014}\natexlab{}.
\newblock \showarticletitle{Receiver configuration and testbed development for
  underwater cognitive channelization}. In \bibinfo{booktitle}{{\em 2014 48th
  Asilomar Conference on Signals, Systems and Computers}}.
  \bibinfo{pages}{1594--1598}.
\newblock
\showDOI{%
\url{https://doi.org/10.1109/ACSSC.2014.7094734}}


\bibitem[\protect\citeauthoryear{Srinivasan, Rajesh, Ramesh, Babu, Abraham,
  Raphael, Ramadass, and Atmanand}{Srinivasan et~al\mbox{.}}{2007}]%
        {tracking5}
\bibfield{author}{\bibinfo{person}{Ramji Srinivasan}, \bibinfo{person}{S.
  Rajesh}, \bibinfo{person}{N.R. Ramesh}, \bibinfo{person}{S.M. Babu},
  \bibinfo{person}{Raju Abraham}, \bibinfo{person}{Deepak Raphael},
  \bibinfo{person}{G.A. Ramadass}, {and} \bibinfo{person}{Ma Atmanand}.}
  \bibinfo{year}{2007}\natexlab{}.
\newblock \showarticletitle{Design and testing of Control and Positioning
  System for Underwater mining Machine}. \bibinfo{pages}{1 -- 5}.
\newblock
\showISBNx{978-0933957-35-0}
\showDOI{%
\url{https://doi.org/10.1109/OCEANS.2007.4449146}}


\bibitem[\protect\citeauthoryear{Sur, Wei, and Zhang}{Sur
  et~al\mbox{.}}{2014}]%
        {sur2014autodirective}
\bibfield{author}{\bibinfo{person}{Sanjib Sur}, \bibinfo{person}{Teng Wei},
  {and} \bibinfo{person}{Xinyu Zhang}.} \bibinfo{year}{2014}\natexlab{}.
\newblock \showarticletitle{Autodirective audio capturing through a
  synchronized smartphone array}. In \bibinfo{booktitle}{{\em Proceedings of
  the 12th annual international conference on Mobile systems, applications, and
  services}}. \bibinfo{pages}{28--41}.
\newblock


\bibitem[\protect\citeauthoryear{Tan, Diamant, Seah, and Waldmeyer}{Tan
  et~al\mbox{.}}{2011}]%
        {tracking7}
\bibfield{author}{\bibinfo{person}{Hwee Tan}, \bibinfo{person}{Roee Diamant},
  \bibinfo{person}{Winston Seah}, {and} \bibinfo{person}{Marc Waldmeyer}.}
  \bibinfo{year}{2011}\natexlab{}.
\newblock \showarticletitle{A Survey of Techniques and Challenges in Underwater
  Localization}.
\newblock \bibinfo{journal}{{\em Ocean Engineering - OCEAN ENG\/}}
  \bibinfo{volume}{38} (\bibinfo{date}{10} \bibinfo{year}{2011}),
  \bibinfo{pages}{1663--1676}.
\newblock
\showDOI{%
\url{https://doi.org/10.1016/j.oceaneng.2011.07.017}}


\bibitem[\protect\citeauthoryear{Tan, Gabor, Eu, and Seah}{Tan
  et~al\mbox{.}}{2010}]%
        {tracking1}
\bibfield{author}{\bibinfo{person}{H.-P. Tan}, \bibinfo{person}{A.~F. Gabor},
  \bibinfo{person}{Z.~A. Eu}, {and} \bibinfo{person}{W.~K.~G. Seah}.}
  \bibinfo{year}{2010}\natexlab{}.
\newblock \showarticletitle{A Wide Coverage Positioning System (WPS) for
  Underwater Localization}. In \bibinfo{booktitle}{{\em 2010 IEEE International
  Conference on Communications}}. \bibinfo{pages}{1--5}.
\newblock
\showDOI{%
\url{https://doi.org/10.1109/ICC.2010.5501950}}


\bibitem[\protect\citeauthoryear{Tomczak}{Tomczak}{2011}]%
        {tracking2}
\bibfield{author}{\bibinfo{person}{Arkadiusz Tomczak}.}
  \bibinfo{year}{2011}\natexlab{}.
\newblock \showarticletitle{MODERN METHODS OF UNDERWATER POSITIONING APPLIED IN
  SUBSEA MINING}.
\newblock \bibinfo{journal}{{\em AGH Journals of Mining and Geoengineering\/}}
  \bibinfo{volume}{35} (\bibinfo{date}{01} \bibinfo{year}{2011}),
  \bibinfo{pages}{381--394}.
\newblock


\bibitem[\protect\citeauthoryear{Ullah, Liu, Su, and Kim}{Ullah
  et~al\mbox{.}}{2019}]%
        {ullah2019efficient}
\bibfield{author}{\bibinfo{person}{Inam Ullah}, \bibinfo{person}{Yiming Liu},
  \bibinfo{person}{Xin Su}, {and} \bibinfo{person}{Pankoo Kim}.}
  \bibinfo{year}{2019}\natexlab{}.
\newblock \showarticletitle{Efficient and accurate target localization in
  underwater environment}.
\newblock \bibinfo{journal}{{\em IEEE Access\/}}  \bibinfo{volume}{7}
  (\bibinfo{year}{2019}), \bibinfo{pages}{101415--101426}.
\newblock


\bibitem[\protect\citeauthoryear{Vickery}{Vickery}{1998}]%
        {vickery1998acoustic}
\bibfield{author}{\bibinfo{person}{Keith Vickery}.}
  \bibinfo{year}{1998}\natexlab{}.
\newblock \showarticletitle{Acoustic positioning systems. A practical overview
  of current systems}. In \bibinfo{booktitle}{{\em Proceedings of the 1998
  Workshop on Autonomous Underwater Vehicles (Cat. No. 98CH36290)}}. IEEE,
  \bibinfo{pages}{5--17}.
\newblock


\bibitem[\protect\citeauthoryear{Walls and Gagnepain}{Walls and
  Gagnepain}{1992}]%
        {walls1992environmental}
\bibfield{author}{\bibinfo{person}{Fred~L Walls} {and} \bibinfo{person}{J-J
  Gagnepain}.} \bibinfo{year}{1992}\natexlab{}.
\newblock \showarticletitle{Environmental sensitivities of quartz oscillators}.
\newblock \bibinfo{journal}{{\em IEEE transactions on ultrasonics,
  ferroelectrics, and frequency control\/}} \bibinfo{volume}{39},
  \bibinfo{number}{2} (\bibinfo{year}{1992}), \bibinfo{pages}{241--249}.
\newblock


\bibitem[\protect\citeauthoryear{Wang and Gollakota}{Wang and
  Gollakota}{2019a}]%
        {millisonic}
\bibfield{author}{\bibinfo{person}{Anran Wang} {and} \bibinfo{person}{Shyamnath
  Gollakota}.} \bibinfo{year}{2019}\natexlab{a}.
\newblock \showarticletitle{MilliSonic: Pushing the Limits of Acoustic Motion
  Tracking}. In \bibinfo{booktitle}{{\em Proceedings of the 2019 CHI Conference
  on Human Factors in Computing Systems}} {\em (\bibinfo{series}{CHI '19})}.
  \bibinfo{publisher}{Association for Computing Machinery},
  \bibinfo{address}{New York, NY, USA}, \bibinfo{pages}{1–11}.
\newblock
\showISBNx{9781450359702}
\showDOI{%
\url{https://doi.org/10.1145/3290605.3300248}}


\bibitem[\protect\citeauthoryear{Wang and Gollakota}{Wang and
  Gollakota}{2019b}]%
        {wang2019millisonic}
\bibfield{author}{\bibinfo{person}{Anran Wang} {and} \bibinfo{person}{Shyamnath
  Gollakota}.} \bibinfo{year}{2019}\natexlab{b}.
\newblock \showarticletitle{Millisonic: Pushing the limits of acoustic motion
  tracking}. In \bibinfo{booktitle}{{\em Proceedings of the 2019 CHI Conference
  on Human Factors in Computing Systems}}. \bibinfo{pages}{1--11}.
\newblock


\bibitem[\protect\citeauthoryear{Wang, Nguyen, Sridhar, and Gollakota}{Wang
  et~al\mbox{.}}{2021}]%
        {cardiac}
\bibfield{author}{\bibinfo{person}{Anran Wang}, \bibinfo{person}{Dan Nguyen},
  \bibinfo{person}{Arun Sridhar}, {and} \bibinfo{person}{Shyamnath Gollakota}.}
  \bibinfo{year}{2021}\natexlab{}.
\newblock \showarticletitle{Using smart speakers to contactlessly monitor heart
  rhythms}.
\newblock \bibinfo{journal}{{\em Communications Biology\/}}
  \bibinfo{volume}{4} (\bibinfo{date}{03} \bibinfo{year}{2021}),
  \bibinfo{pages}{319}.
\newblock
\showDOI{%
\url{https://doi.org/10.1038/s42003-021-01824-9}}


\bibitem[\protect\citeauthoryear{Wang, Sunshine, and Gollakota}{Wang
  et~al\mbox{.}}{2019}]%
        {infant}
\bibfield{author}{\bibinfo{person}{Anran Wang}, \bibinfo{person}{Jacob~E.
  Sunshine}, {and} \bibinfo{person}{Shyamnath Gollakota}.}
  \bibinfo{year}{2019}\natexlab{}.
\newblock \showarticletitle{Contactless Infant Monitoring Using White Noise}.
  In \bibinfo{booktitle}{{\em The 25th Annual International Conference on
  Mobile Computing and Networking}} {\em (\bibinfo{series}{MobiCom '19})}.
  \bibinfo{publisher}{Association for Computing Machinery},
  \bibinfo{address}{New York, NY, USA}, Article \bibinfo{articleno}{52},
  \bibinfo{numpages}{16}~pages.
\newblock
\showISBNx{9781450361699}
\showDOI{%
\url{https://doi.org/10.1145/3300061.3345453}}


\bibitem[\protect\citeauthoryear{Wen, Huang, and Zhang}{Wen
  et~al\mbox{.}}{2006}]%
        {wen2006cazac}
\bibfield{author}{\bibinfo{person}{Yang Wen}, \bibinfo{person}{Wei Huang},
  {and} \bibinfo{person}{Zhongpei Zhang}.} \bibinfo{year}{2006}\natexlab{}.
\newblock \showarticletitle{CAZAC sequence and its application in LTE random
  access}. In \bibinfo{booktitle}{{\em 2006 IEEE Information Theory
  Workshop-ITW'06 Chengdu}}. IEEE, \bibinfo{pages}{544--547}.
\newblock


\bibitem[\protect\citeauthoryear{Wilson}{Wilson}{1960}]%
        {wilson1960equation}
\bibfield{author}{\bibinfo{person}{Wayne~D Wilson}.}
  \bibinfo{year}{1960}\natexlab{}.
\newblock \showarticletitle{Equation for the speed of sound in sea water}.
\newblock \bibinfo{journal}{{\em The Journal of the Acoustical Society of
  America\/}} \bibinfo{volume}{32}, \bibinfo{number}{10}
  (\bibinfo{year}{1960}), \bibinfo{pages}{1357--1357}.
\newblock


\bibitem[\protect\citeauthoryear{Xue, Yu, Lyu, and Li}{Xue
  et~al\mbox{.}}{2020}]%
        {xue2020push}
\bibfield{author}{\bibinfo{person}{Hua Xue}, \bibinfo{person}{Jiadi Yu},
  \bibinfo{person}{Feng Lyu}, {and} \bibinfo{person}{Minglu Li}.}
  \bibinfo{year}{2020}\natexlab{}.
\newblock \showarticletitle{Push the limit of multipath profiling using
  commodity WiFi devices with limited bandwidth}.
\newblock \bibinfo{journal}{{\em IEEE Transactions on Vehicular Technology\/}}
  \bibinfo{volume}{69}, \bibinfo{number}{4} (\bibinfo{year}{2020}),
  \bibinfo{pages}{4142--4154}.
\newblock


\bibitem[\protect\citeauthoryear{Yonggang, Jiansheng, Yue, and Li}{Yonggang
  et~al\mbox{.}}{2008}]%
        {underwater-cognitive1}
\bibfield{author}{\bibinfo{person}{Wang Yonggang}, \bibinfo{person}{Tang
  Jiansheng}, \bibinfo{person}{Pan Yue}, {and} \bibinfo{person}{Huangfu Li}.}
  \bibinfo{year}{2008}\natexlab{}.
\newblock \showarticletitle{Underwater communication goes cognitive}. In
  \bibinfo{booktitle}{{\em OCEANS 2008}}. \bibinfo{pages}{1--4}.
\newblock
\showDOI{%
\url{https://doi.org/10.1109/OCEANS.2008.5151898}}


\bibitem[\protect\citeauthoryear{Youssef, Youssef, Rieger, Shankar, and
  Agrawala}{Youssef et~al\mbox{.}}{2006}]%
        {youssef2006pinpoint}
\bibfield{author}{\bibinfo{person}{Moustafa Youssef}, \bibinfo{person}{Adel
  Youssef}, \bibinfo{person}{Chuck Rieger}, \bibinfo{person}{Udaya Shankar},
  {and} \bibinfo{person}{Ashok Agrawala}.} \bibinfo{year}{2006}\natexlab{}.
\newblock \showarticletitle{Pinpoint: An asynchronous time-based location
  determination system}. In \bibinfo{booktitle}{{\em Proceedings of the 4th
  international conference on Mobile systems, applications and services}}.
  \bibinfo{pages}{165--176}.
\newblock


\bibitem[\protect\citeauthoryear{Zhang, Xue, Waghmare, Jain, Pu, Hersek, Lyons,
  Cunefare, Inan, and Abowd}{Zhang et~al\mbox{.}}{2017}]%
        {zhang2017soundtrak}
\bibfield{author}{\bibinfo{person}{Cheng Zhang}, \bibinfo{person}{Qiuyue Xue},
  \bibinfo{person}{Anandghan Waghmare}, \bibinfo{person}{Sumeet Jain},
  \bibinfo{person}{Yiming Pu}, \bibinfo{person}{Sinan Hersek},
  \bibinfo{person}{Kent Lyons}, \bibinfo{person}{Kenneth~A Cunefare},
  \bibinfo{person}{Omer~T Inan}, {and} \bibinfo{person}{Gregory~D Abowd}.}
  \bibinfo{year}{2017}\natexlab{}.
\newblock \showarticletitle{Soundtrak: Continuous 3d tracking of a finger using
  active acoustics}.
\newblock \bibinfo{journal}{{\em Proceedings of the ACM on Interactive, Mobile,
  Wearable and Ubiquitous Technologies\/}} \bibinfo{volume}{1},
  \bibinfo{number}{2} (\bibinfo{year}{2017}), \bibinfo{pages}{30}.
\newblock


\bibitem[\protect\citeauthoryear{Zhang, Huang, Yang, Wang, Chen, You, Huang,
  Xue, and Yu}{Zhang et~al\mbox{.}}{2020}]%
        {zhang2020endophasia}
\bibfield{author}{\bibinfo{person}{Yongzhao Zhang}, \bibinfo{person}{Wei-Hsiang
  Huang}, \bibinfo{person}{Chih-Yun Yang}, \bibinfo{person}{Wen-Ping Wang},
  \bibinfo{person}{Yi-Chao Chen}, \bibinfo{person}{Chuang-Wen You},
  \bibinfo{person}{Da-Yuan Huang}, \bibinfo{person}{Guangtao Xue}, {and}
  \bibinfo{person}{Jiadi Yu}.} \bibinfo{year}{2020}\natexlab{}.
\newblock \showarticletitle{Endophasia: Utilizing acoustic-based imaging for
  issuing contact-free silent speech commands}.
\newblock \bibinfo{journal}{{\em Proceedings of the ACM on Interactive, Mobile,
  Wearable and Ubiquitous Technologies\/}} \bibinfo{volume}{4},
  \bibinfo{number}{1} (\bibinfo{year}{2020}), \bibinfo{pages}{1--26}.
\newblock


\bibitem[\protect\citeauthoryear{Zhang, Wang, Wang, Wang, and Liu}{Zhang
  et~al\mbox{.}}{2018}]%
        {infocom2018}
\bibfield{author}{\bibinfo{person}{Yunting Zhang}, \bibinfo{person}{Jiliang
  Wang}, \bibinfo{person}{Weiyi Wang}, \bibinfo{person}{Zhao Wang}, {and}
  \bibinfo{person}{Yunhao Liu}.} \bibinfo{year}{2018}\natexlab{}.
\newblock \showarticletitle{Vernier: Accurate and Fast Acoustic Motion Tracking
  Using Mobile Devices}. In \bibinfo{booktitle}{{\em INFOCOM}}. IEEE.
\newblock


\bibitem[\protect\citeauthoryear{Zhang, Chu, Chen, and Moscibroda}{Zhang
  et~al\mbox{.}}{2012}]%
        {zhang2012swordfight}
\bibfield{author}{\bibinfo{person}{Zengbin Zhang}, \bibinfo{person}{David Chu},
  \bibinfo{person}{Xiaomeng Chen}, {and} \bibinfo{person}{Thomas Moscibroda}.}
  \bibinfo{year}{2012}\natexlab{}.
\newblock \showarticletitle{Swordfight: Enabling a new class of phone-to-phone
  action games on commodity phones}. In \bibinfo{booktitle}{{\em Proceedings of
  the 10th international conference on Mobile systems, applications, and
  services}}. ACM, \bibinfo{pages}{1--14}.
\newblock


\bibitem[\protect\citeauthoryear{Zhao, Wang, He, and Ding}{Zhao
  et~al\mbox{.}}{2018}]%
        {tracking3}
\bibfield{author}{\bibinfo{person}{Shuang Zhao}, \bibinfo{person}{Zhenjie
  Wang}, \bibinfo{person}{Kaifei He}, {and} \bibinfo{person}{Ning Ding}.}
  \bibinfo{year}{2018}\natexlab{}.
\newblock \showarticletitle{Investigation on underwater positioning stochastic
  model based on acoustic ray incidence angle}.
\newblock \bibinfo{journal}{{\em Applied Ocean Research\/}}
  \bibinfo{volume}{77} (\bibinfo{date}{06} \bibinfo{year}{2018}),
  \bibinfo{pages}{69--77}.
\newblock
\showDOI{%
\url{https://doi.org/10.1016/j.apor.2018.05.011}}


\bibitem[\protect\citeauthoryear{Zhu, Hu, and Li}{Zhu et~al\mbox{.}}{2016a}]%
        {currents1}
\bibfield{author}{\bibinfo{person}{Zhongben Zhu}, \bibinfo{person}{Sau-Lon Hu},
  {and} \bibinfo{person}{Huajun Li}.} \bibinfo{year}{2016}\natexlab{a}.
\newblock \showarticletitle{Effect on Kalman based underwater tracking due to
  ocean current uncertainty}. \bibinfo{pages}{131--137}.
\newblock
\showDOI{%
\url{https://doi.org/10.1109/AUV.2016.7778660}}


\bibitem[\protect\citeauthoryear{Zhu, Hu, and Li}{Zhu et~al\mbox{.}}{2016b}]%
        {soundspeed1}
\bibfield{author}{\bibinfo{person}{Zhongben Zhu}, \bibinfo{person}{Sau-Lon Hu},
  {and} \bibinfo{person}{Huajun Li}.} \bibinfo{year}{2016}\natexlab{b}.
\newblock \showarticletitle{Kalman-based underwater tracking with unknown
  effective sound velocity}. \bibinfo{pages}{1--9}.
\newblock
\showDOI{%
\url{https://doi.org/10.1109/OCEANS.2016.7761086}}


\end{thebibliography}

\end{document}